\documentclass[ oneside, 
                    author={Ruiyong Zhang},
                    degree={MEng},
                     title={EmoTrack: An application to Facilitate User Reflection on Their Online Behaviours},
                  subtitle={To help people develop skills to engage positively with the online world}]{dissertation}

\usepackage{graphicx}
\usepackage{subcaption}
\usepackage{hyperref}
\usepackage{url}
\usepackage{listings}

\begin{document}


\maketitle


\frontmatter




\chapter*{Abstract}

\noindent
With the rapid growth of the internet, all online activities can have both positive and negative effects on human mental health. Online engagement is complex and efforts to regulate online use face challenges in distinguishing between beneficial and harmful content and behaviours. An alternative approach is to help young people develop the skills they need to manage online safety while preserving the benefits of online interactions. This dissertation presents the entire development process and evaluation of an multi-platform application, called EmoTrack that aims to help young people reflect on their online behaviour. It was developed to record their online activities and cultivate strategies for more positive and mindful engagement online. EmoTrack is a personal informatics system, and it is designed to help people track and reflect on their engagement with YouTube videos. The system was evaluated with thirteen participants and it was found that EmoTrack can facilitate them to reflect on their video watching behaviour and the impact on their mood, with reports of different levels of reflections from R0 to R3.

The main contributions of this project are as follows:

\begin{quote}
\noindent
\begin{itemize}

\item I wrote a total of 3000 lines of source code, designing and building the full-stack multi-platform application EmoTrack, a personal informatics system that tracks and automatically categorizes the YouTube videos that users watch and visualize how they impact their mood.

\end{itemize}

Research Contributions:
\begin{itemize}
    \item I evaluated EmoTrack using various approaches with UK university students, with the results showing good usability and acceptability among participants.
    \item The user evaluation found that EmoTrack can help users reflect on their YouTube video watching.
\end{itemize}
\end{quote}


\chapter*{Dedication and Acknowledgements}

\begin{itemize}
    \item I would like to thank my supervisor Dr Jon Bird for his continued support and patient guidance throughout this project.
    \item I would like to thank my family for everything they provided during my life until now.
    \item I would like to thank my friends for their emotional support and engagement in this project.
\end{itemize}



\makedecl



\tableofcontents
\listoffigures
\listoftables


\chapter*{Ethics Statement}

This project fits within the scope of ethics application 0026, as reviewed by my supervisor, Jon Bird.


\chapter*{Supporting Technologies}

\noindent

\begin{quote}
\noindent
\begin{itemize}
\item I used Python to establish all back-end functions, which satisfies all data processing requirements.
\item I built a Python server with Quart RESTful API endpoints and deployed it to Google App Engine, which handles all HTTP requests between the front-end and the back-end.
\item I used the Visual Studio Code IDE to write the code.
\item I used OpenAI's API and Python library to integrate the ChatGPT model to categorize videos. Available from: \url{https://platform.openai.com/docs/api-reference/introduction?lang=python}.
\item I used Flutter to design and build the User Interface of EmoTrack, which provides a multi-platform application and satisfies different requirements from users. Available from: \url{https://flutter.dev/}.
\item I used Firebase Hosting to deploy the Front-end for the Web application version of EmoTrack.
\item I used Firebase Firestore Database for data storage. Available from: \url{https://firebase.google.com/}.
\item I used Firebase Cloud Storage for file storage.
\item I used GitHub to version control the code for my project.
\end{itemize}
\end{quote}


\chapter*{Notation and Acronyms}

\begin{quote}
\noindent
\begin{tabular}{lcl}
API                 &:     & Application Programming Interface                                         \\
AI                 &:     & Artificial Intelligence                                             \\
IDE                &:     & Integrated Development Environment                                            \\
URL    &:     & Uniform Resource Locator                                                          \\
JSON               &:     & JavaScript Object Notation
                       \\
CORS           &:         &  Cross-Origin Resource Sharing
                   \\
UI             &:      & User Interface
                  \\
UX             &:      & User Experience
\end{tabular}
\end{quote}


%

\mainmatter

\chapter{Introduction}
\label{chap:context}

Young people are the most frequent consumers of digital media platforms in the UK, and globally, most of them are smartphone owners who tend to engage in a wide range of activities online. The most recent data indicates YouTube, for instance, a popular social media platform among younger audiences, had an ad reach equivalent to 82.8\% of the UK's total population in early 2024. Similarly, Instagram had 33.1 million users, which was equivalent to 57.1\% of the eligible audience aged 13 and above, indicating a high level of engagement among younger demographics \cite{r3}. Additionally, Ofcom's reports \cite{r37} on media usage highlight that digital engagement among young people continues to evolve, reflecting broader changes in how digital platforms are used.

Several scientific studies \cite{r4, r5}, users and policy makers have acknowledged the well-documented contradiction that internet activity can have both positive and negative effects on mental health. Social media can be considered as a ‘double-edged sword’. Research indicates the advantages of enabling people to communicate their ideas and emotions as well as obtain social support online \cite{r5}. Positive effects also include chances to ask for assistance, get linked with peers, obtain their support, and find the distraction.

However, studies have linked using social media with psychological issues \cite{r38}. Many young people report that they have had a negative online experience that had affected their mental health in recent research \cite{r44}. These kinds of encounters were particularly common in vulnerable groups, such as those who already had mental health issues. A wide range of behaviours, social issues, and health issues have been linked to internet use, including anxiety, depression, body dissatisfaction, self-harm, disordered eating, suicidality, bullying, gambling, and pornography, as well as academic achievement, attention, and concentration \cite{r45, r46}. According to a comprehensive evaluation, there is a slight but statistically significant correlation between using social media and depressed symptoms in children and adolescents. Problematic Facebook usage among adolescents and young adults has been linked to psychological distress \cite{r38}.

The inability to completely distinguish between beneficial and detrimental online information and activities, as well as the propensity to place more emphasis on online content than behaviour, make it difficult to control or restrict the use of the internet. Studies show that because online interactions are so complicated, negative and positive effects commonly coexist \cite{r5}. Furthermore, the influence varies over time across and among people based on mood, external factors, other users’ actions, and the capacity to create and implement personalized safety plans \cite{r47,r48}. Prior research conducted by Dr Lucy Biddle from Bristol Medical School suggests an alternative method: give young people the tools they need to control their online safety and preserve the advantages of the internet \cite{r49}. She discovered that some young people gained metacognition abilities through recording their online behaviours on a paper-based research diary. This led to enhanced digital literacy, more control, and altered responses to online content.

This project will develop an application to enable people to record their online activities and reflect on their online behaviour and its impact on their mood. In particular, it will focus on tracking and reflecting on YouTube video watching.

This project aims to create a brand new beneficial tool that can help people to maintain the advantages of engaging online and minimize any negative effect of using YouTube. Although there are applications on the market that help parents monitor their children's online behaviours and content, there are no apps created for young people that help them reflect on their online behaviour. EmoTrack takes into account the initiative of users who want to improve their online engagement skills and protect their privacy.

This project aims to evaluate the usability and its acceptability among UK University students and investigate whether the system helps them to reflect on their YouTube video watching and its positive and negative impacts.

The main aims and objectives of this project are as follows:
\begin{enumerate}

    \item Design and develop the EmoTrack application that help people record their online behaviours and reflect on how their impact.
    \item Evaluate the usability and acceptability of EmoTrack among UK University students.
    \item Investigate whether EmoTrack facilitates reflection and, if so, what level of reflection.
    \item Collect feedback about EmoTrack and determine how it could be improved in the future.

\end{enumerate}


\chapter{Contextual Background}
\label{chap:contextual}

\noindent
\section{Introduction}
This chapter introduces the influence of social media on young people today and the motivation for this project. It also provides a contextual background for the importance of establishing it with a personal informatics system and facilitates the reflection of users. It provides an overview of the model of the personal informatics system and discusses the barriers encountered during each stage. Along with the importance of self-reflection, which will include all the different perspectives to be evaluated after the whole application is completed. In addition, it discusses relative research on the collection of different human behaviours, mood and their purposes, as well as individual privacy issues.

\section{Social Media}
Social media has been defined as services, websites and associated tools that enable users to produce and share content \cite{r4}. It relies on large groups of people to participate rather than centre-controlled content providers, collect and mix content from various sources. These days, people can access social media on a variety of platforms and devices. Social media thus provides new opportunities for interactions, expectations and activities, including content generated by users, combination and distribution, which is linked to the emergence of ``participatory culture’’ and forms the expectations of children and adolescents as active participants in media and their environment \cite{r47}.

\begin{figure}[h]
    \centering
    \includegraphics[width=0.9\textwidth]{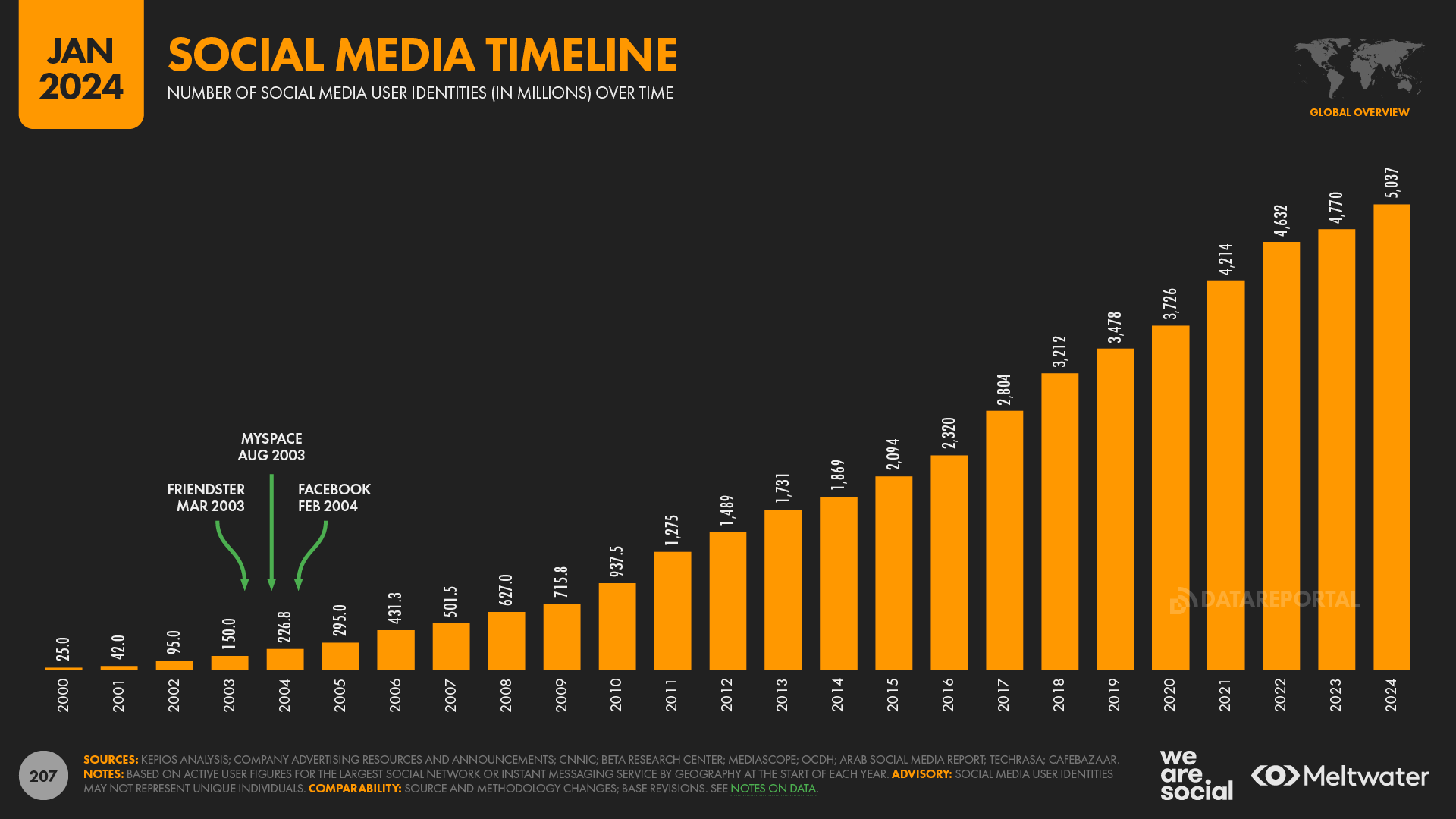}
    \caption{Number of Social Media User Identities Over Time \cite{r3}}
    \label{fig:social-media-1}
\end{figure}

As stated in Figure \ref{fig:social-media-1}, social media users have increased year on year, since the turn of the country. There are 5.35 billion internet users worldwide according to the most recent report (2024) \cite{r3}, which represents more than 66\% of all humans on the earth and has increased by 1.8\% over the last 12 months. As reported by Kepios \cite{r3}, there are more than 5 billion active social media user identities\footnote{note: Social media user identities may not represent unique individuals.} worldwide, which equals 62.3\% of the global population. According to GWI’s latest research \cite{r3}, the ``typical'' social media user spends 2 hours and 23 minutes on average each day on social media nowadays.

\begin{figure}[h]
    \centering
    \begin{minipage}{0.49\textwidth}
        \centering
        \includegraphics[width=\textwidth]{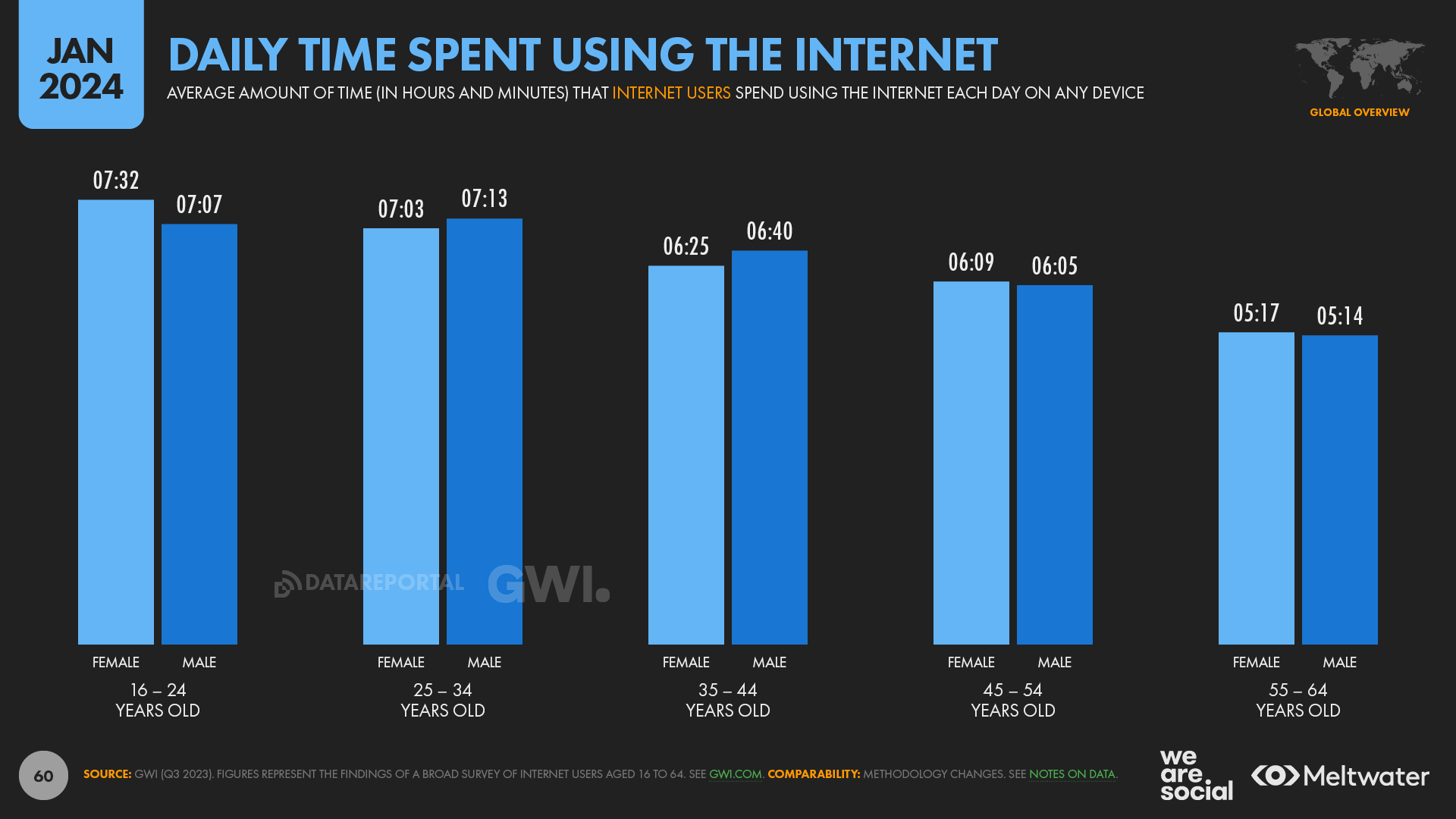}
        \caption{Daily Time Spent Using The Internet \cite{r3}}
        \label{fig:social-media-2}
    \end{minipage}
    \hfill
    \begin{minipage}{0.49\textwidth}
        \centering
        \includegraphics[width=\textwidth]{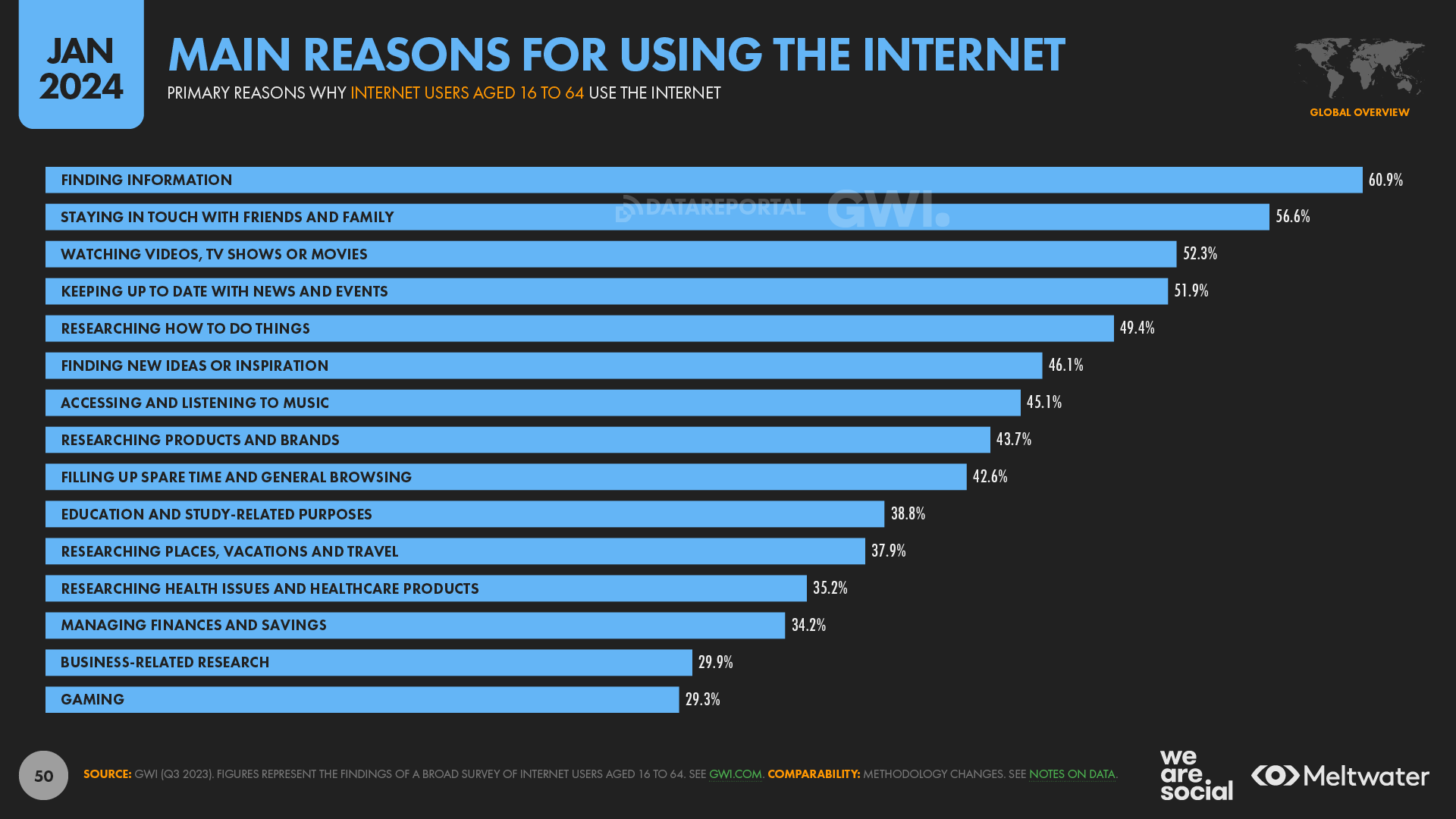}
        \caption{Main Reasons For Using The Internet \cite{r3}}
        \label{fig:social-media-3}
    \end{minipage}
\end{figure}

\begin{figure}[h]
    \centering
    \begin{subfigure}{0.49\textwidth}
        \centering
        \includegraphics[width=\textwidth]{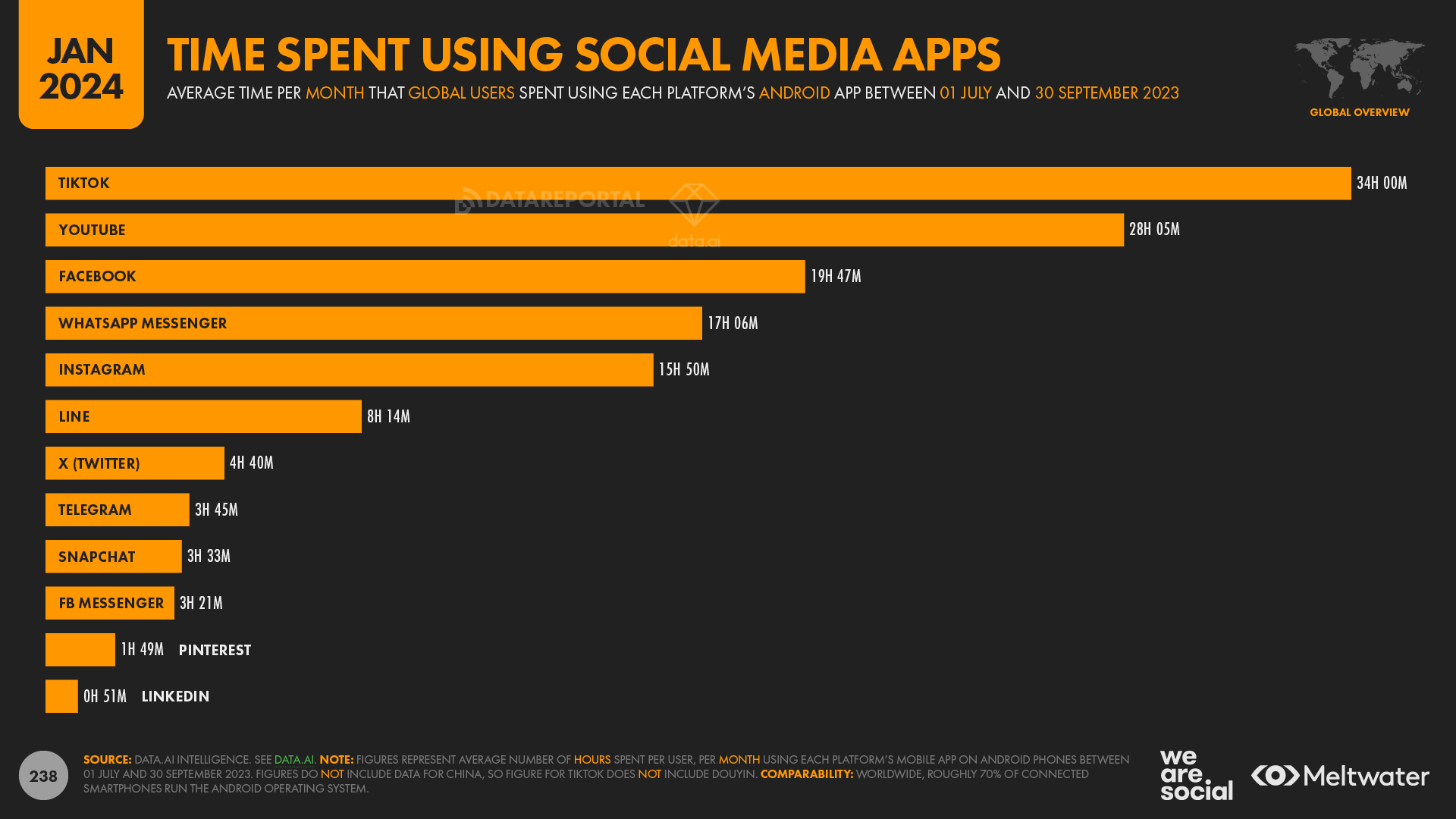}
        \caption{Time Spent Using Social Media Apps \cite{r3}}
        \label{fig:social-media-4}
    \end{subfigure}
    \hspace*{\fill}
    \begin{subfigure}{0.49\textwidth}
        \centering
        \includegraphics[width=\textwidth]{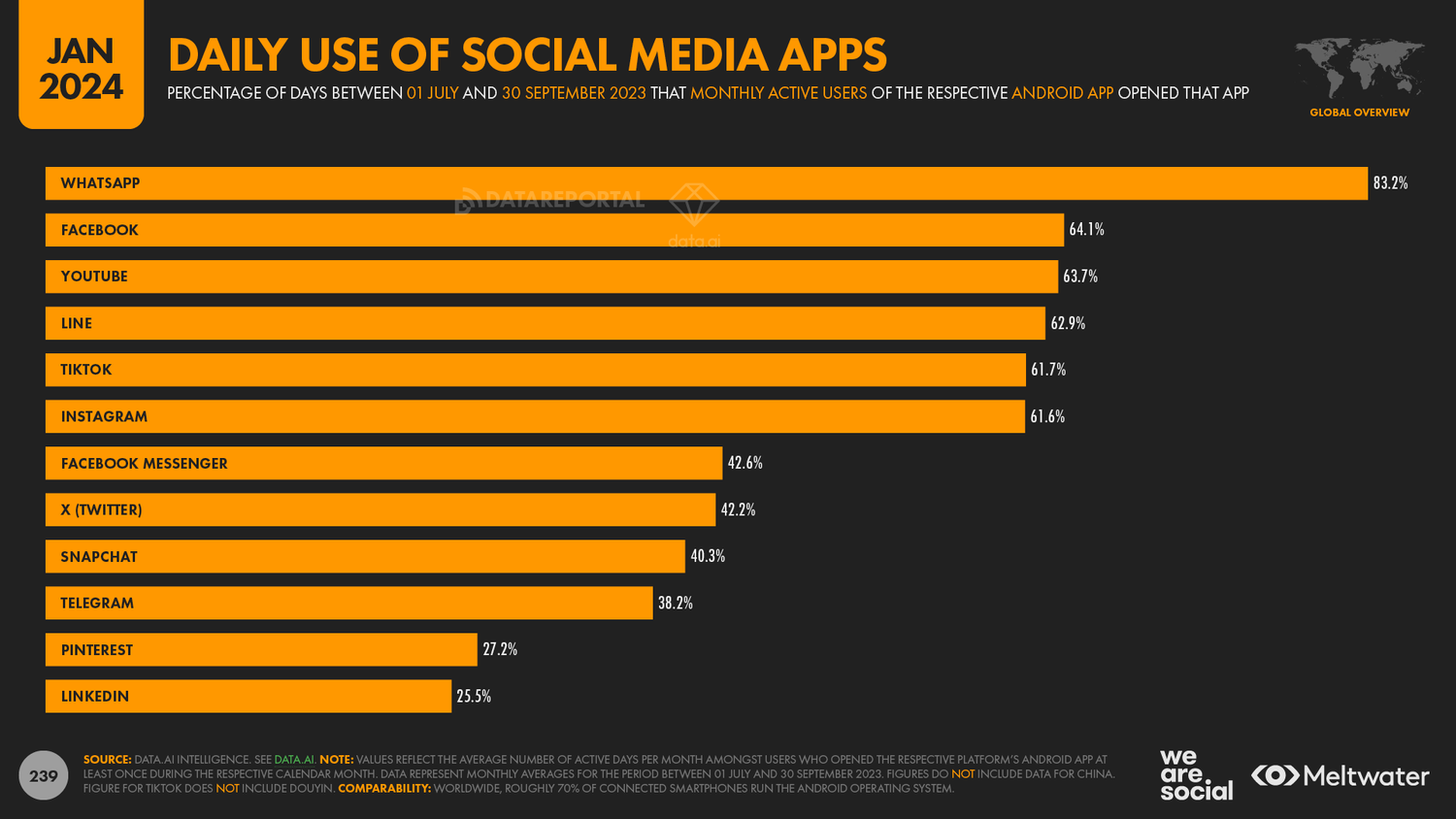}
        \caption{Daily Use of Social Media Apps \cite{r3}}
        \label{fig:social-media-5}
    \end{subfigure}
    \caption{Usage Amount of Social Media \cite{r3}}
    \label{fig:social-media-45}
\end{figure}

It can be seen from Figure \ref{fig:social-media-2} that even people aged range from 55 to 64 spent more than five hours per day using the internet, not to mention other younger people. For example, people in the age range of 16 to 24 and 25 to 34, spent more than seven hours on the internet. It is worth noting that among the main reasons for using the internet, ``Watching videos, TV Shows or Movies'' is in the third place. Compared with the first two, it is more for entertainment, which can explain why TikTok and YouTube have ranked one and two and make up the majority of time spent using social media apps (Figure \ref{fig:social-media-4}) and rank only second to WhatsApp among the daily use of social media apps (Figure \ref{fig:social-media-5}). However, keep in mind that there is a considerable amount of YouTube activities that occurs in the web browsers, even if these are not counted in the total amount of app activities, YouTube continues to be one of the most popular social apps due to its larger user base \cite{r3}.

Social media apps use a range of designed strategies from the technical perspective, including attention-grabbing notification icons, attractive clickbait, and endless auto-playing to maximally occupy users’ time in the attention economy, which takes advantage of psychological weaknesses and works against users’ best interests, making them feel that they have no control over the time they spend on apps \cite{r6}.

\subsection{YouTube}

YouTube has rapidly expanded to become the biggest website for sharing videos since it was founded in 2005. It is used to find, view, and share videos \cite{r8}. The majority of videos are brief and less than 10 minutes, so the next video is played automatically based on its recommendation list if users do nothing, which means YouTube might have some control over what users watch \cite{r8}.

YouTube is a significant example of comprehending attention-capture design strategies, with over 2 billion monthly users around the world \cite{r6}, 70\% of the time spent watching on Google’s huge video platform is drawn in by one of the AI-driven recommendations. When watching on mobile devices, the typical viewing session usually lasts longer than 60 minutes due to the content that recommendation engines present \cite{r7}. It also makes use of a variety of design features, such as notifications, auto-play, and recommendations, that could compromise users’ feeling of control over how they spend their time using YouTube. More than 70\% of YouTube watching time is driven particularly by algorithmic recommendations \cite{r6}. As analyzed for YouTube recommendation, users watch more, the recommendations for new videos increase \cite{r6}.

Behavioural analytics is used by YouTube to improve the user experience. Studying the user's viewing history determines their preferences and shows them only the content they are likely to find interesting \cite{r35}. According to the research on YouTube’s recommendation system \cite{r9}, two neural networks are used to perform filtering, one for candidate video creation and the other for ranking \cite{r10}. Collaborative filtering is looking at what other viewers who are similar to you have viewed and a user’s YouTube watching history, which are two main sources for filtering to generate a list of recommended videos \cite{r10}. The ranking filter then uses an abundant set of attributes that describe the user and video to provide a score for each video based on a ``desired objective function'', which is a function to analyze the relevance between videos depending on various factors \cite{r10}. Users are shown the videos with the highest scores, arranged according to their scores in order \cite{r10}.

Furthermore, the recommended content on YouTube can be biassed towards the platform's advantages rather than the welfare of its viewers. YouTube, for example, can recommend certain channels that pay them. Also, users can receive recommendations from YouTube for videos that might change users’ thoughts or viewpoints about certain issues, political parties, or even musical tastes while inadvertently altering their attitudes or actions \cite{r8}.

YouTube officially has fifteen different categories for videos, for example, music, comedy, education, and so on. To upload videos, users are required to select the category that best fits their videos \cite{r8}. Categories of videos are an important part to be analyzed in my application, EmoTrack.

\subsubsection{YouTube Shorts}
Short video content has been increasingly popular in recent years. Shortly after the success of TikTok, sites such as YouTube debuted their own short-form video content capabilities, which follow a similar format across all platforms. YouTube particularly launched its ``Shorts’’ format in March 2021. Ever since its debut on YouTube, an increasing number of content providers have begun to create content in this format \cite{r43}.

Short video material is a hot topic right now, with many different viewpoints. From one perspective, it offers content producers chances to interact and amuse their followers with brief and visually appealing content. Furthermore, short videos can provide content producers with fresh professional views. However, there are some potentially worrisome characteristics for its users, including reliance, an increase in daytime weariness, and a loss in prospective memory \cite{r43}.

Research showed that YouTube Shorts and YouTube regular videos are not uniformly divided. Shorts were primarily for entertainment, whereas regular videos included a wide range of categories. This suggests that the two sorts of videos coexist on YouTube rather than address the same topics \cite{r43}.

\subsection{Effects of Social Media}

Social media has the greatest impact on teenagers, as they are among the most active users, with 87\% reporting using at least one social platform every day \cite{r4}. Teenagers spend a significant portion of their lives on social media. Despite the complexity of social media behaviours, the majority of research about social media has focused on negative outcomes, offensive content, and online antisocial behaviour. The amount of research on its positive effects is rather low, and it is unclear how commonplace the positive effects are \cite{r4}.

Adolescents are particularly sensitive to the effects of social media, especially during the early stages of adolescence. Biological, psychological, and social changes might make them particularly vulnerable to the potential risks of social media \cite{r5}. According to a longitudinal study of youth data from the United Kingdom, life happiness is associated with the amount of social media use. During one-year observation, more social media use results in lower life happiness, whereas lower use of social media is associated with higher life satisfaction \cite{r5}. 

However, there are some benefits of social media use, the main one being social interaction. Teenagers can locate people who have similar identities and interests, which is relevant for them in establishing new relationships or reestablishing friendships with friends \cite{r5}. When teenagers are under stress and alone, engaging in positive online socialization might help them stay in touch with others in person and develop better social skills, especially for LGBTQ+ adolescents, who might find it difficult and uncomfortable to talk about themselves with other people \cite{r5}. Social media will help them in these situations \cite{r5}.

\section{Motivations}
Since the rapid development of social media has affected many people’s daily lives seriously, especially efficiency regarding study or work, the motivation of this project is to develop an application that can record and help people pay attention to their daily online behaviours, comprehend how social media will affect their mood, and provides the integration of various types of information for users to reflect on. In order to start, this project focuses on YouTube, one of the most popular social media, and utilizes AI, particularly ChatGPT, to perform accurate data processing and better analysis.

Also, this project aims to provide a new perspective on life that people take for granted and assist in their creative reflection.

\section{Personal Informatics and Reflection}
People have understood the value of self-awareness since ancient times. Until now, people are still making an effort to know themselves, which can be achieved through gathering and analyzing personal data, including thoughts, habits and behaviours. This is called ``Personal Informatics'' \cite{r1}.

Personal informatics is an interesting field of study in human-computer interaction that aids people in obtaining better self-knowledge by helping them collect and reflect on personal information, including exploring and understanding this information. With developments in sensor technology, the widespread availability of information on the Internet better facilitates personal informatics. To create and implement such a personal information system, it is important to make it effective and simple enough for people to use and reflect insightfully. Furthermore, pure self-reflection is frequently flawed due to the fact that people's memory is limited, some behaviours cannot be observed directly, for example, sleep apnea and some behaviours cannot be observed continuously and consistently, such as counting steps throughout the day, so it is hard to identify patterns and trends only reflect on the memory. People may also lack relative skill or knowledge to draw appropriate inferences from observations \cite{r1}, which is the reason for the existence of the personal informatics system.

\subsection{The Model of Personal Informatics System}

\begin{figure}[h]
    \centering
    \includegraphics[width=0.6\textwidth]{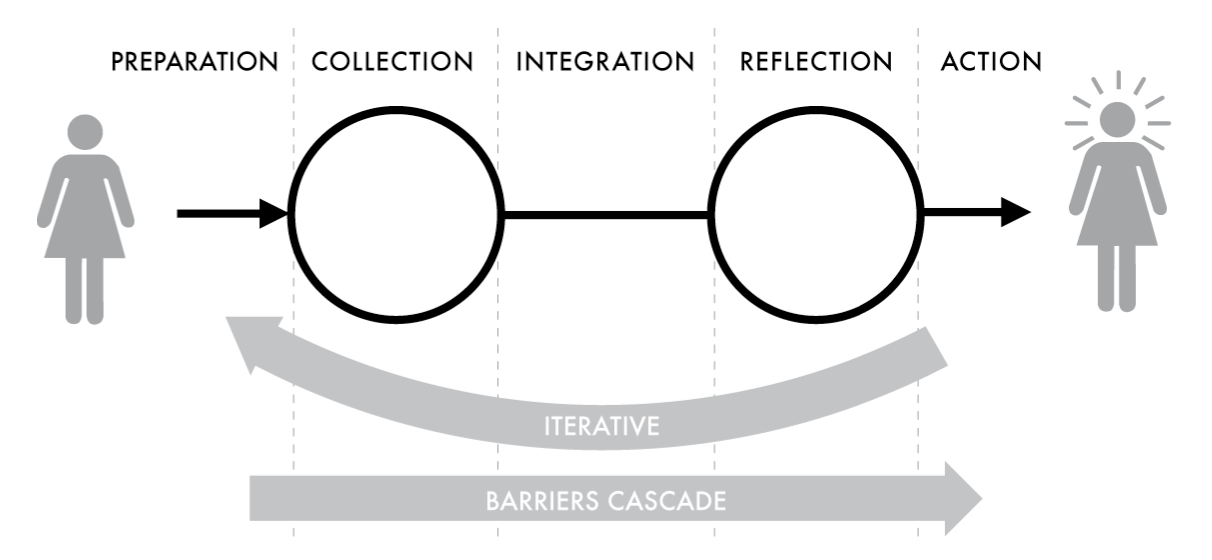}
    \caption{The Stage-Based Model of Personal Informatics Systems \cite{r1}}
    \label{fig:personal-informatics-1}
\end{figure}

There is a mature stage-based model of personal informatics (fig) introduced by Li et al. (2010) \cite{r1}, which consists of five stages: Preparation, Collection, Integration, Reflection, and Action. The Preparation stage is for users to choose what kind of information they want to collect and how to collect it \cite{r1}. The Collection stage is for users to collect information about themselves. Several difficulties were encountered throughout this stage, such as users’ lack of time and motivation or forgetting to record their information, as well as data may depend on subjective ratings without a standard rule, subjective estimation or difficult to locate \cite{r1}. Integration is the stage that occurs between the Collection and the Reflection stages, where the information is collected, merged, and changed for users to reflect on \cite{r1}. The Reflection stage is where users reflect on their personal information by exploring lists of collected data or interacting with visualizations of their data. Research has found that there are short-term and long-term reflections, while the former helps users become conscious of their current status and make changes timely, the latter enables people to compare between different periods and highlights patterns and trends \cite{r1}. In this stage, lack of time or miscomprehending information might prevent people from understanding themselves \cite{r1}. The Action stage is the final stage, where people make their decisions on what to do with their newly acquired self-awareness based on their comprehension of the information, using the information to track their progress towards objectives by reflecting on it \cite{r1}.

\subsection{Reflection}
\label{sec:reflection}

Reflection is the last stage in the stage-based model of the personal informatics system. It facilitates a type of issue-solving by constructing an understanding and re-framing of the situation that helps professionals utilize and advance their knowledge and abilities \cite{r18}.

Fleck and Fitzpatrick (2010) have developed a framework that incorporates a variety of reflection-related characteristics gleaned from existing research, which include the goal, prerequisites and stages of reflection. Then they used this to demonstrate how technology had been utilized to support reflection in the past and also in the future \cite{r18}. Understanding the goal of the reflection and directing thinking towards it are crucial to offering a structure for it. If there is no clear goal, technology may simply be used to facilitate and stimulate reflection rather than to organize and promote it. Opportunities for introspection could be missed in this approach \cite{r18}. Reflection could have a variety of purposes, such as learning and the information for additional reflection, reflection on the learning process, action or other presentation of learning, critical review, theory construction, self-development, decisions for uncertainties, and so on \cite{r18}. Time is also an important condition for reflection \cite{r18}.

Fleck and Fitzpatrick (2010) raised five different reflection levels, from the lowest R0 to the highest R4:

\begin{itemize}
    \item \textbf{R0} stands for \textbf{Description}: reviewing a summary or assertion concerning events not expanded upon or explained, not thoughtful \cite{r18}.
    \item \textbf{R1} stands for \textbf{Reflective Description}, a review of the explanatory description, including a narrative or descriptive explanation of action or interpretation, with reasons related to personal experience with basic analysis. No other theories were investigated, and little analysis but no viewpoints switched \cite{r18}.
    \item \textbf{R2} stands for \textbf{Dialogic Reflection}: searching for connections between bits of information or experience, proof of cycles of interpretation and inquiry, evaluation of various theories, hypotheses and other viewpoints \cite{r18}.
    \item \textbf{R3} stands for \textbf{Transformative Reflection}: reviewing something with the intention of organizing it better or approaching it in a different way, which viewing it from an alternative viewpoint by asking important queries and challenging individual hypotheses would change in practice or understanding \cite{r18}.
    \item \textbf{R4} stands for \textbf{Critical Analysis}: taking ethical and social concerns into account and achieving broader consequences, which means a broad view of the situation. This is the ultimate degree of reflection \cite{r18}.
\end{itemize}

Furthermore, beyond the definition of reflection, Fleck and Fitzpatrick (2010) presented the characteristics of reflection, which are structured as purpose, conditions and levels. This emphasizes the concerns to take into account while comprehending the enabling role that technology can play in supporting reflection \cite{r18}. For example, annotation technologies are encouraged to promote users’ reflection, with reflective questions such as ``What emotions do you have when you see this image? positive or negative?'', the idea can then be used to tell a story \cite{r40}. 

The techniques to support different levels of reflection are as follows:
\begin{itemize}
    \item \textbf{R0 (Descriptive)}: Technology can provide a detailed record of events, such as video or data logs, acting as a foundation for future reflection \cite{r18}.
    \item \textbf{R1 (Descriptive Reflection)}: People frequently reflect when reviewing their knowledge or recalling past memories. For more of this reflection, technology prompts users to explain their actions through guided questions or interactive learning tools, helping structure their thoughts \cite{r18}.
    \item \textbf{R2 (Dialogical Reflection)}: Technology enables viewing experiences from multiple perspectives, such as through collaborative tools or visualizations that encourage questioning. Additionally, technology can expand the field of vision. For example, sensor technologies have the ability to collect, detect and portray information that is not normally visible to the human eye \cite{r18}.
    \item \textbf{R3 \& R4 (Transformative Reflection)}: These two levels build on the processes of levels R0 to R2, by engaging the resources available for reflection at a deep level. Technology is limited at this stage, people mostly aid themselves in deeply analyzing assumptions, promoting transformation in practice and challenging the original interpretations \cite{r18}.
\end{itemize}

As a technique, what EmoTrack can provide to users and the extent to which it can help users reach these different levels of reflection will be discussed in Section \ref{sec:2-discuss}.

\subsection{Self-Reflection (Autoethnography)}
Autoethnography is a self-study approach that allows researchers to comprehend and sympathize with the challenges in various environments, despite the fact that it can be disruptive at times. By using this technique, HCI researchers can gain a deeper understanding of how users interact with mobile devices, particularly during non-routine times that are challenging to investigate in person using conventional user studies \cite{r19}.

Design and evaluation benefit greatly from empathy \cite{r20}. Understanding users is an aim shared by HCI design, evaluation, and research projects \cite{r21}. Self-study enables developers to capture some of the complexities and subtle differences of the circumstances that can influence the adoption and usage of the device, which are difficult to do without first-hand experience. This method could be especially beneficial as an initial step in user research to help with the design of user studies, and as a low-tech tool to work through the interactive design process. Although the method is disruptive and the results are subjective and personal, the self-reflection of developers and other researchers on the results can yield interesting insights into the use of mobile technologies and it can be useful for testing hypotheses, generating empathy for user experiences, and organizing future user studies \cite{r19}.

\section{Relative Research}
The following sections provide some existing related research about collecting people's behaviours and mood, either in reality or in virtual environments. It also discusses their advantages and potential drawbacks.

\subsection{Parental Control Applications}
Parents can get various information about their children, such as illness prediction, mental health ratings, or unfavourable behaviour ratios, by monitoring their children’s phone usage. Nowadays, there are indeed some applications created to enable parents to do so. For example, KidsGuard Pro, FamilyTime, etc. \cite{r33}. There are different opinions regarding these kinds of apps. Baldry et al. (2019) indicated that parents can help their children avoid cyberbullying or becoming cyber-victims by actively supervising and monitoring online activities, restricting their children's behaviour and making them aware of whether they are participating in cyberbullying in some way \cite{r34}.

However, there are still some problems with these apps. Children are bothersome because they think these parental control apps are extremely intrusive and observant over their lives, particularly those teenagers, which might have a negative impact on family relationships \cite{r33}. The additional layer of complication arises from parental supervision when children are required to make judgements and mistakes in front of their parents, who are vigilant and frequently critical \cite{r39}. However, children frequently have no choice. When parents choose to utilize technical surveillance on their children’s mobile devices, sensitive information disclosures become mandatory, in contrast to many privacy theories that typically presume users have some influence over their disclosure decisions \cite{r39}. This may cause specific difficulties between parents and children because there are explicit trade-offs between the child's, particularly adolescents’, digital privacy requirements and online safety \cite{r39}.

\subsection{Internet of Behaviours (IoB) \cite{r35}}
Global data traffic is growing exponentially due to the rapid growth of data and information technologies including big data, cloud computing, 5G, artificial intelligence, and the Internet of Things. With the emergence of the digital world, data may now be used to characterize every aspect of existence. These uniform definitions can serve as a foundation for sensible judgement calls and well-informed choices \cite{r35}. The relevance of human behaviour data on the Internet cannot be emphasized, even if it makes up a very small percentage of the vast data. Since humans are the primary agents of all social interactions, it is crucial to do the study to better understand people's intentions, forecast behavioural trends, and modify behaviour in desired directions \cite{r35}. According to Sun et al. (2023), the applications of IoB can help people in many different fields, such as Medicine, Business, Education, Transportation, Intelligent environments and so on. For instance, the trend of behaviour can be fairly predicted using the results of analysing data on human behaviour in conjunction with techniques related to data mining and machine learning. The forecast can be used to create early warning systems (health, social security, etc.), modify services pertaining to the movement of people, and create commercial plans, management techniques, and other areas. Also, it can give learners recommendations and techniques that are more supportive of their learning based on the analysis of this data \cite{r35}.

\subsection{Web Search Behaviour and Critical Online Reasoning (COR)}
This research \cite{r36} was conducted by recording students’ search behaviour and students’ responses to the questions, aiming to investigate whether students have the ability to use online resources critically and reflectively. It emphasizes the importance of "Critical Online Reasoning" (COR) when it comes to navigating and evaluating information found online. COR involves the ability to critically evaluate online content, distinguish reliable sources from unreliable ones, and ultimately make informed decisions. This study shows that students in higher education need to have targeted support, which should be immediately addressed by putting the right policies in place. This is because the capacity to use online resources and think critically online competently forms a vital foundation not only for academic success but also for lifelong learning \cite{r36}.

\subsection{Mobile phone mood charting for adolescents \cite{r50}}
This study describes an application that enables teenagers to record their mood on the phone \cite{r50}. Mood charting is used in therapeutic treatments for a variety of conditions. Several studies have demonstrated that consistent and trustworthy self-charting by clients themselves improves therapy results. Using mood charting, clients note their feelings on a daily basis in order to pinpoint the causes of their emotional states and behaviours. This might aid clients to become more self-aware, which would give them more control and comprehension over their behaviour \cite{r50}. 

Mobile mood charting can precisely record the time-stamp when users enter the mood data and users can do so at practically any time and from any location, which offers higher data quality than paper-based charting. It can also remind users to chart automatically and send tracked data to practitioners. It is easy for practitioners to visualize the daily, monthly or annual graphical charts \cite{r50}.

\section{Personal Privacy}
The concept of networked privacy has gained popularity in academic studies, which means privacy not only depends on an individual’s control or disclosure but also on their social networking, particularly on social media platforms where sharing content with friends \cite{r39}. Many studies have stated that privacy is the negotiation of information between two or more parties based on presumptions, conventions, biases, and cultural factors. Communication Privacy Management theory has also been used to describe privacy as a process of boundary negotiation in which people decide which sensitive disclosures are subsequently co-shared with their confidants. The impact of exposure to social media has also been studied recently using networked privacy \cite{r39}.

These ideas of privacy revolve around the concepts of visibility and information leaks. Our behaviours’ visibility influences how we perceive or think about each other. Visibility, in turn, is a tool that our social networks and different audiences use to socially monitor us and create opinions \cite{r39}. Young people are creating their online personas and learning how to deal with complexity, like knowing when and how to provide personal information to others when communicating with them in virtual environments \cite{r39}.

\section{Discussion}
\label{sec:2-discuss}
In order to face the impact of the increasing growth of social media on people, EmoTrack is a tool informed by the personal informatics system, that can help people better track their personal information, especially their YouTube watching behaviour and the impact on their mood, which can help people make reflections and actions.

Here is how EmoTrack aligns with the model of the personal informatics system:
\begin{enumerate}
    \item \textbf{Preparation}: EmoTrack is a tool for users to collect and manage personal data. In addition, YouTube is the objective to be monitored and the YouTube watch history is stored in Google Takeout.
    \item \textbf{Collection}: Users can record their mood using EmoTrack by selecting the mood before and after watching YouTube. Users need to download their history from Google Takeout and upload it to EmoTrack. The app gathers this information systematically over time, building a comprehensive dataset for analysis.
    \item \textbf{Integration}: EmoTrack integration data from multiple sources, such as mood tracking and video categorization, to provide users with a holistic view of their YouTube video watching habits. By integrating different types of data, the app enables users to gain deeper insights into their video watching behaviour and emotional responses.
    \item \textbf{Reflection}: This stage allows users to reflect on their  YouTube watching habits with regard to their mood patterns. Through features like mood tracking, and data visualizations. Users can analyze their data and understand how their behaviour impacts their mood.
    \item \textbf{Action}: EmoTrack facilitates action by providing users with data visualizations and detailed reports based on their data. This actionable feedback empowers users to make positive changes in their behaviour and improve online engagement skills.
\end{enumerate}

In order to explore the extent to which EmoTrack can be useful as a tool to support users’ reflection during the reflection stage, and the level of reflection that users will have that will be in line with the theories presented by Fleck et al. (2010) \cite{r18}, I have made the following predictions about the possible results, which will be validated in the final evaluation Section \ref{sec:interview}.

\begin{itemize}
    \item \textbf{R0 (Descriptive)}: EmoTrack can log users' YouTube viewing habits and mood changes, enabling data to be revisited for later reflection. By tracking these, the app provides a foundational record that users can reflect upon. Users can view the records and review their online behaviours without further insight or analysis.
    \item \textbf{R1 (Descriptive Reflection)}: EmoTrack provides bar charts and lists, linking specific video genres to emotional responses. Promote users to identify which content might trigger positive or negative emotions, providing a baseline understanding of how media affects them.
    \item \textbf{R2 (Dialogic Reflection)}: EmoTrack provides analysis reports and encourages users to explore multiple perspectives. Users actively reflect on their viewing habits, engaging in a mental dialogue about why certain videos impact their mood. They might consider different perspectives, questioning the reasons behind their responses.
    \item \textbf{R3 (Transformative Reflection)}: Users revisit the events to have different actions in the same or similar situations. Users in this stage might make transformative changes. They will acknowledge that a great deal of their actions are unconsciously absorbed from specific environments, which means users will be aware that they are trapped by the algorithm of YouTube \cite{r18}.
    \item \textbf{R4 (Critical Reflection)}: This level involves a thorough analysis, beyond the local setting and takes ethical and social concerns into account. Reaching this level of reflection is extremely difficult.
\end{itemize}

\section{Summary}
This chapter provides an overview of how social media can have both positive and negative effects on people, especially young people. Next, it discusses the motivation of this project, which is to create an application which can help people track and reflect on their online activities on YouTube and the impact on their mood. Then it introduces the theory of the personal informatics system and various levels of reflection, which also discusses how EmoTrack will be developed following the model of the personal informatics system and how each level of reflection will be structured with EmoTrack. In addition, it describes the theory of self-reflection (Autoethnography) that will be used in the evaluation chapter (details in Section \ref{sec:self-reflection}). Finally, it explains the relative research on the collection of human behaviours and mood, as well as the issue of personal privacy while using social media.



\chapter{Technical Background}
\label{chap:technical}

\noindent

\section{Introduction}
In this chapter, I introduce the technical background necessary to understand the implementation in the Project Execution chapter. It provides an overview of the major programming languages, AI models, and other techniques, as well as all of the Google services that will be integrated into this project. This overview is used to justify the selection of each tool and its contribution to the efficient execution of the project.

\section{Technology Used}
\subsection{Web Scraping}
\label{sec:web-scraping}
\begin{figure}[!h]
    \centering
    \includegraphics[width=0.4\textwidth]{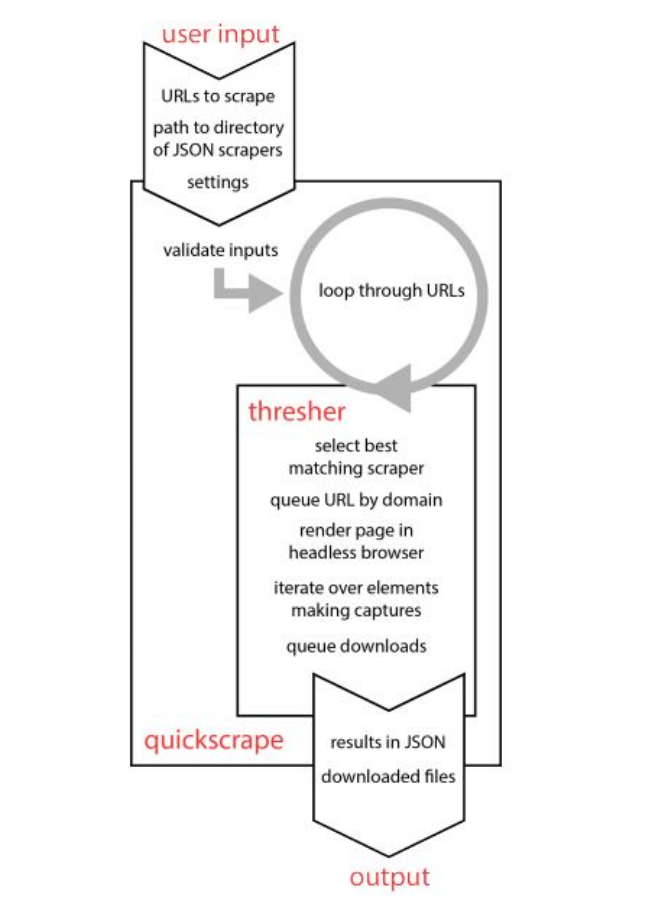}
    \caption{Overview of ContentMine Scraper \cite{r2}}
    \label{fig:contentmine}
\end{figure}

Web scraping converts unstructured data into structured data that can be saved and evaluated in a database, where its goals are to collect, store, and validate data \cite{r2}. The next critical stage is data analysis, which allows validated data to be turned into more useful information \cite{r2}. Figure \ref{fig:contentmine} is used by Lusiana et al. (2019) to introduce scraperJSON’s scraper ContentMine, which sends various URLs that need to be scraped. Only the validated URLs filtered by the technique looping through them will be sent to the most appropriate scraper. The scraper renders each URL in sequence for a page in a headless browser, iterates over all elements, captures the required items and stores them in the format of JSON.

\subsection{API}
Application Programming Interface (API) is a collection of rules or protocols that permit software programs to interact with one another and share features, data and other functionalities. APIs make it easier and faster to develop applications and software by enabling developers to integrate data, services and capabilities from other applications. APIs can be classified by their use case, such as data APIs, remote APIs, operating system APIs, and web APIs \cite{r32}. In EmoTrack, web APIs are used to transfer data and functionality over the Internet using HTTP protocol, including YouTube Data API and calling functions in the back-end from the front-end. The latter uses gRPC, an open source, high-performance RPC framework first developed by Google. The Protocol Buffers data format and the HTTP/2 network protocol buffers are used by gRPC, which is often used to connect services in microservices architectures \cite{r32}.

\subsubsection{YouTube Data API}
Using the YouTube Data API can incorporate features often used on the YouTube website into the developer’s application or website, which can obtain various resources through the API, such as information of YouTube videos. Many of these resources can also be updated, deleted or inserted using approaches supported by the API \cite{r31}. It can be accessed through the client library in multiple programming languages, including Python.

\subsection{ChatGPT}
ChatGPT is an intelligent chatting robot that can deliver a thorough response to a prompt. According to the official document, it has demonstrated powerful functions in a variety of language interpretation and generating activities such as multilingual machine translation, code debugging, story creation, admitting faults, and even rejecting unsuitable requests \cite{r22}. ChatGPT is different from earlier chatbots in that it can recall users’ comments, facilitating continuous communication \cite{r23}. With the release of OpenAI’s creation GPT-4 in March 2023 [24], ChatGPT also benefited from a significant update that added new features. Users can now send texts and pictures to ChatGPT at the same time, solving increasingly demanding multimodal tasks including summarizing paragraphs, reasoning charts and captioning images \cite{r22}.

ChatGPT’s fundamental methodologies include pre-trained large-scale language models (PLM), in-context learning and reinforcement learning based on human feedback [22]. Language models are statistical models that explain how natural language is distributed probabilistically \cite{r25}. The primary task of natural language processing (NLP) is assessing the probability of a given sentence, or the probability of creating alternative contents given a portion of the sentence, which may be practically used for all downstream NLP tasks \cite{r22}. The technical level of NLP is denoted by several statistical language modelling methodologies. Using self-learning tasks, such as predicting masked words, recognizing the sequence of sentences, filling in the blanks, and generating text, PLM acquires large-scale general language models on texts. It offers a uniform modelling foundation for NLP activities in addition to enhancing word semantics by transitioning from static to context-aware dynamic representation \cite{r22}.

\section{Google Services Integration}

\subsection{Flutter}
Flutter is an open-source UI software development toolkit created by Google, designed primarily for building pretty, natively compiled applications for mobile, web, and desktop from a single codebase \cite{r27}. Launched initially in 2017, Flutter leverages the Dart programming language, which itself was designed to offer fast app development and performance.

It has some key features that differentiate it from other similar development frameworks, such as Hot Reload, which allows developers to see the effects of their code changes immediately without restarting the application, speeding development and iteration. Flutter is widely used to develop visually appealing and high-performance mobile applications in a variety of domains, including social media, retail, finance, and productivity tools. Adoption by companies such as Alibaba, BMW, and Google Ads has established its reputation in the industry \cite{r28}.

\subsection{Firebase}
Firebase is a famous platform for developing mobile and web applications, that allows people to create apps and games that people enjoy \cite{r11}. It was originally developed by Firebase, Inc., an independent company founded in 2011. It was acquired by Google to be a Google product in 2024 and was incorporated into its cloud service platform. Since then, Firebase has grown to be a key component of Google Cloud, making it expand its services and better integrate with other Google Cloud services and APIs.

Several powerful features of Firebase are particularly used in this project:

\begin{enumerate}
    \item \subsubsection{Firebase Authentication}

Most applications require some type of user identification, which enables them to configure preferences, save data and offer personalized experiences consistent with all their devices. To achieve this, they must have the ability for new users to sign up and existing users to sign in, manage account information and securely store all data, which is an extremely challenging and drawn-out procedure. It is also very hard to get right in terms of user experience since users are hesitant to give out their personally identifying information, for example, their passwords, user names, and anything else that could be dangerous if leaked. Consequently, people frequently prefer to use login credentials they have already given to a third party to handle the sign-in process. For instance, people who have Facebook accounts would prefer to log in to your app through Facebook, without giving their information to one more new app \cite{r12}.

Firebase Authentication provides an end-to-end identification process and supports phone auth, email and password accounts and Google, Twitter, Facebook and GitHub login and more, which was created by the same team as Google Sign-in, Chrome Password Manager and Smart Lock, leveraged Google’s in-house knowledge to ensure security comprehensively \cite{r13}. EmoTrack allows users to log in with their email or phone number, which takes only a few lines of code to complete the setup and is flexible enough to be placed on the UI.

    \item \subsubsection{Firebase Cloud Firestore Database}
    \label{sec:firestore-db}

\begin{figure}[h]
    \centering
    \includegraphics[width=0.5\textwidth]{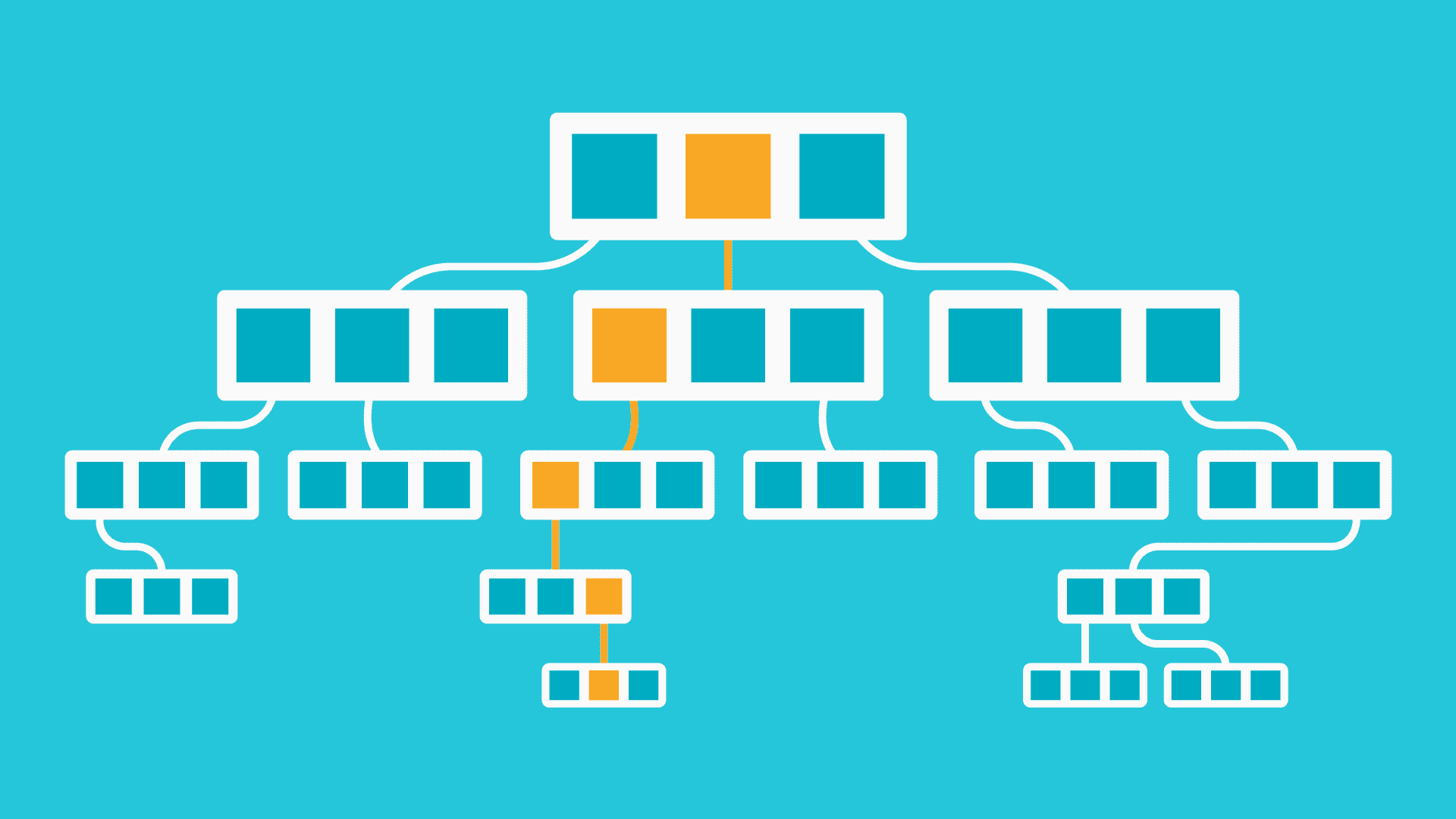}
    \caption{The Structure of Firestore Database \cite{r14}}
    \label{fig:firestore-1}
\end{figure}

There are two kinds of databases provided by Firebase, one is the Firestore Database, and the other is the Firebase Real-Time Database. I used the Firestore Database in my project because it is more robust in handling complex data structures and queries, its document model allows more complex hierarchies and richer data structures compared to the flat structure of the Real-Time Database \cite{r14}.

Cloud Firestore is a NoSQL document database, data can be structured into collections and documents clearly and easily (\ref{fig:firestore-1}), which builds hierarchies to organize relevant data and expressive queries are used to retrieve the data needed quickly \cite{r14}.

    \item \subsubsection{Firebase Cloud Storage}

Cloud Storage is designed to store and serve content generated by users easily and quickly, such as files, photos and videos. Integrated with Firebase Authentication, it provides security to allow access control based on the user’s identity and the file's properties, such as name, size, content type and other metadata, ensuring security for users \cite{r15}.

    \item \subsubsection{Firebase Hosting}

Firebase Hosting is used to build and deploy websites and apps without any infrastructure, which can preview the site before going live. I used Firebase Hosting to deploy the UI on websites.

    \item \subsubsection{App Check}

App Check is an extra security layer that helps guard against unauthorised access to my services by verifying that incoming traffic originates from my app and preventing access to it with invalid credentials, which assists in guarding against abuse of my back-end, including data poisoning, phishing, billing fraud, and app impersonation. App Check works with other Firebase and Google Cloud products, as well as other back-end APIs \cite{r16}.

\end{enumerate}

\subsection{Google App Engine}
Google App Engine (GAE) is a Platform-as-a-Service (PaaS) offering by Google Cloud that provides a fully managed environment to build and host applications. Launched in 2008, it abstracts much of the infrastructure concerns required to deploy applications, enabling developers to focus on writing code without worrying about server provisioning or management \cite{r29}. It has several advantages, firstly, it supports various popular languages, including Python, which is used in the development of EmoTrack. Second, it can use logs to monitor application performance and report errors for developers to diagnose and fix bugs in a timely manner. Last but not least, it can version control the hosting application and easily create development, test, staging and production environments \cite{r29}.

\subsection{Google Takeout}
Google Takeout is a tool provided by Google for users to download their data, enabling users to export data from a wide range of Google services, including YouTube, in a structured and standardized way. Data can be exported in various common formats, such as JSON, HTML, and CSV, ensuring compatibility with the system. Launched in 2011 by the Google Data Liberation Front, it was developed to empower users by giving them control over their data and allowing them to download a copy for personal backup or transfer to other platforms \cite{r30}.

\section{Discussion}
In this project, Flutter was used to design the User Interface of EmoTrack and develop the same code for multi-platform, including iOS, android, and web applications. The web application was deployed on the Firebase Hosting. Firebase Authentication and App Check were used to ensure the safety of users and developers while using and developing.

The technique of web scrapping and YouTube Data API were both used for acquiring information on YouTube videos, such as their titles, descriptions, categories, etc., I would explain the reason for the selection between these two methods in the following Section \ref{sec:scrape}. Google Takeout was the platform for users to download their Google data, including YouTube history.

ChatGPT was used to categorize YouTube videos, since it performs well in natural language comprehension and generation, showing good assessment capabilities for factual consistency of text summaries without additional training data \cite{r26}. However, ChatGPT can exhibit lexical similarity bias, judging document-summary relationships based on lexical similarity, leading to misinterpretation of semantic implications. It can also produce incorrect reasoning, leading to misjudgment of textual relations, and may not accurately follow instructions, resulting in unsatisfactory output. Evaluation of the performance of ChatGPT is discussed in Section \ref{sec:self-reflection}.

Google App Engine was used to deploy the back-end functions written in Python and the application’s API endpoints by Quart. These API endpoints are called by different functions in front-end whenever users press the particular buttons, enabling the corresponding operations to be processed in the back-end functionality.

Firebase Cloud Storage was used for storing the files uploaded by users and Firestore Database was used for all important data collection, including users’ YouTube watch history, mood records and necessary data for generating analysis reports.

\section{Summary}
This chapter explores different technologies used in the project, such as ChatGPT, YouTube Data API, Google Services, etc. Every technology is essential to the development of EmoTrack.

In the following chapter, how these technologies were used and integrated to create the EmoTrack will be explained comprehensively.


\chapter{Project Execution}
\label{chap:execution}

\noindent
\section{Introduction}
This chapter provides a comprehensive development process of EmoTrack, including an overview of how the application works, the design of the User Interface, the development of the back-end and how it can be deployed on the Cloud as a server, as well as the integration between them and the database. It also explains the challenges encountered during the process and how they were resolved.

\section{Overview of How the Application Works}
\begin{figure}[h]
    \centering
    \includegraphics[width=\textwidth]{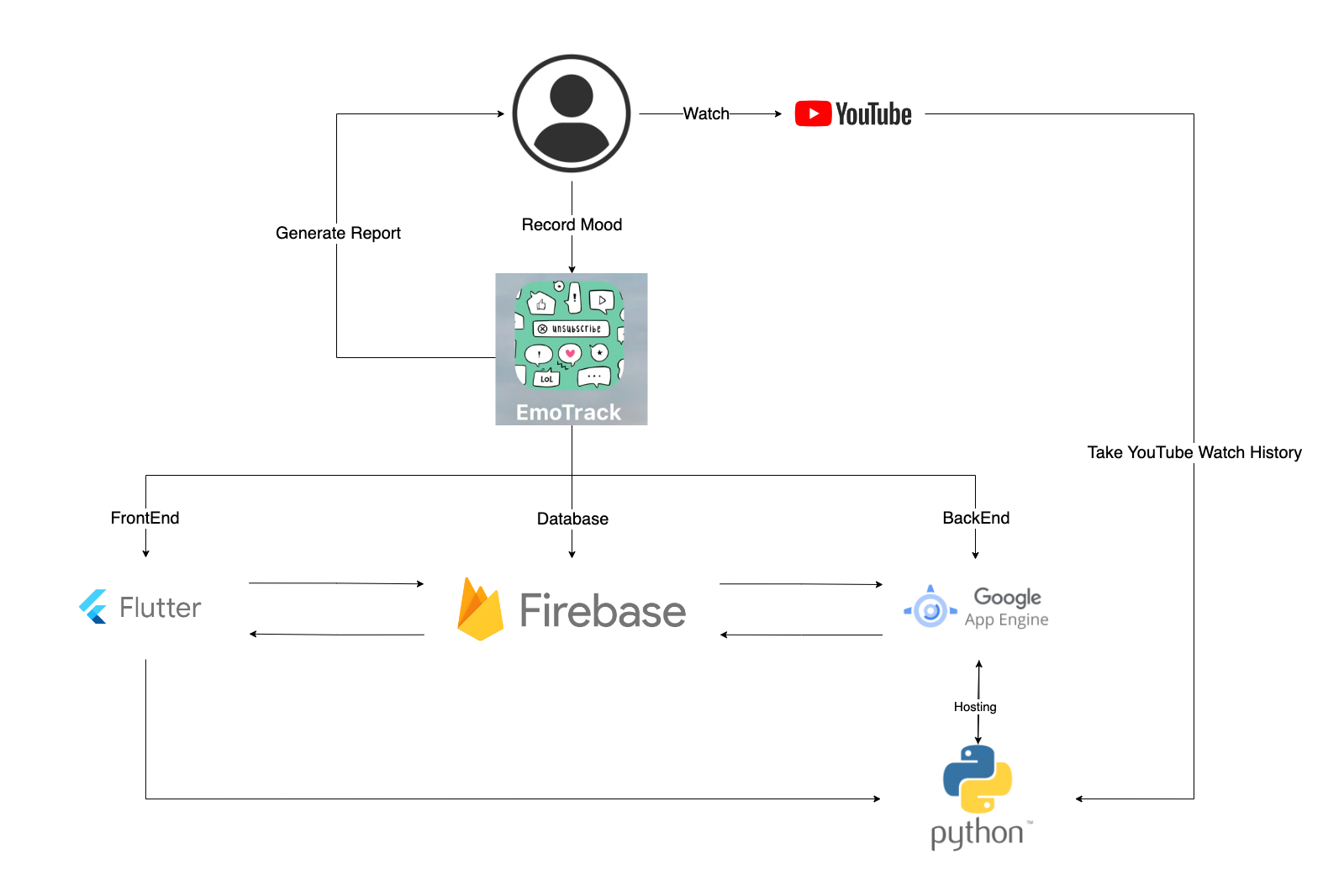}
    \caption{The Work Flow of EmoTrack}
    \label{fig:flow-chart-1}
\end{figure}

EmoTrack is designed to serve three main purposes:
\begin{enumerate}
    \item Record users’ mood before and after watching YouTube videos each time.
    \item Organize users’ YouTube watch history by day and categorize videos with ChatGPT.
    \item Generate analysis report for users to present the relationship between the changes in mood and YouTube videos.
\end{enumerate}

The simple example workflow of EmoTrack is shown in Figure \ref{fig:flow-chart-1}, and there is a detailed workflow in the appendix (Figure \ref{fig:flowchart-details}).

From the user’s perspective:
\begin{enumerate}
    \item When users initially register with EmoTrack, a personal account with a unique user UID is generated.
    \item Every time users plan to watch YouTube, they open EmoTrack and record their mood first, and then they start watching. After users stop watching YouTube, they go back to EmoTrack and record their latest mood again.
    \item Users have to download their YouTube watch history from Google Takeout and upload it to EmoTrack
    \item Users can view the analysis report on EmoTrack, including the bar chart showing how many videos they watch and how their mood changes every day, and the detailed categories of videos showing the relevant changes in mood.
\end{enumerate}

From the back-end perspective:
\begin{enumerate}
    \item EmoTrack records users’ selections for their mood each time.
    \item EmoTrack stores users’ YouTube watch history and filters the required videos by time ranges in particular.
    \item EmoTrack categorizes videos with ChatGPT.
    \item EmoTrack analyzes the impact on users’ mood with all data collections.
\end{enumerate}

Thus, this section introduces the general process of how EmoTrack works using a combination of front-end and back-end components, and the detailed implementation process will be explained in the following sections.

\section{User Interface Design}
\label{sec:ui}
This section describes the UI design of EmoTrack, written in Dart, the programming language used by Flutter. It also discusses the problem of building the front-end on different platforms and the solution for it.

There are three main user requirements of EmoTrack on the user level:
\begin{enumerate}
    \item Safe for the users to have a private account and data collection.
    \item Easy for the users to record their mood in the moment.
    \item Easy for the users to see their YouTube watching statistics.
\end{enumerate}
	
\subsection{User Login System and Firebase Authentication}

Users only need to create their accounts and log in for once. When using EmoTrack for the first time, users can choose to register a new account on the main entry page with their email or phone number, which uses the authentication function of Firebase and generates a unique ``User UID'' for each user. Throughout the whole running process, ``User UID'' is quite important and binds all data and files with the user deeply.

\subsection{Record the mood of users and send them to Database}

\begin{figure}[!h]
    \centering
    \begin{subfigure}{0.49\textwidth}
        \centering
        \includegraphics[width=0.7\textwidth]{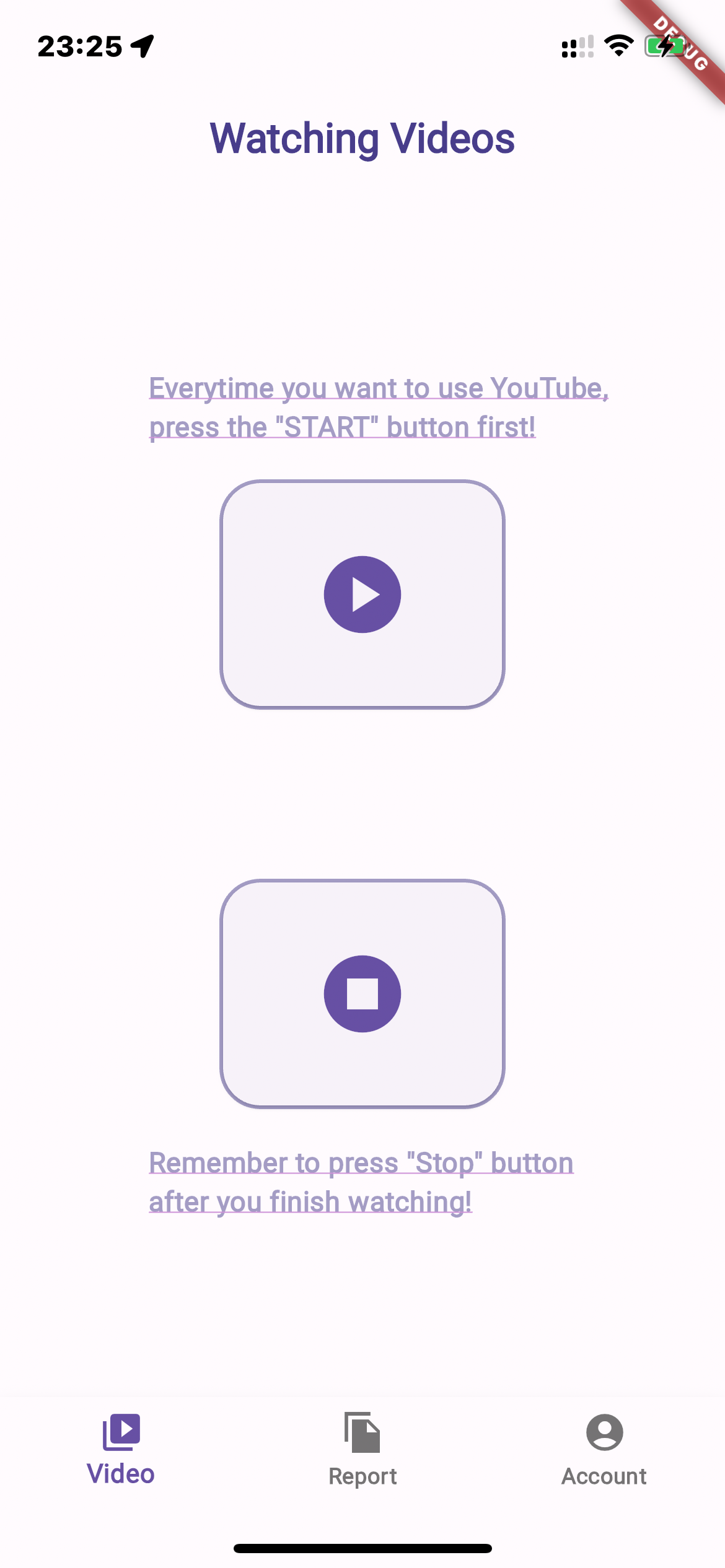}
        \caption{}
        \label{fig:mood-1}
    \end{subfigure}%
    \hspace*{\fill}
    \begin{subfigure}{0.49\textwidth}
        \centering
        \includegraphics[width=0.7\textwidth]{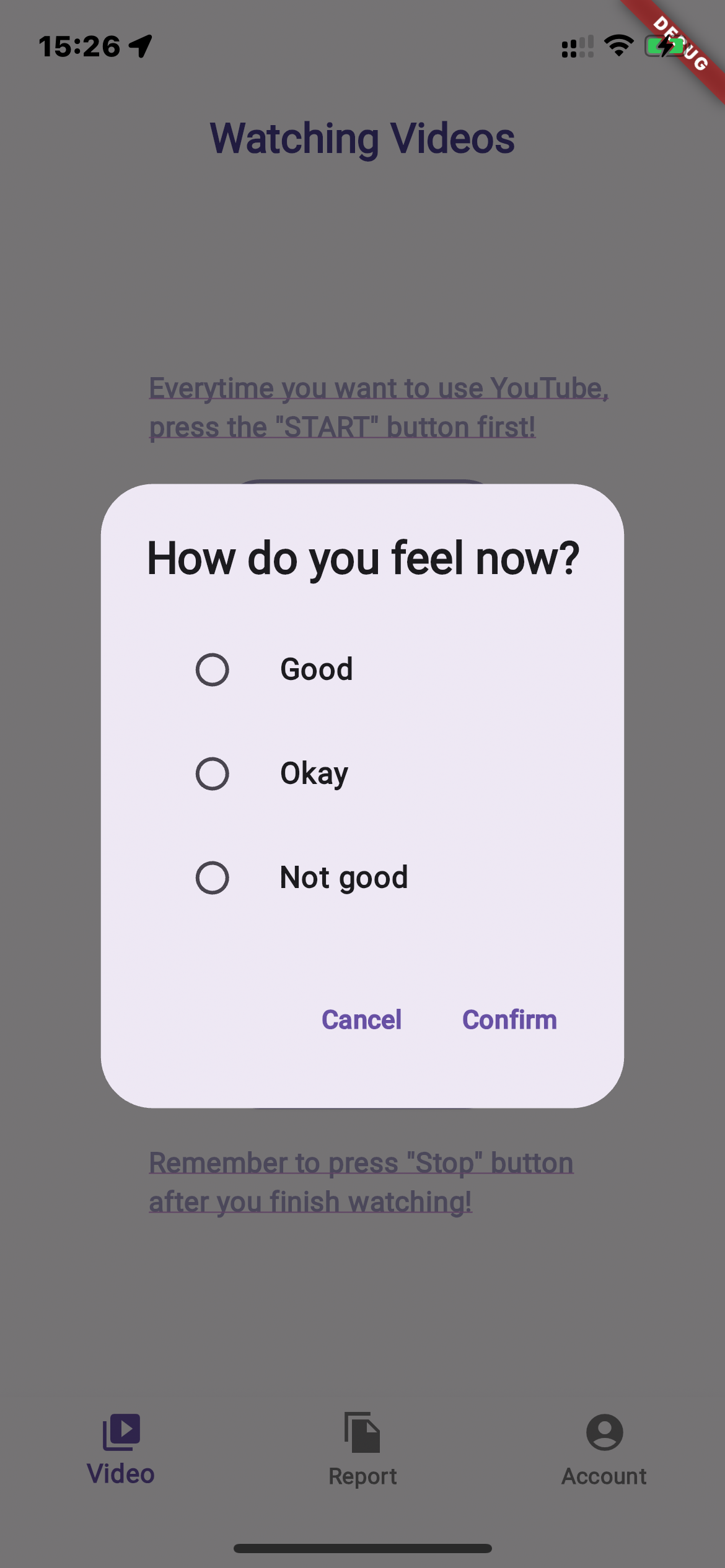}
        \caption{}
        \label{fig:mood-2}
    \end{subfigure}
    \caption{User Interface for Mood Recording}
    \label{fig:mood}
\end{figure}

Every time users want to watch YouTube videos, they open the app and press the ``Start'' button first (see Figure \ref{fig:mood-1}) and then they record their current mood by selecting one of the three options: `Good', `Okay' and `Not good' (see Figure \ref{fig:mood-2}). Then they can open YouTube and watch videos as long as they want. When they finish watching YouTube videos, users press the ``Stop'' button and select their current mood.

How these mood data are to be stored in the database is explained in Section \ref{sec:db-mood}.

\begin{figure}[!h]
    \centering
    \begin{subfigure}{0.49\textwidth}
        \centering
        \includegraphics[width=0.7\textwidth]{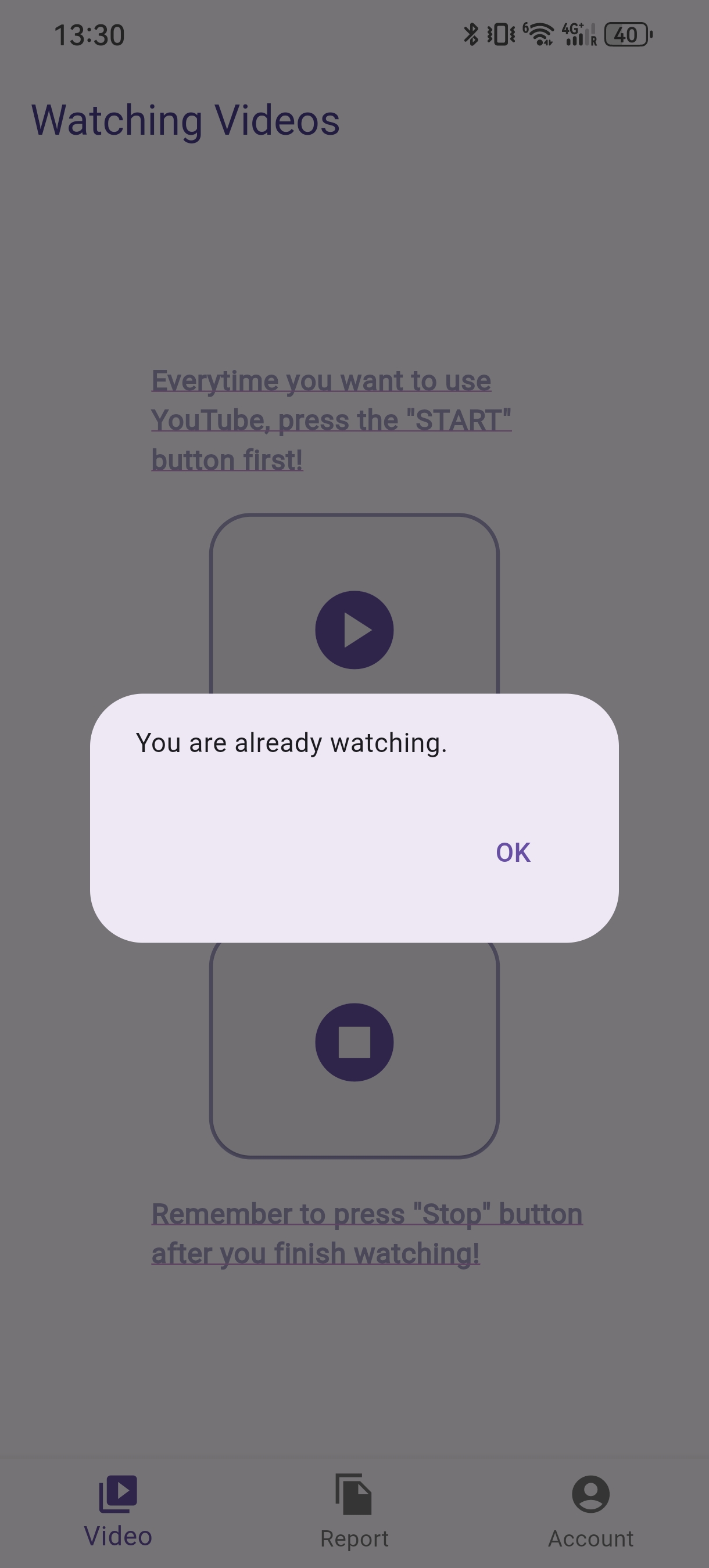}
        \caption{Message pops up for ``Start'' button}
        \label{fig:start-msg}
    \end{subfigure}%
    \hspace*{\fill}
    \begin{subfigure}{0.49\textwidth}
        \centering
        \includegraphics[width=0.7\textwidth]{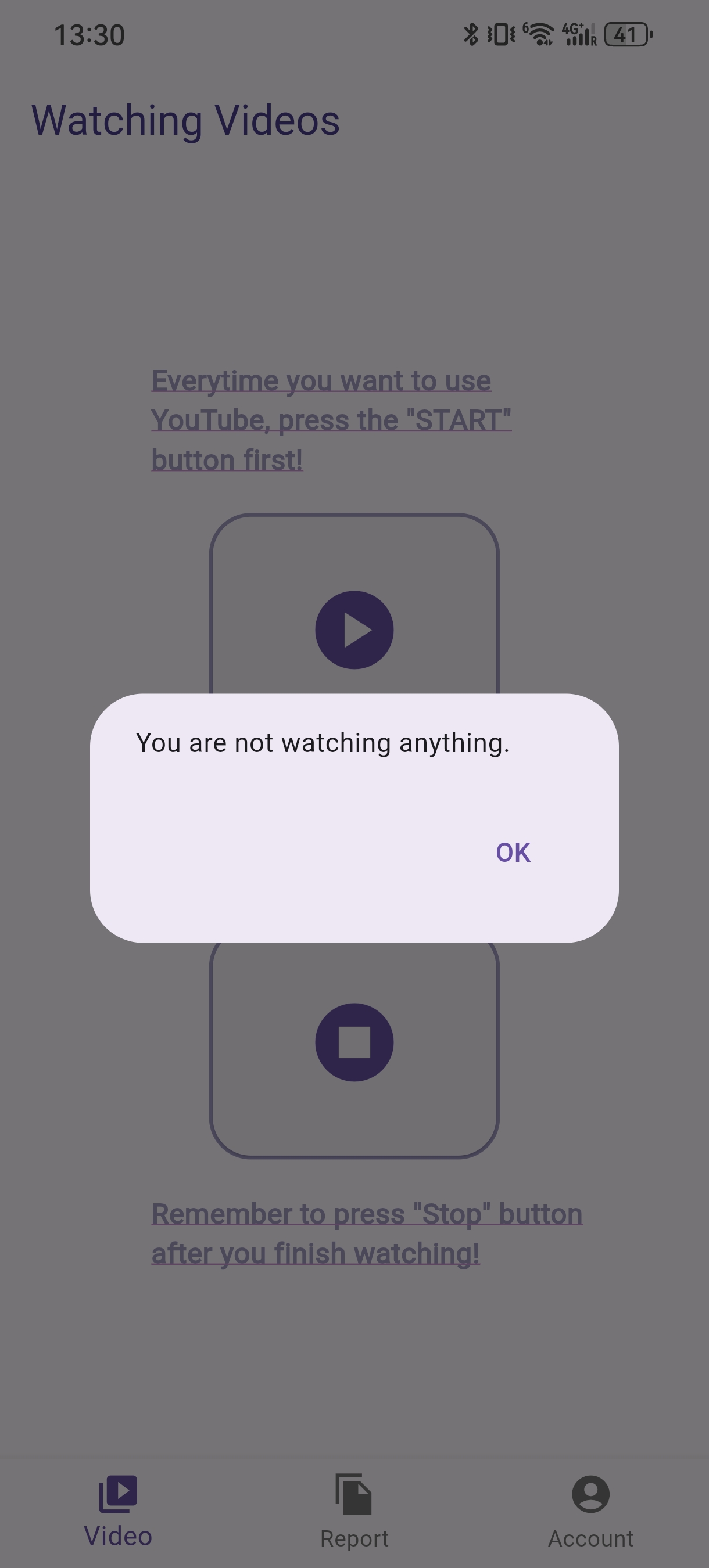}
        \caption{Message pops up for ``Stop'' button}
        \label{fig:stop-msg}
    \end{subfigure}
    \caption{Messages pop up for alerting}
    \label{fig:start-stop}
\end{figure}

Here is the setting to ensure the buttons record the correct data for each `Mood record’:
\begin{enumerate}
    \item In the beginning, users can only press the ``Start'' button. If users press the ``Stop’’ button first, the message `You are not watching anything’ will pop up (Figure \ref{fig:start-msg}).
    \item After pressing the ``Start'' button, users select their mood and press the ``Confirm'' button. The ``Start'' button will be disabled with the message `You are already watching’ pops up (Figure \ref{fig:stop-msg}), until users have pressed the ``Stop'' button.
    \item The existence of the ``Confirm’’ button is to ensure users have confirmed their selections of mood, and in case users accidentally touch the buttons. The selections and the current timestamp are transmitted to the database at the same time.
    \item Whenever users select their mood with the ``Start’’ button, it creates a new document for the `Mood Records’ collection in the database, which is explained in detail in Section \ref{sec:db-mood}. The mood selected with the ``Stop’’ button is updated to the same file.
\end{enumerate}

\subsection{Users upload their YouTube watch history to EmoTrack, which is processed and transferred to the database}
\label{sec:ui-upload}

\begin{figure}[!h]
    \centering
    \includegraphics[width=0.35\textwidth]{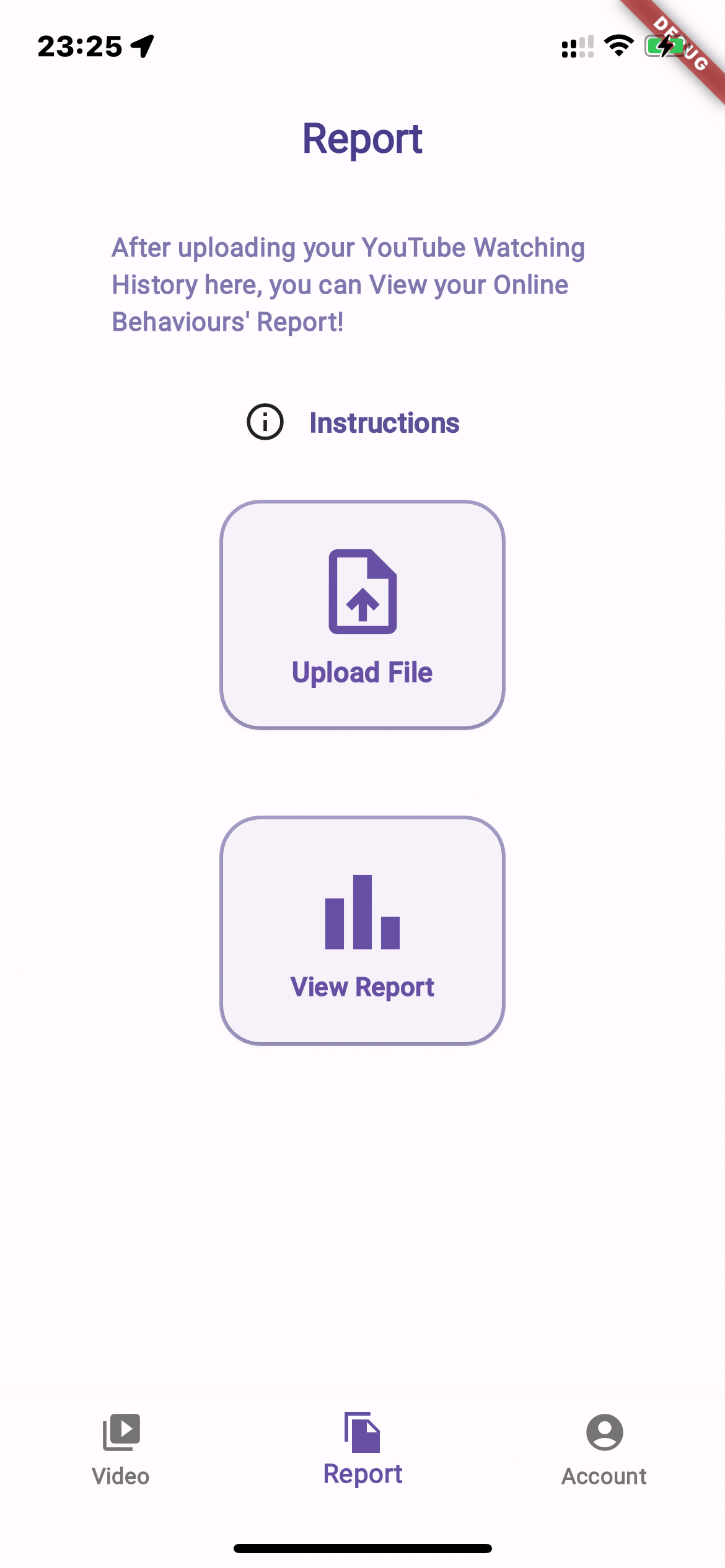}
    \caption{User Interface of the Report Page}
    \label{fig:upload-report}
\end{figure}

On the ``Report'' page of EmoTrack (Figure \ref{fig:upload-report}), users can upload their YouTube watching history to the application and visualize their changes in mood with YouTube watching history.

Firstly, users need to download their YouTube watching history from Google Takeout in JSON file format. The uploading function of the ``Upload File'' button is asynchronous and waits for the uploading operation to be completed. There is an uploadTask function provided by Flutter that can monitor whether the uploading process succeeds or fails, as shown in the code snippets \ref{lst:upload}. If the file is uploaded successfully, it then makes an HTTP request to the server, through the specific API endpoint, data will be processed by the corresponding function in the back-end. The Python function iterates over the file completely and extracts the URL and `watched time', transmitting them to the database as strings. A detailed description of how the JSON file is processed and converted is given in Section \ref{sec:scrape}.

\begin{lstlisting}[caption={Example Code for Listen for state changes, errors, and completion of the upload}, label={lst:upload}]
    uploadTask!.snapshotEvents.listen((TaskSnapshot taskSnapshot) {
      switch (taskSnapshot.state) {
        case TaskState.running:
          final progress =
              100.0 * (taskSnapshot.bytesTransferred / taskSnapshot.totalBytes);
          print("Upload is $progress% complete.");
          break;
        case TaskState.paused:
          print("Upload is paused.");
          break;
        case TaskState.canceled:
          print("Upload was canceled");
          break;
        case TaskState.error:
          // Handle unsuccessful uploads
          print('Upload failed');
          showHint(context, 'Upload Failed!');
          break;
        case TaskState.success:
          // Handle successful uploads on complete
          print('Upload succeed');
          showHint(context, 'Upload Succeed!');
          break;
      }
    }
\end{lstlisting}

\subsection{Analyze the data and generate the report}

\begin{figure}[!h]
  \centering
  \begin{minipage}{0.5\textwidth}
    \centering
    \includegraphics[width=0.7\textwidth]{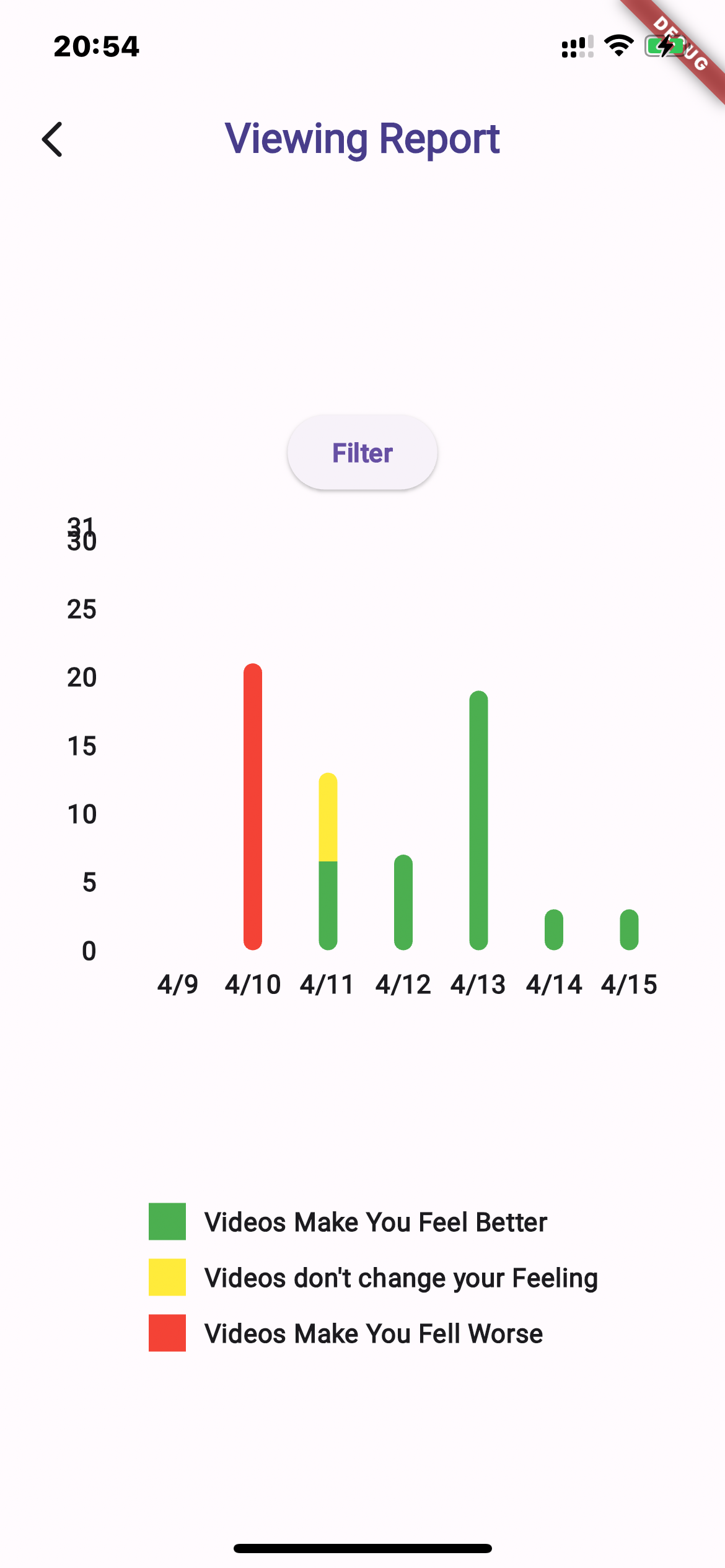}
    \caption{YouTube Watching Report}
    \label{fig:report-1}
  \end{minipage}\hfill
  \begin{minipage}{0.5\textwidth}
    \centering
    \includegraphics[width=0.7\textwidth]{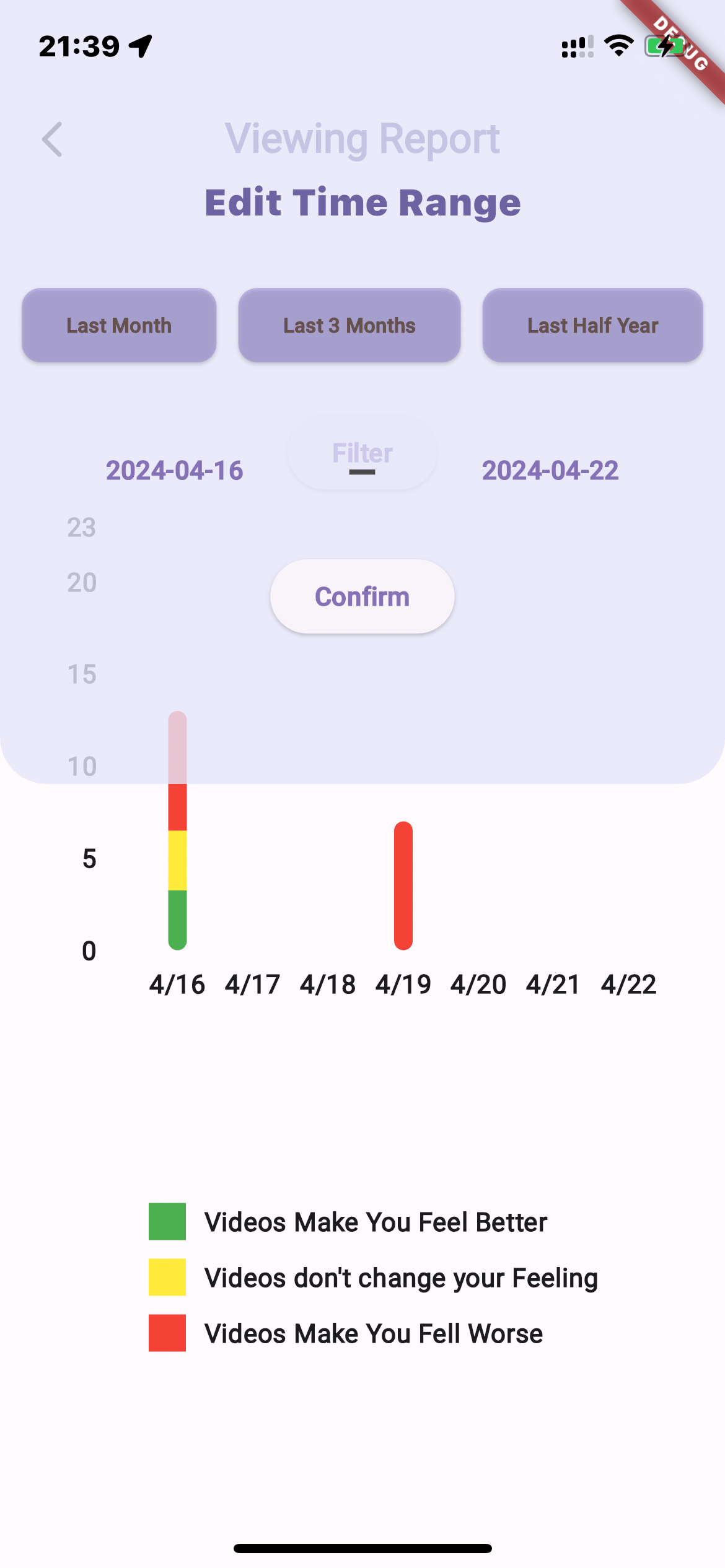}
    \caption{Filter Function to Select Time Range}
    \label{fig:report-2}
  \end{minipage}
\end{figure}

After a user presses the ``View Report'' button (Figure \ref{fig:upload-report}), EmoTrack first runs an asynchronous function that makes an HTTP request and sends a command to the server, calling the corresponding function (see Section \ref{sec:filter-scrape}) in the back-end to process the data, filter relative videos' information according to the time range and then analyze the mood records. When this processing is completed, a new UI page pops up, which shows the report for the user’s YouTube viewing behaviours of the previous week by default (Figure \ref{fig:report-1}).

\subsection{Report Content and Selectable Time Range}

\begin{figure}[!h]
  \centering
  \begin{subfigure}{0.49\textwidth}
    \centering
    \includegraphics[width=0.7\textwidth]{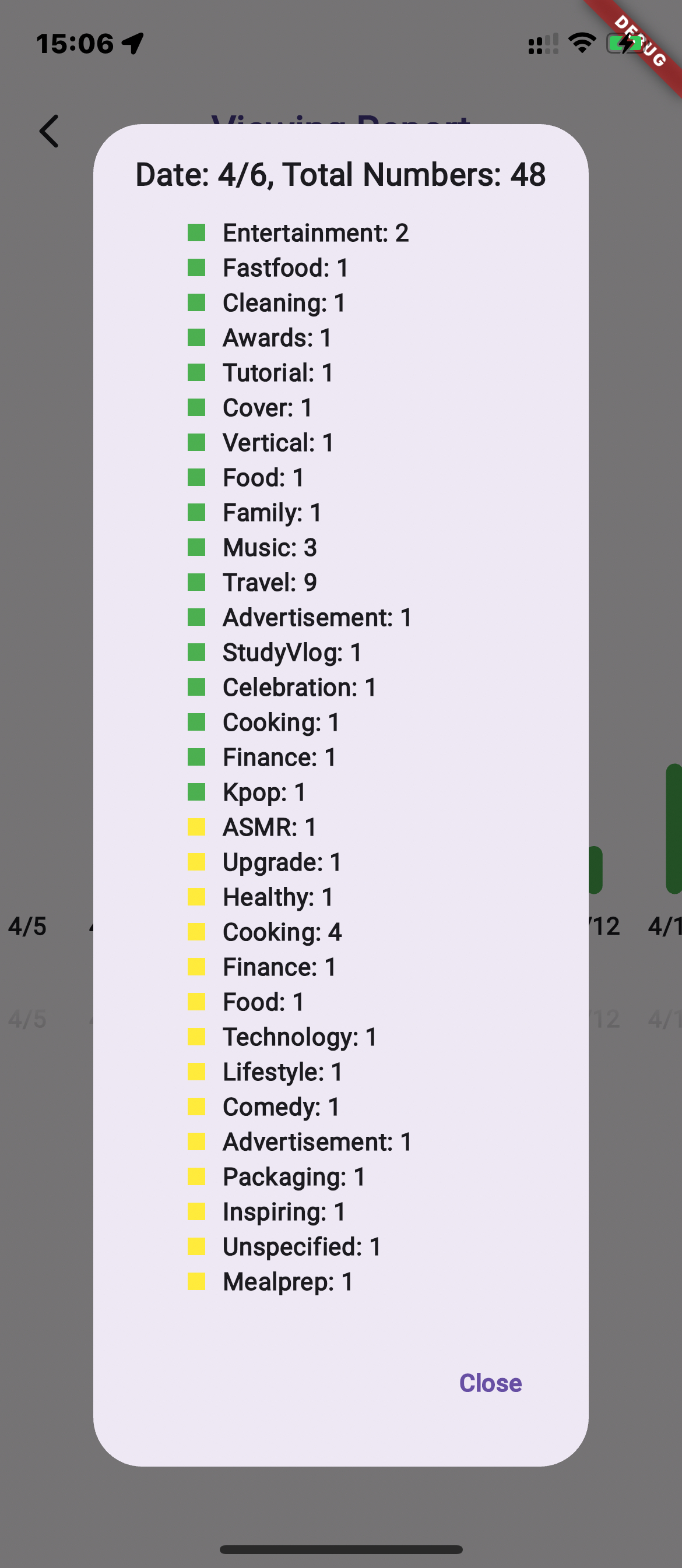}
    \caption{}
    \label{fig:report-3}
   \end{subfigure}
   \hspace*{\fill}
   \begin{subfigure}{0.49\textwidth}
    \centering
    \includegraphics[width=0.7\textwidth]{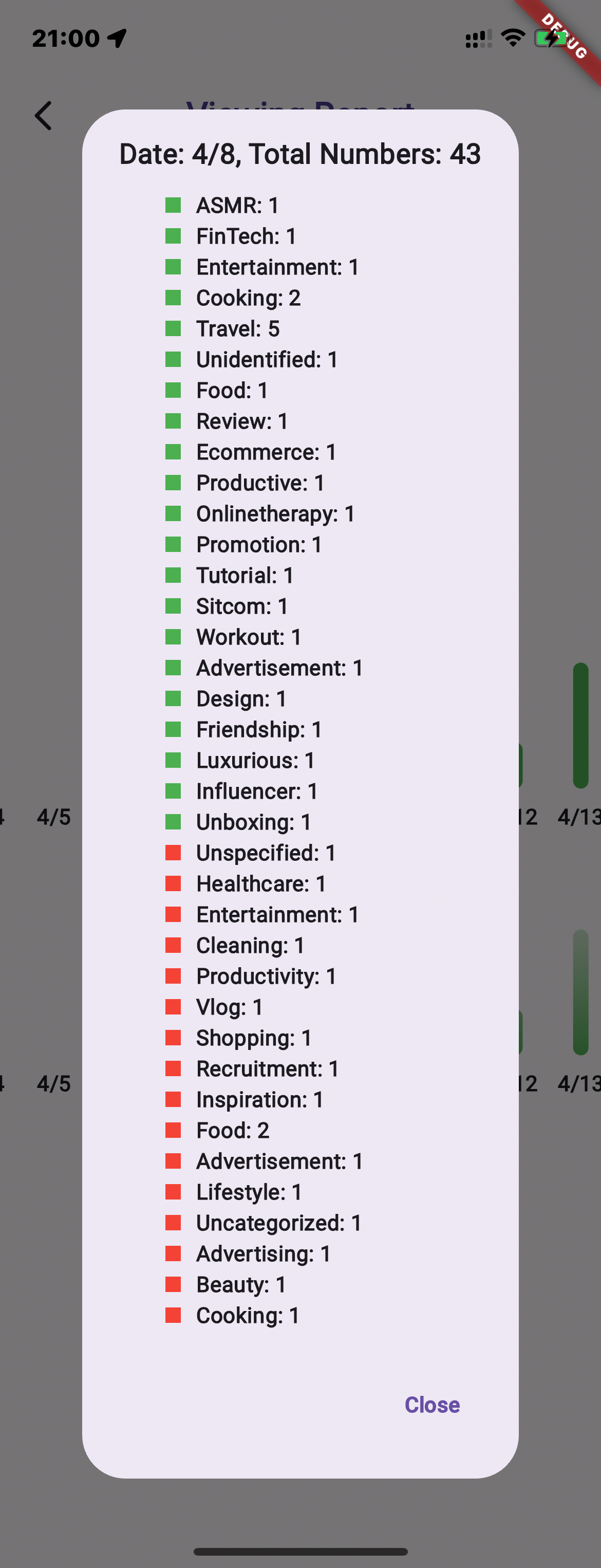}
    \caption{}
    \label{fig:report-4}
   \end{subfigure}
   \caption{Example User Interface for viewing video categories watched during a specific time period}
   \label{fig:report-34}
\end{figure}

To visualize the analysis results, I used the chart library from Flutter \cite{r42}. It was decided to use a bar chart because it can clearly show how each date relates to videos and mood changes. While the X-axis is the date, the Y-axis represents the number of videos watched on that day, with the height of the Y-axis determined by the highest number of videos watched every day among all dates. The colour of each column shows how a user's mood changes while they are watching YouTube videos, as recorded by EmoTrack. It has been specified in the interface (Figure \ref{fig:report-34}), Green means the user’s mood gets better, yellow implies the user’s mood does not change, and red represents the user’s mood getting worse.

Clicking on the column, a more detailed report is presented (Figure \ref{fig:report-34}), which shows on the specific date how many videos the user has watched in total, and all the categories of videos that were watched are listed below, as well as the tiny cube in the corresponding colour of how mood changes.

This report is generated using data from the `Details' collection (see Section \ref{sec:report}). In the `Report' collection in the database, there are several documents named by the date. Each date document contains two sub-collections, `Details' and `Summary'. The document in the `Summary' collection stores the total watched number for that date, and the number of each status of changing in mood. For example, if a user watches three times of YouTube in one day, the changes in his/her mood are `Feel Better' twice and `Not Change' once, then the document in the `Summary' collection should be written as `Better: 2' and `Same: 1', that is the reason that the column is a combination of green and yellow colours. Documents in the `Details' collection are each mood record in one day. Using the same example to explain is that there are three documents in the `Details' collection, two of them have the status as `Better' and one as `Same'. In each document, there should be various categories of watched videos during that time period. All categories in each document are presented in the detailed report, and the colour of tiny cubes in front of each category is decided by the status, green for `Better', yellow for `Same' and red for `Worse' as usual.

By default, the time range of the report is the previous week, including the current date. If users want to view the report for another time range, they can click the ``Filter'' button (Figure \ref{fig:report-2}) to see whether they would prefer ``Last Month'', ``Last Three Months'' or ``Last Half Year''. For other specific date range requirements, users can press the text button showing the current date range and edit it to the date they want.

\subsection{Web Application Hosting}
After building the EmoTrack for iOS and Android, the same codebase can be further deployed on Firebase Hosting for web applications. Use Flutter's build system to compile the application into a web-optimized format. This generates static assets that are ready to deploy, including HTML, CSS, and JavaScript files. After modifying the generated `firebase.json` configuration file to define the settings for Firebase Hosting, and specifying the folder where the web application is built. Once deployed, EmoTrack can be assessed from any browser on mobile phones, computers or pads.

However, there is a new problem coming out. Web applications operate differently because they run within browsers, which enforce the same-origin policy as a security measure. The same-origin policy restricts scripts running in a browser from making requests to a domain different from the one that served the web page. Cross-Origin Resource Sharing (CORS) is a mechanism that allows web pages to request resources from a different domain. To allow these requests, the target server must include specific CORS headers in its response.

\begin{lstlisting}[language=Python, caption={Example Code for CORS Configuration}, label={lst:cors}]
    app = cors(app, 
           allow_origin="https://sms-app-project-415923.web.app", 
           allow_methods=["GET", "POST", "OPTIONS"],
           allow_credentials=True)
\end{lstlisting}

After reading the logs provided by Google App Engine, I added a few lines of code as shown in the code snippets \ref{lst:cors}, including the CORS configured to set the allowed origin, which was the website of EmoTrack's domain. The allowed methods were because I found that the browser always wrapped the POST method with the OPTIONS method. As well as allowing the "assess control credentials" to be true. As a result, adding these lines fixed the problem.

\section{Backend Implementation}
\label{sec:back-end}
This section will discuss all the back-end functions written in Python, the results they achieved, as well as the problems I encountered during implementation, and the modifications or reconstructions I made to solve them.

\subsection{Get Users’ YouTube Watch History}
\label{sec:scrape}

The approach to obtain users' YouTube watch history developed in this project is to acquire users to download their YouTube watch history from Google Takeout manually and upload it to EmoTrack. Here I will explain how to read the file uploaded by the user and dump the data into the database in the back-end development.

The file uploading function is included in the front-end of EmoTrack, whereas Flutter is well integrated with Firebase and it can handle the task of uploading files to Firebase Cloud Storage (see Section \ref{sec:ui-upload}). The function in the back-end needs to access the Cloud Storage, find the particular file uploaded by the user according to the user’s UID, read each line in order and take out specific data. Here the data I required to be transferred to the database is the ``Watched Time'' and URL of each video. Information for each video will be stored in an individual document, the names of documents in the Firestore Database are unique and will be the ID of the document as well. I use the watch time to be the name, as it is impossible to have more than one video to be watched at the same time on YouTube. Before writing the ``Watched Time'' and URL into the document, I wrote a Python function to convert the time in the format of ISO 8601 to the conventional format as ``Year-Month-Day Hour:Minute:Second'' with the data type of String.

The problem that occurred here was the time zone Google used in the YouTube watch history is UTC (Universal Time), as we all know there is daylight saving time in Britain, during that period Britain is in a time zone that is one hour different from UTC, also the start and end date of daylight saving are different every year. While in EmoTrack, the time that users start or stop watching and select the mood is recorded depending on the local time. To solve the time mismatching, I converted the time in YouTube watch history to the local time zone with a library \verb|tz| in Python \cite{r41}, as shown in the code snippets \ref{lst:time-zone}.

\begin{lstlisting}[language=Python, caption={Example code for converting time formats and time zones}, label={lst:time-zone}]
    def convert_time_str(iso_time_str):
        time_utc = parser.isoparse(iso_time_str)
        london_tz = tz.gettz('Europe/London')
        dt_uk = time_utc.astimezone(london_tz)
        dt_uk_str = dt_uk.strftime('%Y-%m-%d %H:%M:%S')
        return dt_uk_str
\end{lstlisting}

\subsection{Filter videos and scrape information}
\label{sec:filter-scrape}

Firstly, each time when users click the ``View Report'' button, a specific time range is sent to the back-end through the particular API endpoint (details in Section \ref{sec:api}). There is a Python function that receives the start and end date of the time range and filters the videos watched within this time scope. Then only videos that meet this requirement will be scraped for their titles, descriptions and categories selected by the author. 

\subsubsection{Initial Attempted Method}
Initially, I attempted the method called web scraping to obtain videos' information (described in Section \ref{sec:web-scraping}). I leveraged Python with Selenium, which is a tool for automating web browsers to collect video information. The process begins with the initialization of a Selenium-driven browser session, where Python scripts programmatically control a Chrome browser to interact with YouTube’s user interface. Notably, the method handles dynamic web elements and JavaScript-driven content, making it possible to extract data that is not readily accessible through YouTube's official Data API \cite{r51}.
The procedure outlined includes several critical steps:

\begin{enumerate}
    \item Setting up the project environment and installing necessary Python libraries such as Selenium and webdriver-manager.
    \item Navigating to the target YouTube video page using the browser controlled by Selenium.
    \item Managing cookies and user consent dialogues to ensure uninterrupted access to the page content.
    \item Employing CSS selectors and XPaths to locate and extract specific data points from the web page, such as video titles, channel information, and viewer statistics.
    \item Closing the browser session and outputting the scraped data into a structured format like JSON for further analysis or integration into data systems.
\end{enumerate}
 
However this method did not work after I packaged the back-end functions and deployed it as the server to Google App Engine, since App Engine does not contain the Chrome web browser, it is impossible to drive Chrome and complete the web scraping, so I used YouTube Data API instead for the development of EmoTrack in the current version.

\subsubsection{Alternative Method}

Here are the code snippets \ref{lst:api} showing how I used the YouTube Data API to get the information of videos:

\begin{lstlisting}[language=Python, caption={Example Code for Getting Information using YouTube API}, label={lst:api}]
    async def new_scrape_info(url):
      parsed_url = urlparse(url)
      if parsed_url.netloc == 'www.youtube.com':
        parsed_qs = parse_qs(parsed_url.query)
        video_id = parsed_qs.get("v", [None])[0]
        
        request = youtube.videos().list(
            part="snippet",
            id=video_id
        )
        response = request.execute()
        
        if response['items']:
          video_title = response['items'][0]['snippet']['title']
          video_description = response['items'][0]['snippet']['description']
          video_category_id = response['items'][0]['snippet']['categoryId']
          
          category_request = youtube.videoCategories().list(
              part="snippet",
              id=video_category_id
          )
          category_response = category_request.execute()
          video_category = category_response['items'][0]['snippet']['title'] if category_response['items'] else 'null'
          return video_title, video_description, video_category
        else:
          video_title = 'null'
          video_description = 'null'
          video_category = 'null'
          
        if not video_title or not video_description or not video_category:
          return 'null', 'null', 'null'
        
        return video_title, video_description, video_category
      return 'null', 'null', 'null'
\end{lstlisting}

\subsection{Categorize videos with ChatGPT}
\label{sec:categorize}

Although YouTube mandatorily requires users to choose the category for videos before they upload them, there are only fifteen categories, Cars and Vehicles, Comedy, Education, Entertainment, Film and Animation, Gaming, How-to and Style, Music, News and Politics, Non-profits and Activism, People and Blogs, Pets and Animals, Science and Technology, Sport, Travel and Events.

\begin{lstlisting}[language=Python, caption={Example Code for Categorising videos with ChatGPT}, label={lst:chatgpt}]
    content =  "The YouTube video has the title: " + str(video_title) + ", and the description: " + str(video_description) + ". Categorise this YouTube video in only ONE word strictly."
    chat_completion = client.chat.completions.create(
      model="gpt-3.5-turbo",
      messages=[{"role": "user", "content": content}]
    )
    video_category = chat_completion.choices[0].message.content
\end{lstlisting}

To have more fine grained categories. I used ChatGPT to determine the category of videos according to their titles and descriptions. The specific version of the ChatGPT model I used here is GPT-3.5(gpt-3.5-turbo). The example code is shown as \ref{lst:chatgpt}, sent ChatGPT the title and description of the video and asked it to return the category limited to one word. However, sometimes it would still return more than one word, which would cause some problem when transferring data between database, to solve this problem, I wrote one more line code (\ref{lst:sub}) to remove any spaces and any special symbols if the category contains.

\begin{lstlisting}[language=Python, caption={Example code to remove any spaces and special symbols}, label={lst:sub}]
    category = re.sub(r'\W+', '', category)
\end{lstlisting}

\subsection{Generate Both the Detailed and Summative Reports}
\label{sec:backend-4}
This section introduces the report generation feature in the back-end, the data processed will be stored in the corresponding location in the database, which will be explained in Section \ref{sec:report}.
\subsubsection{The Detailed Report}
With the time range sent from the front-end through the particular API endpoint, all mood records within this time range will be filtered. Inside each mood record, there are the ``Mood Before Watch'' and ``Mood After Watch'', after calculating the relationship between ``Good'', ``Okay'' or ``Not Okay'', the status shows how mood changes such as ``Feel Better'', ``Not Change'' or ``Feel Worse'' will be stored in the same location in the database as each mood record.

According to the ``Start Watching Time'' and the ``Stop Watching Time'' of each mood record, it will go through each video’s ``Watch Time'' in the entire YouTube watch history and filter out all eligible videos. Then the number of these videos, as well as all videos’ categories, will be stored in the database, along with each corresponding mood record.

\begin{lstlisting}[language=Python, caption={Example Code for Summing Categories and Video Number}, label={lst:report}]
mood_ref = db.collection('Users').document(uid).collection('Mood Records')
mood_docs = mood_ref.stream()
for mood_doc in mood_docs:
    mood_data = mood_doc.to_dict()
    if 'After Watch Mood' in mood_data and 'Before Watch Mood' in mood_data:
        history_ref = db.collection('Users').document(uid).collection('YouTube Watch History')
        history_docs = history_ref.stream()
        for history_doc in history_docs:
            video_data = history_doc.to_dict()
            watch_time = video_data.get('time', 'null')
            watch_time = datetime.strptime(watch_time, "%Y-%m-%d %H:%M:%S")
            
            if start_time <= watch_time <= end_time:
              watch_total_number += 1
              video_category = video_data.get('category', 'null')
              if str(video_category) in category_counts:
                category_counts[str(video_category)] += 1
              else:
                category_counts[str(video_category)] = 1
\end{lstlisting}

\subsubsection{The Summative Report}
The summative report will be generated after all detailed reports have been produced, As shown in the code sample \ref{lst:report}, it calculates the number of the same change in mood within each day, as well as the sum of the total number of watched videos in each detailed mood record.

\section{Database Structure}

\begin{figure}[!h]
    \centering
    \includegraphics[width=\textwidth]{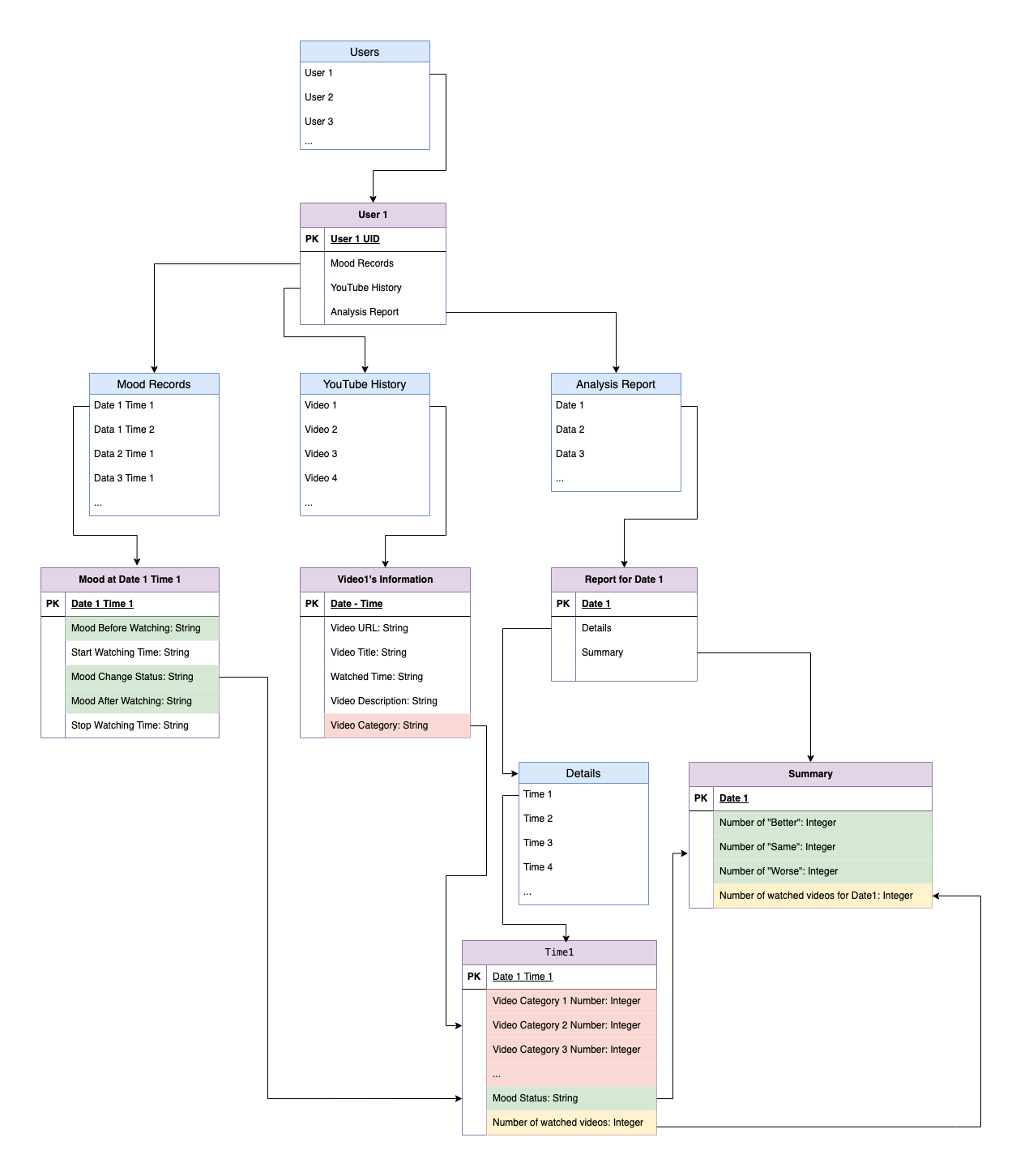}
    \caption{EmoTrack Database Structure}
    \label{fig:DB-1}
\end{figure}

As stated in Section \ref{sec:firestore-db}, the Firestore Database is a NoSQL document database, consisting of several collections and documents.
As shown in Figure \ref{fig:DB-1}, I use different colours to present different features in the database: textboxes in blue are for collections and textboxes in purple are for documents. It can be seen that all documents have a component called ``PK''. I used this to represent the unique document ID for each document. Other fields and variables with the same colour mean they are variables with the same attributes, taken from other documents or calculated by variables in the same type in other documents.

Various unique users’ UIDs are under the biggest collection ``Users''. For each user, there are three sub-collections:

\begin{enumerate}
    \item \subsubsection{Mood Records} \label{sec:db-mood}

Every time users plan to watch videos on YouTube. They go to EmoTrack first, press the ``Start'' button and record their mood at that moment. After users finish watching, they come back to EmoTrack, press the ``Stop'' button and record their mood at that moment as well.

\begin{lstlisting}[caption={Example Code for Sending Mood Records to Database}, label={lst:mood}]
void saveWatchRecord(String value, DateTime time) {

    final userDocRef = FirebaseFirestore.instance.collection('Users').doc(FirebaseAuth.instance.currentUser!.uid).collection('Mood Records');

    if (_addNewDoc) {
      fileName = DateFormat('yyyy-MM-dd HH:mm:ss').format(time);
      userDocRef.doc(fileName).set({
        'Before Watch Mood': value,
        'Start Watch Time': DateFormat('yyyy-MM-dd HH:mm:ss').format(time),
      });
    } else {
      userDocRef.doc(fileName).update({
        'After Watch Mood': value,
        'Stop Watch Time': DateFormat('yyyy-MM-dd HH:mm:ss').format(time),
      });
    }
}
\end{lstlisting}

Once they press the ``Confirm'' button on the mood-selecting floating window, the mood they have chosen is recorded and sent to the database as well as the String converted from the real-time and date. As illustrated by the code snippets \ref{lst:mood}, I named the document in the ``Mood Records'' collection with the start time and set a new document to store the ``Mood Before Watching'' and ``Start Watching Time'' data. The ``Mood After Watching'' and ``Stop Watching Time'' data are updated to the same document created when `started', using the same document ID (name).
The last field ``Mood Change Status'' stores the data processed through the Python function (see Section \ref{sec:backend-4}).

    \item \subsubsection{YouTube (Watch) History}

Documents in this collection are filtered and saved from the JSON files of YouTube watch history uploaded by users. Only the URLs and watch time of videos will be taken out from the whole file, the watch time reported in the watch-history file provided by Google Takeout is formatted by ISO 8601 standard so I used it as the name of the document. The time stored in the document as a field is converted to a string in the conventional format through the Python function (see code snippets \ref{lst:time-zone}).

    \item \subsubsection{Analysis Report} \label{sec:report}

Documents in this collection are named with date, formatted by ``Yead-Month-Day'', each document has two sub-collections:

\begin{itemize}
    \item Details
    
    In this collection, documents with the names of all start times that happened on that date, are taken from the corresponding documents in the ``Mood Records'' collection.

    Various ``Video Category'' and the ``Number of watched videos'' in each document are processed by a Python function (see Section \ref{sec:backend-4}). As shown in the code snippet \ref{lst:report}, it goes through each document in ``YouTube History'' collection and filters by each video’s watched time according to the time range of users’ periods of watching YouTube (i.e. from the ‘Start Time’ to the ‘Stop Time’). After calculating the total number of watched videos in that period and video categories, these data are saved to the corresponding locations in each document.

    \item Summary
    
    Each document is named with the date in the format of ``Yead-Month-Day'' and will be generated after all documents in the ``Details'' collection of that date have been completed. The Python function (in Section \ref{sec:backend-4}) calculated the number of ``Mood Change Status'' (``Better'', ``Same'' and ``Worse'') and summed the total number of watched videos from each document in the `Details' collection. For example, if the user watched four times YouTube videos on 1st April at 9, 12, 19 and 23 o’clock, with four ``Mood Change Status'' as two ``Better'', one ``Same'' and one ``Worse'', then ``Better: 2'', ``Same: 1'' and ``Worse: 1'' will be stored into his/her summary report of 1st April, as well as its total watched videos number will be summed and saved.

\end{itemize}

\end{enumerate}

\section{Server}
The approach to transfer data between the front-end and the back-end in this project is to transmit data in JSON scheme and make HTTP requests to send through API endpoints from the client (front-end) to the server (back-end). This section will explain how to construct the server and how HTTP requests work.

\subsection{Initial Thinking}
Initially, the Flask RESTful API framework was used by the Python server to implement the back-end of EmoTrack. It is responsible for handling multiple types of requests from the front-end application and includes the following functionalities:

\begin{itemize}
    \item Accept and process commands sent from the front-end, serving as the primary point of communication between the client and server.
    \item Call the corresponding Python functions according to the commands and API endpoints accepted.
\end{itemize}

Flask primarily follows a synchronous request handling model, meaning that it processes each request sequentially. This design choice implies that each incoming request is handled one at a time by the server's worker thread until it completes, potentially impacting performance for I/O-bound applications or scenarios that require high concurrency.

However, asynchronous functions are usually chosen when it comes to taking or putting data from the database to prevent blocking the main thread, causing the application to lag or stop responding, which seriously affects the user experience. As has been mentioned in Section \ref{sec:report}, the detailed and summative reports are generated after the data of video information and mood records have been processed. To solve the issue here, I replaced the Flask with Quart.

\subsection{Improved Framework}
Quart is an asynchronous web framework that extends the API of Flask to provide full support for asynchronous programming. By leveraging Python's asyncio library, Quart allows the use of async and await syntax to enable non-blocking I/O operations and highly concurrent request handling. This enables the framework to efficiently handle multiple simultaneous connections, improving performance for I/O-bound tasks such as database queries or external API calls. Quart's native support for asynchronous functions allows developers to implement complex workflows while maintaining optimal server responsiveness.

After switching the framework from Flask to Quart, most of the codes can be used without modification since Quart is the asynchronous framework extension from Flask, the improved framework implemented the back-end as shown in Figure \ref{fig:server}:

\begin{figure}[h]
    \centering
    \includegraphics[width=\textwidth]{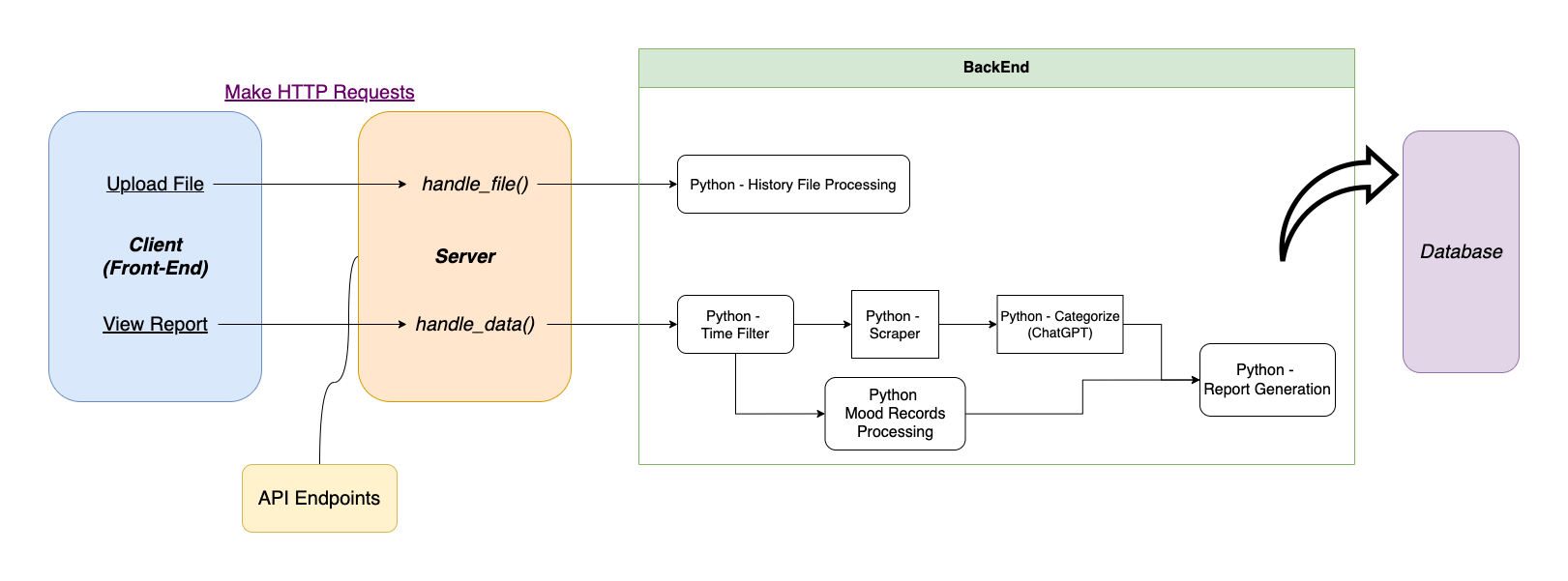}
    \caption{Architecture Diagram of Server}
    \label{fig:server}
\end{figure}

Every time users press the buttons, the client (front-end) makes the HTTP requests to the server (back-end):

\begin{enumerate}
    \item When ``Upload File'' Button is pressed: the server receives a Boolean value from the client to confirm the success or failure of the user file uploads. When the Boolean value is true, initiate the processing of the corresponding Python function that uploads the required information in the file uploaded by users to the database. Performing tasks including data validation, format conversion, etc.
    \item When ``View Report'' Button is pressed, the server accepts commands from the client to generate analytical reports based on the user’s data. It Invokes specified Python functions to fetch and analyze data stored in the database. The first is to process the data from video information and mood records. Then, it generates the reports.
\end{enumerate}

\subsection{API Endpoints}
\label{sec:api}
The API endpoints contained inside the buttons of EmoTrack are as follows:
\begin{itemize}
    \item \verb|handle_file()|
    \item \verb|handle_data()|
\end{itemize}

The \verb|handle_file()| function handles the POST request from the client. Due to the specificity between Flutter and Firebase Cloud, the front-end can transfer files to the database directly without making requests to the back-end. While it is still necessary to use the Python functions from the back-end to execute the required information from the file in the Firebase Cloud Storage and transmit it to the Firestore database. As mentioned in Section \ref{sec:ui-upload}, the uploadTask function monitors whether the file is uploaded successfully or not. A Boolean value showing the status of the upload, as well as the user-UID will be sent through the particular API endpoint (\verb|'/api/handle_file'|) to the \verb|handle_file()| function. When the Boolean value is true, the corresponding Python function will be called to process and convert the data from the JSON file to the database (see Section \ref{sec:scrape}).

The \verb|handle_data()| function also handles the POST request from the client. After the users press the ``View Report'' button (Figure \ref{fig:upload-report}) or the ``Confirm'' button inside the ``Filter'' pop-up window (Figure \ref{fig:report-2}), the client sends the user-UID and a time range through the particular API endpoint (\verb|'/api/handle_data'|) to the \verb|handle_data()| function. The first step is to filter out all the videos in the user's entire YouTube watch history within that time range, then use the YouTube data API to obtain the titles and descriptions of those videos, and use those titles and descriptions to categorize the videos with ChatGPT (see sections \ref{sec:filter-scrape} and \ref{sec:categorize}). Awaiting everything in the first step has been done, another Python function will be called to generate the detailed and summative reports, as explained in Section \ref{sec:backend-4}.

\subsection{Hosting}
After the server and API endpoints are established and work well locally, they can be further deployed to Google App Engine. Once deployed, these endpoints can be set up to integrate with the front-end of EmoTrack, which means that each user assessed the UI of EmoTrack will be directed to these endpoints in Google App Engine for further processing.

\section{Summary}
In summary, this chapter explained the complete development process of EmoTrack from different perspectives including the front-end UI design, the server hosting on the Cloud established by the back-end functionalities, and the integration with the database. Ensuring the usability of EmoTrack on multiple platforms. As well as the challenges encountered during the execution and how I resolved them.

Through long and thorough editing and iterations, EmoTrack has achieved the initially desired functionalities of tracking changes in users' mood and their YouTube activities, and analyzing the relationship between them. In addition, EmoTrack is extensible for future development.


\chapter{Critical Evaluation}
\label{chap:evaluation}

\noindent

\section{Introduction}

This chapter describes the evaluation of EmoTrack application from both the technical development and research perspectives. The evaluation consists of quantitative and qualitative approaches to analyze EmoTrack in terms of its user interface usability, functionalities, and how well it meets user requirements from the human-interaction computer area. EmoTrack was evaluated using a number of techniques, each of which is described in detail in the sections below. User Testing aims at investigating the usability and acceptability among the objectives, students, and collect their data to summarize the general effects that social media, especially YouTube have on students. User Feedback Survey Analysis introduces the ``quick and dirty'' reliable tool \cite{r17} System Usability Survey (SUS) and utilize it to find out whether EmoTrack has met users’ requirements. User Interview Analysis aims at achieve users’ in-depth thoughts about EmoTrack with all aspects and their valuable suggestions. The objective of Self Reflection is to anaylze the application overall from both the developer's and user's viewpoints.

\section{User Testing}
\subsection{Introduction}
The user testing aimed to verify the usability and effects of EmoTrack on real users, especially young people. Thirteen participants (11 females and 2 males) aged 20 to 25, all UK university undergraduate and postgraduate students, were invited to use EmoTrack for one week for testing and their data were analyzed quantitatively and qualitatively.

During the one week of user testing, I first showed users how to use EmoTrack at the beginning, including what function each button has, how they can download and upload their watching history, and so on. I also wrote a detailed step-by-step guide for participants to read, which is included in appendix \ref{appx:EmoTrack-guide}. The most important thing that was emphasised to users was that they needed to record their mood before and after watching YouTube.

The major goal of the one-week testing period was to see if users understood how to use EmoTrack, to measure how frequently they used the app and to determine whether it led them to reflect on their YouTube watching habits.

\subsection{User Data Analysis}
\label{sec:user-data}
Each participant used EmoTrack for at least a week. I sampled the data stored in the database on 4/30, and chose the week from 4/22 to 4/28 when each user was using EmoTrack, to conduct a preliminary analysis of the user's behaviour in watching YouTube, which mainly involves the following aspects:

\begin{enumerate}
    \item The frequency of users using YouTube
    
\begin{figure}[!h]
    \centering
    \includegraphics[width=0.5\textwidth]{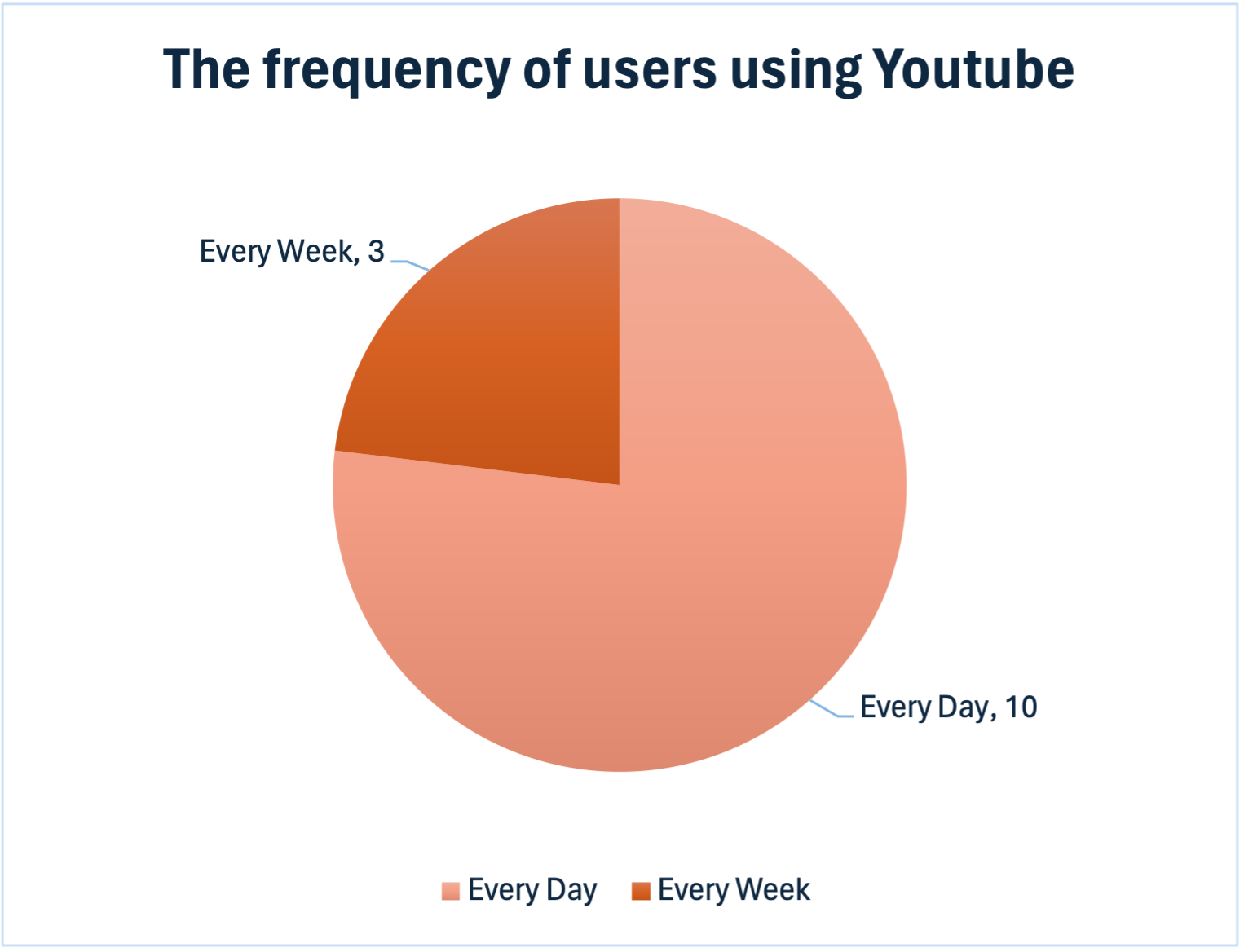}
    \caption{The frequency of users using YouTube}
    \label{fig:eva-1}
\end{figure}
During the one-week of user testing, the number of days that participants watched YouTube were calculated follows: 
\begin{itemize}
    \item Participant who watched YouTube for 5 days or more was considered as ``Every Day''.
    \item Participant who watched YouTube for 2 days or less was considered as ``Every Week''.
\end{itemize}
Figure \ref{fig:eva-1} shows that 10 of the 13 participants use YouTube every day and 3 of them use YouTube every week.

    \item Time spent on YouTube Every Day

\begin{figure}[!h]
    \centering
    \includegraphics[width=0.5\textwidth]{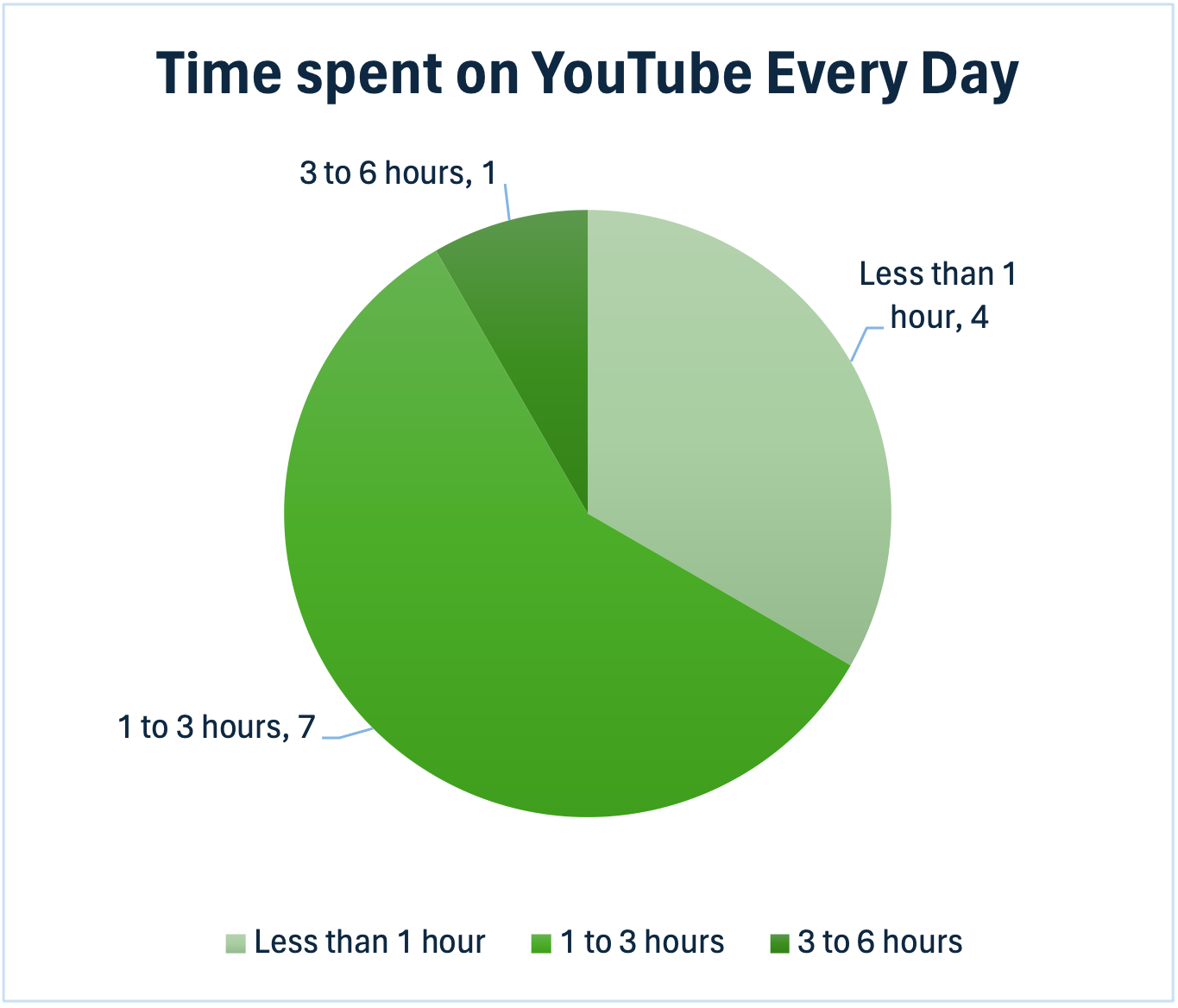}
    \caption{Time spent on YouTube Every Day}
    \label{fig:eva-2}
\end{figure}
These results were calculated by dividing the total watch minutes by each user's total days of watching YouTube, so those users who watched for 2 days or less were also concluded.

According to Figure \ref{fig:eva-2}, I could classify the participants as light users, moderate users, and heavy users. 4 of the 13 participants were light users, 7 were moderate users and 1 was heavy users.

    \item Time periods users mostly watched YouTube

      \begin{figure}[!h]
        \centering
        \includegraphics[width=0.5\textwidth]{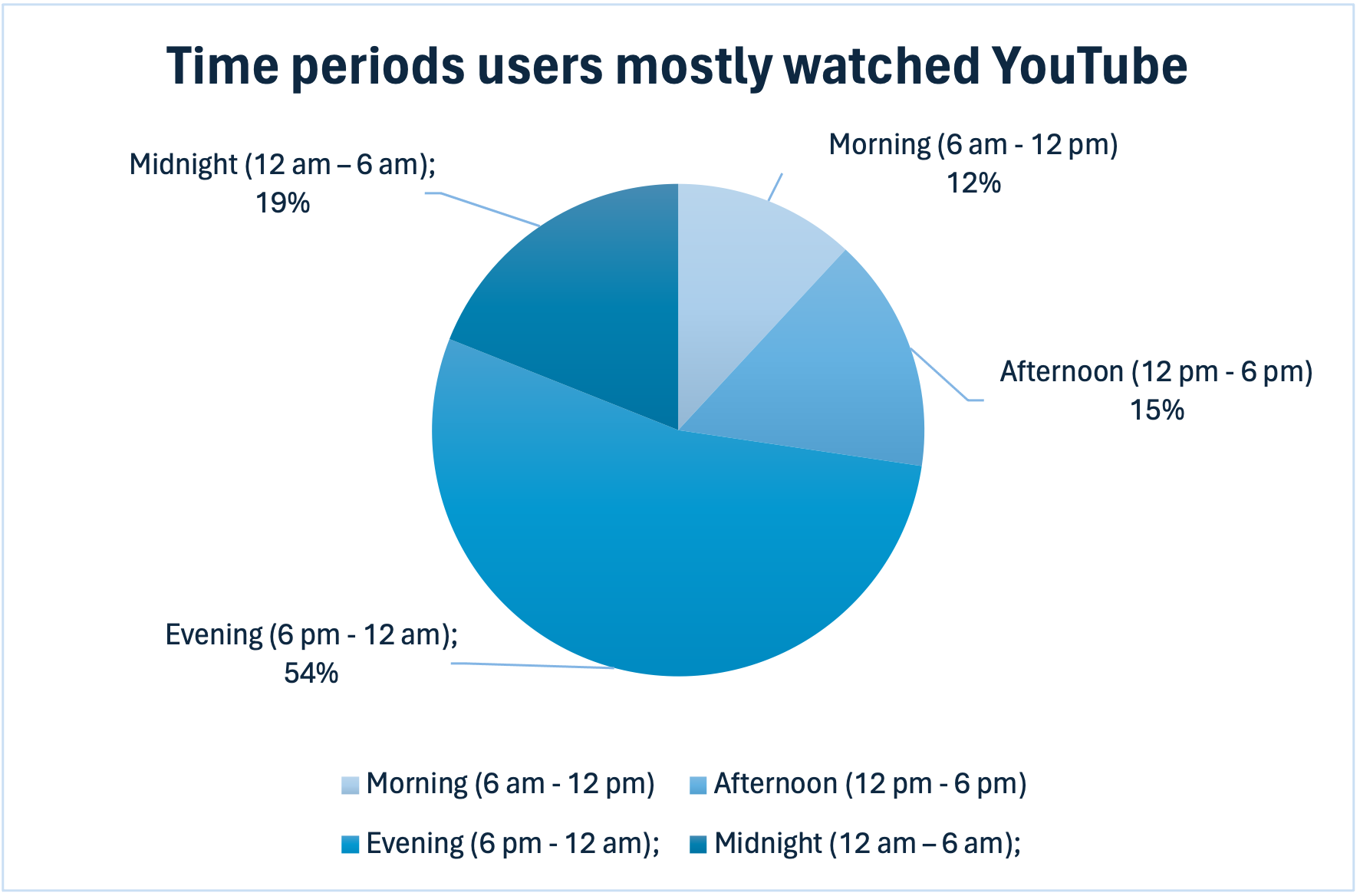}
        \caption{Time periods users mostly watched YouTube}
        \label{fig:eva-3}
      \end{figure}
      The results for this part were calculated with all the data during that week, not differentiated by the user, which represented the overall trend for all participants of at what time period they usually watch YouTube videos.
      
      After counting the total number of videos separated into four different periods using the ``watch time'' of each video stored in the database, Figure \ref{fig:eva-3} illustrated what periods users usually watched YouTube. Users usually watched YouTube in more than one time period, which showed that around 54\% of videos were watched in the evening (between 6 pm and 12 am), mainly because people watch videos after they have finished studying or working and want to relax.

    \item Video types of users prefer (Long vs Short)

      \begin{figure}[!h]
        \centering
        \includegraphics[width=0.5\textwidth]{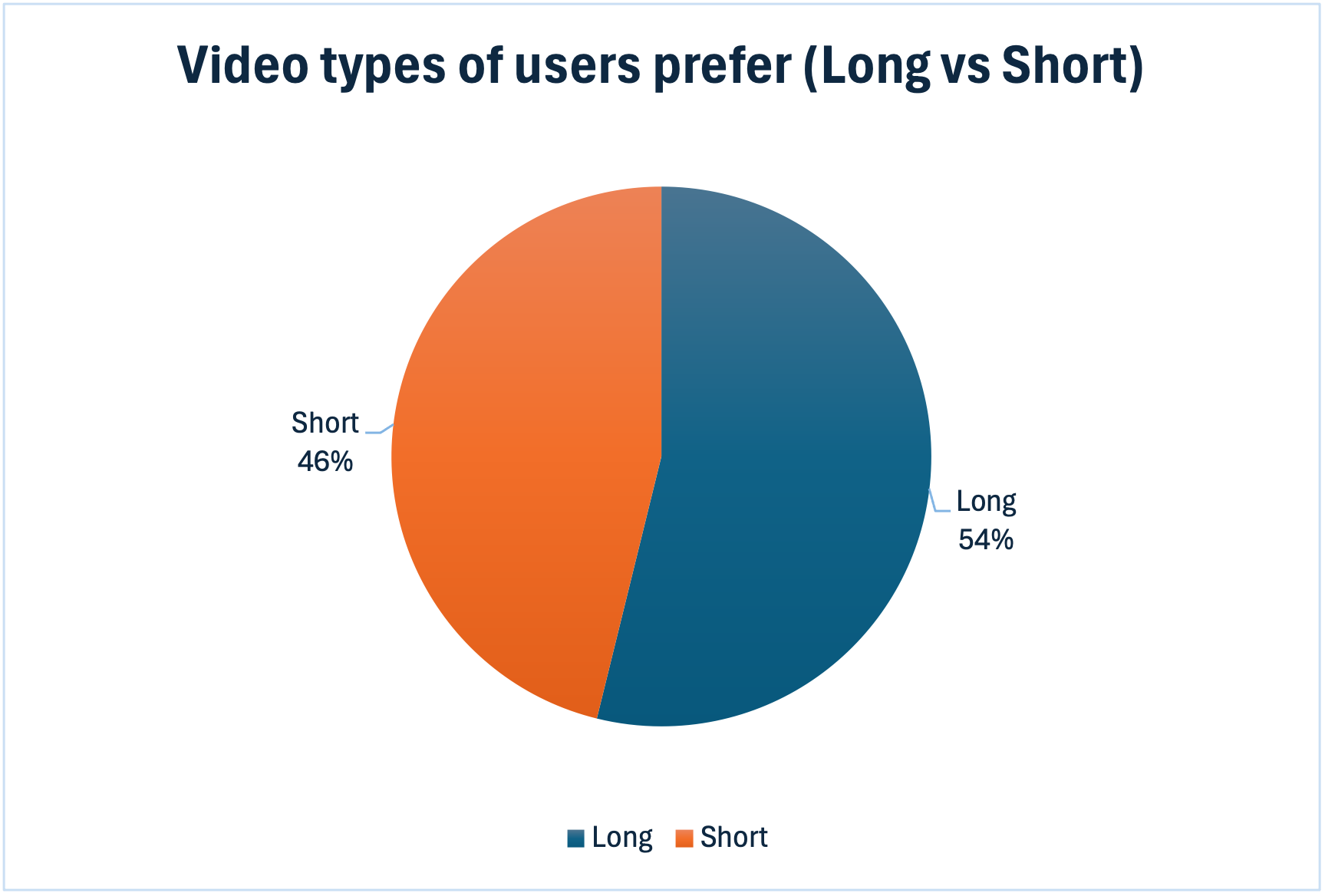}
        \caption{Video types of users prefer (Long vs Short)}
        \label{fig:eva-4}
      \end{figure}
    
    YouTube also has a type called ``YouTube Shorts'', which is specifically created for short videos of less than one minute. To calculate the results, first is to label all the YouTube videos watched in that week as `Long' or `Short', according to the length of the video. Then the number of `Long' and `Short' videos was calculated separately. As Figure \ref{fig:eva-4} shows, users have accepted and grown accustomed to short videos. The proportions of short and long videos on YouTube watched by participants were extremely close.

    \item The relationship between users’ mood and YouTube videos
    
\begin{figure}[!h]
    \centering
    \begin{subfigure}{0.49\textwidth}
        \centering
        \includegraphics[width=\textwidth]{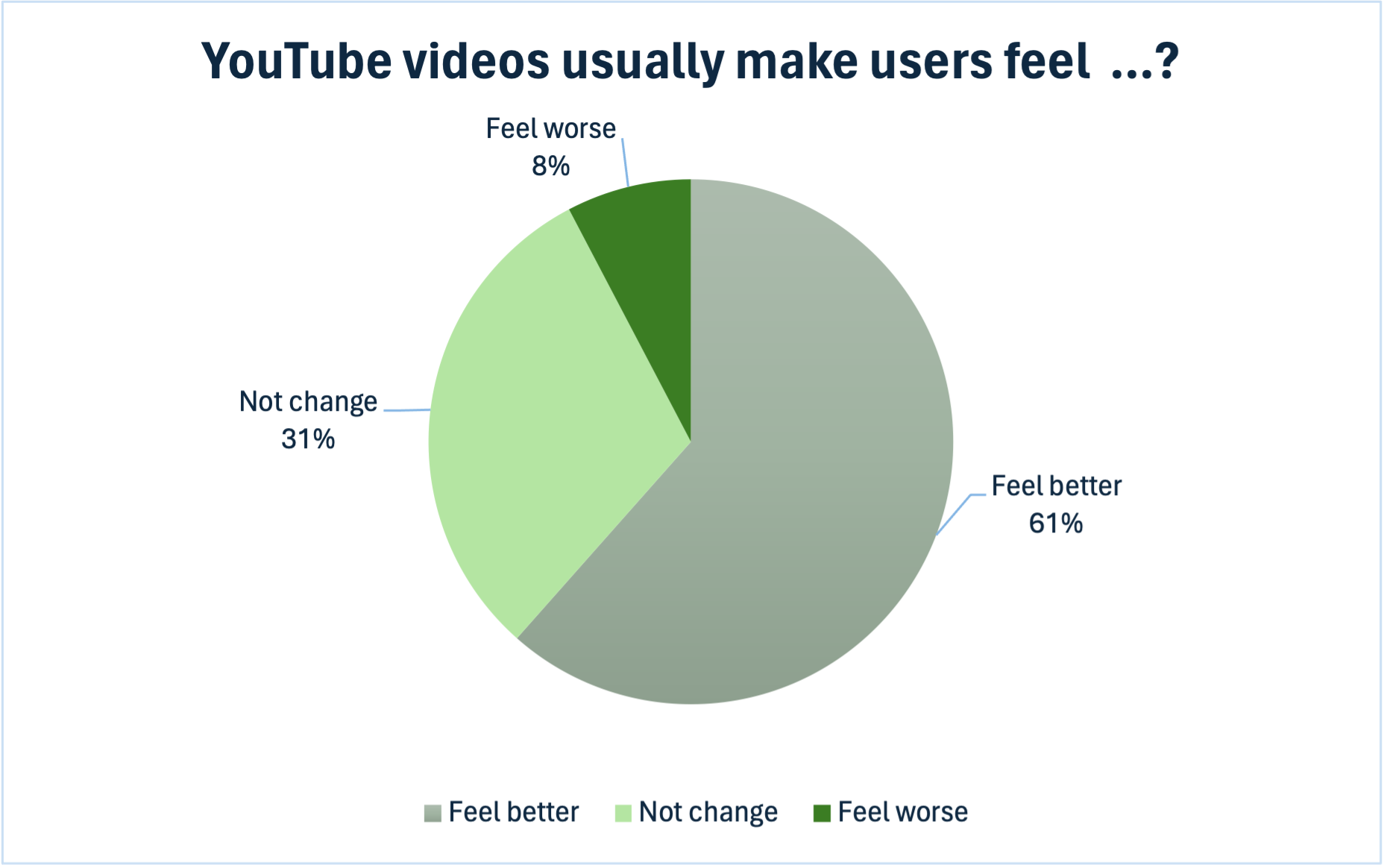}
        \caption{YouTube videos usually make users feel  …?}
        \label{fig:eva-5a}
    \end{subfigure}%
    \hspace*{\fill}
    \begin{subfigure}{0.49\textwidth}
        \centering
        \includegraphics[width=\textwidth]{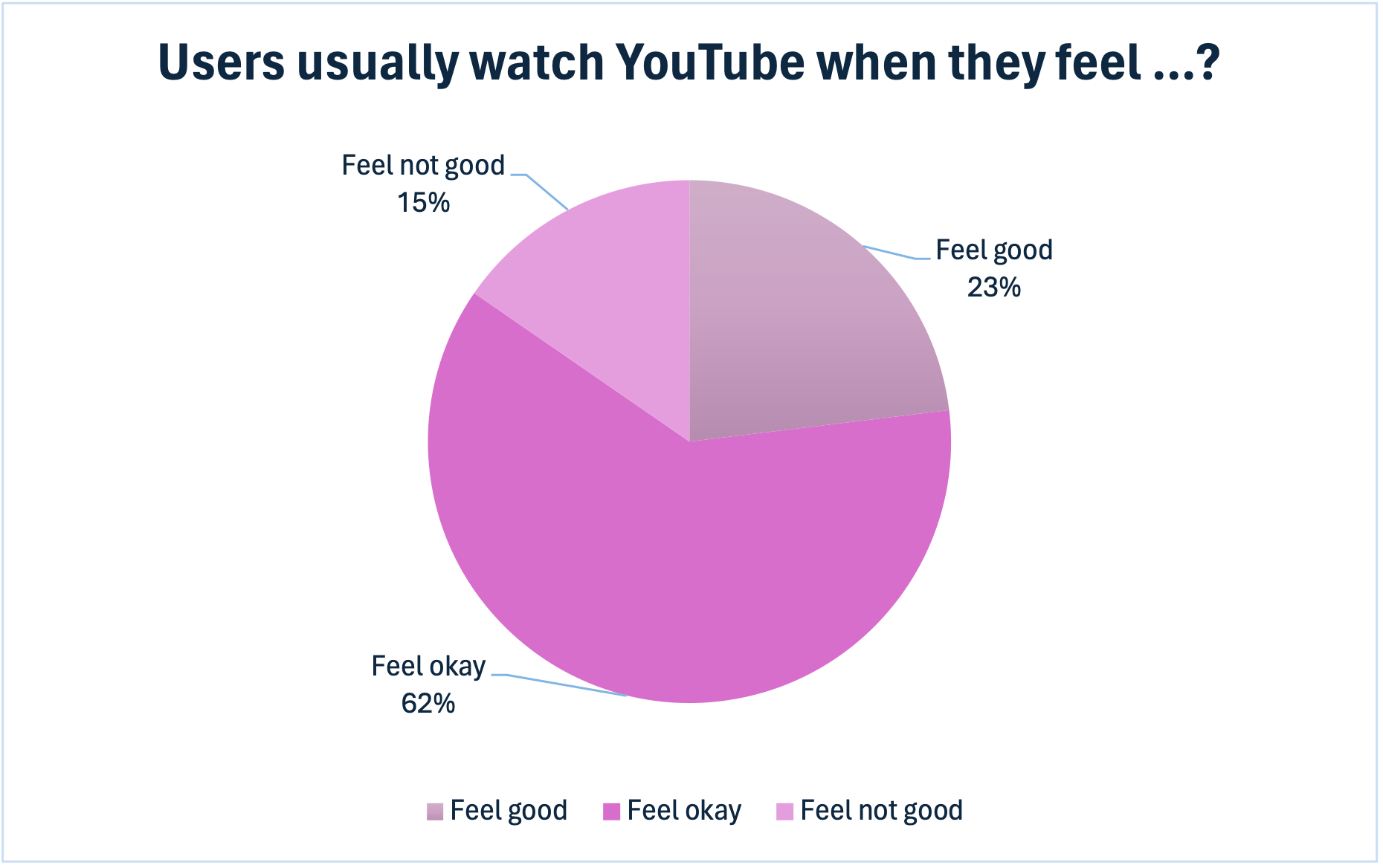}
        \caption{Users usually watch YouTube when they feel …?}
        \label{fig:eva-5b}
    \end{subfigure}
    \caption{The relationship between users’ mood and YouTube videos?}
    \label{fig:eva-5}
\end{figure}

To investigate the relationship between videos participants watched on YouTube and the impact on their mood. All `status' data showing how users' mood changes in the database were counted of each `status' separately as `Feel Better', `Not change' and `Feel Worse'. As shown in Figure \ref{fig:eva-5a}, in the majority, 61\%, of cases, participants felt better after watching YouTube videos. In 31\% of cases, their mood did not change, while in another 8\% cases, YouTube videos made them feel worse.

To find out when users usually start watching YouTube videos was conducted by counting the total number of `Good', `Okay' and `Not good' in the field `Start Watch Mood' for all users in the database for that week. As shown in Figure \ref{fig:eva-5b}, participants usually started watching videos when they felt okay in about 61\% cases. In 23\% of cases, participants started watching when they felt good and in another 15\% cases, they started with feeling not good.
\end{enumerate}

\subsection{Discussion and Summary}
Up to the present, YouTube has had high popularity, especially among young people. However, the results presented in the preceding section could only tell the trend among these thirteen participants. The participants I invited to engage in the user testing used to watch YouTube, at least once every week. Because those people who never or rarely use YouTube cannot provide valuable evaluations regarding the usability and acceptability of EmoTrack. As a result, the results from analyzing the user data during the user testing period can only show the YouTube watching habits among students who usually watch YouTube.

Nowadays, the influence of short videos is rapidly exaggerated. As shown in Figure \ref{fig:eva-4}, among the total number of YouTube videos watched in a week by the 13 participants, short videos accounted for 46 \% of them, which means people generally accept the existence of short videos and their impact may be further extended in the future. Violot et al. (2024) revealed that most channels decreased the number of regular videos they produced while rising and then keeping a steady frequency of uploading YouTube Shorts. They also demonstrated that YouTube Shorts would have increasingly garnered five times more views per video than regular videos by the end of 2022 \cite{r43}.

As a result, the analysis of data extraction from the database would only reveal a simplistic trend among the university students who have the habit of watching YouTube. Furthermore, regarding whether EmoTrack has usability and acceptability among these participants, as well as whether it can facilitate users to reflect on their YouTube watching behaviours according to the impact on their mood, will be presented in subsequent sections.

\section{System Usability Survey (SUS) Analysis}

\subsection{Introduction}

After the one-week trial of EmoTrack, I conducted a surveywith each user to collect feedback and achieve a deep understanding of users’ reflections and viewpoints of EmoTrack.

The survey aimed to measure the usability of EmoTrack, using a widely used questionnaire called System Usability Survey (SUS), which consists of 10 questions with five response options for people to select, ranging from strongly disagree to strongly agree. The SUS has been used to evaluate a wide range of software products and services \cite{r17}. The single number it produces represents a composite measure of the overall usability of the system under study, however, the individual item scores are not relevant on their own. After a more complex calculation than simply summing all the scores, the overall score of SUS ranges from 0 to 100, the testing object with a score higher than 68 would be considered as over average, while anything lower than 68 is below average.

\subsection{SUS Results}

The SUS score was calculated by converting the users’ scores on each question to the range of 1 to 5, where strongly disagree is 1 and strongly agree is 5. Since the SUS questionnaire consisted of both positive statements (the odd numbered questions) and negative statements (the even numbered questions), for each odd-numbered question, the score is calculated by subtracting 1 and for each even-numbered question the score is calculated by subtracting from 5. I calculated the total sum of these calculated scores and multiply it by 2.5, giving an overall SUS score in the range [0 , 100].

The final average score given by all participants that EmoTrack achieved is 79.8, higher than the SUS average, 68, which means EmoTrack is easy to use and understand, and users have a good user experience when using it.

\begin{figure}[h]
    \centering
    \includegraphics[width=\textwidth]{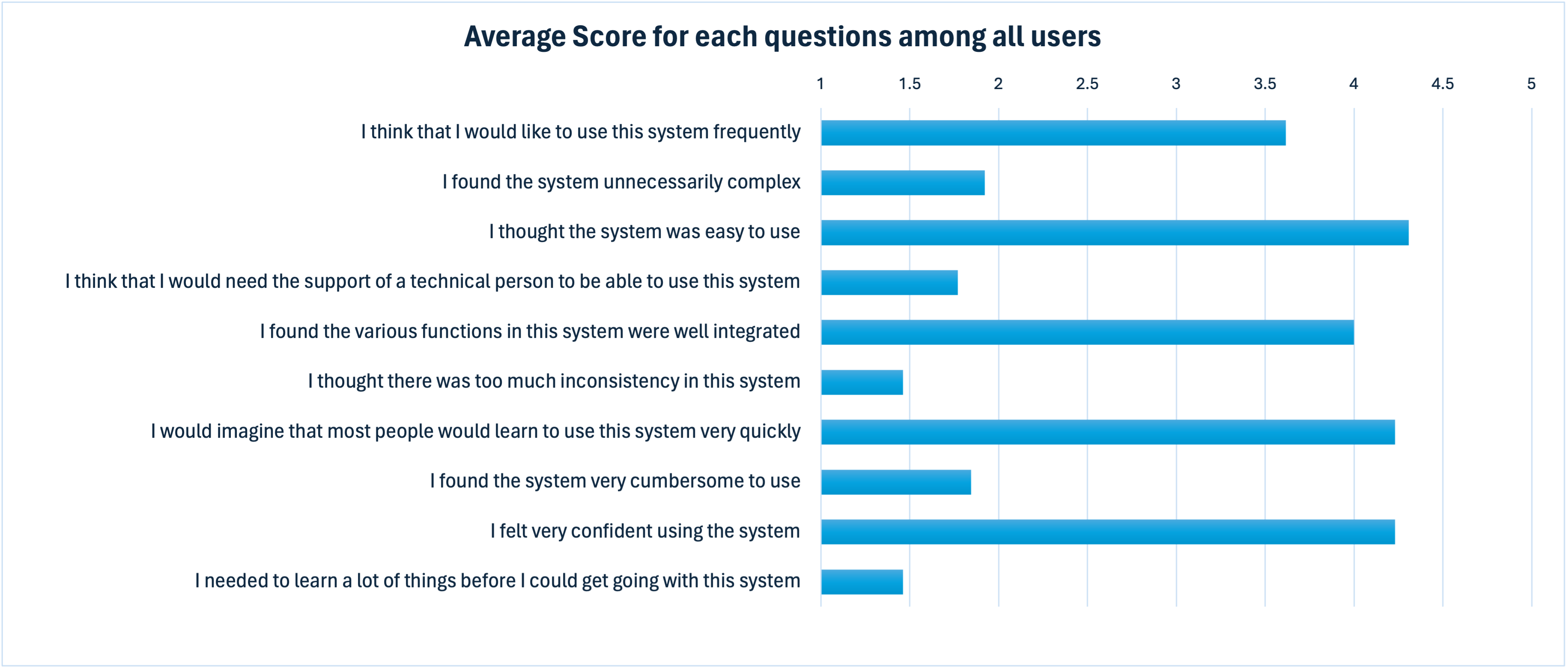}
    \caption{Average Score for each SUS question}
    \label{fig:sus}
\end{figure}

Figure \ref{fig:sus} shows the average score of each question in the SUS questionnaire given by participants, which ranges between 1 and 5, according to the calculation rule for each question explained above, it would be better if the odd-numbered questions gained a higher mark and the even-numbered questions gained a lower mark. It shows that questions ``I thought the system was easy to use'', ``I found the various functions in this system were well integrated'', ``I would imagine that most people would learn to use this system very quickly'' and ``I felt very confident using the system'' achieved the highest average marks, nearly the full score, which means that most participants recognize the usability of EmoTrack and thought it is easy to get started. For all even-numbered questions, the scores are in the range of 1 and 2, which means almost every participants were `Strongly Disagree' or `Disagree' with these negative statements. As a result, both the total SUS score and the average score for each question can tell that EmoTrack is easy for users to comprehend and start using, and EmoTrack only contains the extremely necessary features.

\subsection{Summary}
The SUS has been proven as a cost-effective tool for practitioners to evaluate the general usability of a product or service among a broad range of interface technologies and can be useful for both research administrators and participants \cite{r17}. The results of SUS presented the general idea of how users found EmoTrack. The total SUS score of 79.8 shows the usability and acceptability among these participants.

In the next section, I conducted the face-to-face interview with some participants for more nuanced and richer information about their experience of using EmoTrack. Furthermore, it would be determined whether EmoTrack had led users to reflect on their YouTube watching behaviours and how it impacted their mood.

\section{User Interview Analysis}
\label{sec:interview}

\subsection{Introduction}
I interviewed eight participants who had used EmoTrack for a week, their basic information can be found in the table \ref{tab}. For the purpose of 
acquiring better understanding of users’ feelings about using EmoTrack, as well conducting qualitative analysis on whether EmoTrack had help people to reflect on their personal data, including their YouTube watching behaviours and the impact on their mood. In addition, I also discussed with them about benefits and limitations of EmoTrack to see if they have any suggestions for improving the EmoTrack.

The interview consisted of several open-ended questions, including:

\begin{enumerate}
    \item What do you think about your online behaviours (before using EmoTrack)?
    \item How do you find EmoTrack?
    \item How do you feel about the report (Bar Chart and detailed categories of videos)?
    \item Do you think these kinds of videos really influence your mood (just like what has been shown in the report)?
    \item Does the report make you think of anything you have never thought about before? (Do you find the report useful?)
    \item Do you learn anything when you used the app? And what?
    
    \item Do you find anything that EmoTrack can help with?
    \item Will you make any changes when watching YouTube after using EmoTrack and what?
    \item Do you have any suggestions for improving the app?
\end{enumerate}

Some of the questions might be similar to others, but they are the additional questions that could help me get a more complete picture of what users think about EmoTrack, as well as its feasibility and acceptability. For example, if the interviewee expresses a positive attitude with Question 3, especially with the detailed categories of videos part, then I will ask Question 4 to find out whether the user have more in-depth reflection. For the same reason, Question 6 is the additional question for Question 5. Questions 7 and 8 can be seen as a pair of questions. The results for these `similar' questions will also be summarized together.

I interviewed each of the eight participants for about 20 minutes. All interviews were recorded and transcribed.

\begin{table}[h]
\centering
\begin{tabular}{|c||c|c|}
\hline
Interviewee      & Major      & Status      \\
\hline
$P1     $ & $Computer Science (MEng)  $ & $UG     $ \\
\hline
$P2     $ & $Computer Science with Innovation (MEng) $ & $UG     $ \\
\hline
$P3     $ & $Computer Science $ & $UG $ \\
\hline
$P4     $ & $Marketing and Communications     $ & $PG     $ \\
\hline
$P5     $ & $Computer Science  $ & $PG  $ \\
\hline
$P6     $ & $Law           $ & $UG  $ \\
\hline
$P7     $ & $Computer Science (MEng) $ & $UG   $ \\
\hline
$P8     $ & $Computer Science and Electronics (MEng) $ & $UG $ \\
\hline
\end{tabular}
\caption{Basic Information of Interviewee}
\label{tab}
\end{table}

\subsection{User Reflection}
Here are the responses from interviewees, I have summarize their answers by questions:

\subsubsection{Q1. What do you think about your online behaviours (before using EmoTrack)?}
\textit{Hint: Do you think using social media will affect your study/work efficiency or your daily life?}

Most of the participants indicated that they do have habits of browsing social media, including YouTube, especially during mealtime. Also, between studying they just pick up their phones to reply, but they would unconsciously swipe up social media, and by the time they put it down, a long time had passed, which made them less effective in studying. While there are still people who can balance their lives well daily without external help, P6 said ``\textit{I can balance my schedule with all the things in my daily life, including using social media.}''

\subsubsection{Q2. How do you find EmoTrack?}
Interviewees were required to answer this question from two aspects, one is about the UI appearance, and the other is about the usability and functionality of the app.
\subparagraph{About the UI Appearance}
Most of the participants expressed positive thoughts about the design of EmoTrack’s UI. They said the UI interface is clear and neat, with a comfortable overall colour palette that makes it easy to identify the different functions and easy to operate. However, there were different opinions, P5 and P6 said they were unfamiliar with the start and end emoji, so they didn't recognize them at first, but there are text descriptions above and below the buttons, so there's no chance of mismatching and affecting the use. Furthermore, P4 pointed out that people's emotions are varied, and that sometimes having only three options does not give a very comprehensive view of her mood, so she could only choose the closest one.
\subparagraph{About the Usability and Functionality}
Most affirmed its simplicity and ease of use. Inevitably, however, half of them said that even though they have been reminded many times to choose their mood before and after using YouTube, they still tend to forget to do so. P1 and P3 stated that ``I often watch 2 or 3 videos before remembering that I didn't click ‘Start’, resulting in incomplete analyses.'', whereas P4 said it was usually easy to forget to press ‘Stop’ and make all records useless.

Most of the participants mentioned the complexity of downloading and uploading the YouTube watch history, P1 stated ``When I wanted to access the report I had to download the viewing history and upload it myself, which was a bit time-consuming and required me to switch on my computer, which didn't allow me to get feedback timely.'', while P1 preferred to use EmoTrack on her phone but EmoTrack strictly limited the file format to be JSON, so she can only download the history file through the laptop. P2 said, ``Although it is not difficult to download the history and upload the history, it is tedious to repeat the process every time I want to view the report for a new date''.

\subsubsection{Q3. How do you feel about the report (Bar Chart and detailed categories of videos)?}
Most people agreed that the report could make them realize the length of time they spent on YouTube and they were going to consciously control themselves. For example, P1 said ``The bars on the chart exaggerate how long it took to watch videos, making me subconsciously want a shorter bar for tomorrow'' and ``I’m particularly interested in viewing my weekly report''. P2 said ``It clearly shows the length of time I spent watching videos during the week, helping me to reflect and understand my preferences. I'll find that I actually favour a certain type of video to be able to have a more up-to-date view of myself and I'll adjust my schedule for the rest of the week based on the length of my videos."

When asked about the additional question,

\textbf{Do you think these kinds of videos really influence your mood (just like what has been shown in the report)?}

Half of the interviewees said it was not a surprise, for example, P3 stated that she only watches videos she likes and doesn't watch categories she hasn't watched before out of curiosity. While the other half of the participants were shocked, some of them thought there existed many external factors that would also affect the mood, not just as a result of the videos’ content. P2 and P4 stated that each video may lead to a different mood change, but it seems that the only mood swings brought about by the last video or the most impressive one can be recorded. Moreover, some users said they were shocked by the long list of categories, and upon closer inspection realized it was a breakdown of what he thought was a category of videos into several categories, such as the ``Vlog'' video would be classified into ``Food'' or ``Lifestyle'' by ChatGPT, which would be discussed more detailed in Section \ref{sec:self-discuss}. While P6 said she was shocked because she often watches videos that YouTube recommended to her or those auto-playing videos, after reading this detailed list of categories she realized there was so much variety of videos.

\subsubsection{Q4. Does the report make you think of anything you have never thought about before? (Do you find the report useful?) Did you learn anything when you used the app? And what's that?}
Most interviewees said yes. P2 stated ``I would find that I preferred to watch a certain type of video, and I had a newer step of knowledge about myself''. P4 said, ``It monitors my mood state for a certain period, as I often don't remember past mood, but the report reminds me that I've been pretty down for the last week''. Moreover, P5 said EmoTrack could notice she for better time management, ``It made me notice that sometimes mood changes are caused by browsing social media, and that's when EmoTrack serves as an external reminder to stop this unconscious behaviour, control the time and frequency of watching videos, and find balance in my lives''.

\subsubsection{Q5. Do you find anything that EmoTrack can help with? Will you make any changes when watching YouTube after using EmoTrack and what?}
Most interviewees expressed positive attitudes, P2 said ``I'll have a clearer idea of what kind of video content I should need to watch afterwards, based on what I need for my mood'', P6 said, ``I feel like I'll be more selective about the videos that make me feel better, rather than just blindly swiping to whichever video, which can save a lot of time and energy, because sometimes I'm swiping to videos that are simply not necessary, and I feel like it will be able to change my habit of swiping to social media, and I'll also probably be able to spend so much less time swiping to videos that don't make any sense to me, and go for more of the ones that either make me happier or that I can learn something from''.

\subsection{Suggestions}
\label{sec:suggestion}
All interviewees provided valuable suggestions and I summarized them with different aspects:

\subsubsection{More Detailed Guidance on UI}

P7 and P8 suggested that the instructions about how to download and upload could be more detailed, I explained that I had written another detailed user instruction notes for the whole application but P7 thought they could be combined. And P8 thought ``More instructions can be given on the home page with regards to the aim of the app so that users can know what kind of app they are using.''

\subsubsection{User Consent Form}
From a legal point of view, P6 suggested that a new step directed at obtaining user consent be added before users are asked to upload files, e.g. users need to be informed of what your files will be used for and how the data will be retained, etc.

\subsubsection{Mood Recording Function}
P3 advised that EmoTrack could have a notification function to ask users after one hour of watching, for example, whether they have finished watching, in case users forgot to press the ‘Stop’ button. Moreover, it could also remind users that they have watched more than one hour, which could prevent users from being addicted to YouTube.

P4 and P5 suggested that EmoTrack could show a small translucent hovering window on the YouTube screen for users to record their mood in real-time, so that they could get the change in mood caused by each video. 

Furthermore, P4 said ``It would be better if the phone could automatically detect what social media I am using and automatically record what I have browsed, and proactively ask me what I'm feeling when it detects that I've opened YouTube''. She also suggested that there could be more options for mood selections, such as ‘sad’, ‘anger’ and so on.

\subsubsection{Analysis Report}
P8 said, ``If you can automatically generate weekly summary reports and send notifications that would be great, purely relying on self-consciousness to check the report for a long time will make me forget the existence of EmoTrack''. Since there is no notification function in EmoTrack, users need to view their reports actively.

P3, P4 and P5 advised that the report could be more summarized: "Some categories are similar, they should be summarized", said P5. P3 stated that the application should analyze the results concisely, "not just show everything". They said that sometimes they found themselves watching more than 100 videos in one day, and the categories of those videos were really a long list, so they thought that even though they read the long list, there was nothing that could help them. In addition, P4 advised that when users in the worse mood, it will be better if EmoTrack can provide some advices for users, from the perspective of mental health. Moreover, P5 suggested that ``Video categories can be clustered to analyze the correlation between categories and emotions; Adding tracking of viewing duration and point in time of viewing allows for a more in-depth dimension when analyzing changes in emotions''.

P6 suggested that the presentation of the report can be more diverse, not limited to bar charts, but can be horizontal bar charts, pie charts, tables, etc. ``You can also add a filtering function to filter out the number of dates that you have watched more than 50 videos, for example, so that it will be clearer to view in case there are too many dates, which can meet the diverse needs of users'', she said. Furthermore, P7 suggested that a few days or even a week's worth of data could be used to train the model to produce more accurate results in terms of the correlation between the category of the video and the mood, with more accurate results the longer the number of days used.

\subsubsection{Upload Functions and User Privacy}
Most users hoped the function of acquiring their watch history could be more convenient, P1 said ``Each time I download and upload a complete viewing history, including all my data since signing up the YouTube, and as the days add up, analyzing the length of time gains too much of our personal privacy. Is it possible to find out when the last upload was made and only part of it is intercepted?''. P5 stated that ``User privacy needs to be taken seriously and it can be developed into edge computing instead of being passed to a database for computation''.

\subsubsection{Additional Functions}
During the introduction and interview for EmoTrack, some users were found that they expected that the application could have the video recommendation function, that is, they wished that after analyzing what kinds of videos could make them feel better, EmoTrack could recommend some similar categories of videos for them. ``That will be interesting!'', said P4. And P5 suggested ``Add recommendation algorithms to allow EmoTrack to optimize video recommendations to make the user have a better emotional experience''.

Furthermore, P5 advised that there could be a function for users to set up self-discipline goals, so that EmoTrack can automatically remind itself whether it should stop or recommend watching of certain videos.

\subsection{Discussion}

As mentioned in Section \ref{sec:reflection}, there are different degrees of reflection, while I think which level of reflection different users can reach depends on the users themselves. As a technique, EmoTrack presented data in the same way, while users would have different reflections.

With EmoTrack, some interviewees realized the connection between changes in mood and their YouTube watching behaviour, who should be on the level of R0.

For example, P3 uses EmoTrack as a tool to track how many videos she watches each day and to track her mood changes. As she said above, every time she opens YouTube, she will only watch the videos she is interested in and will not click on the ones she is not interested in, so EmoTrack will only record the situations that make her mood change for the better or not change. She also said that she will only open YouTube in her free time, so she is not worried about becoming addicted to it. In this situation, I think P3 has reached the R1 level of reflection because she read the report that EmoTrack provided and realized that YouTube can at least provide positive mood feedback, and the videos she is interested in can improve her mood sometimes.

While P1 and P4 represent the majority of respondents who would open YouTube and watch short videos and get addicted in between studying or working, resulting in disrupting the pace. Looking at their reports, they were surprised by the number of videos watched per day and the fact that most of the week shows the colour red, which stresses them out. P1 felt that the reasons for this may because these short videos delay her pace of study, made her felt worse, and she said above that she would try to reduce the number of videos per day.

P4 said that she would try to watch less of the categories of videos that the report mentioned as making her feel worse. As a result, I think P1 and P4 are at level R2 because they are informed by EmoTrack that being addicted to short videos is not only slowing down their learning process with the third party perspective, but also make them feel worse most of the time. They were provided with the third party perspective that what categoies of videos had negative impact on their mood, and also daily life study efficiency.

Among the interviewees, I think only P6 has reached the R3 level, because she mentioned that people are not actively addicted to watching videos most of the time, but are trapped by the algorithm of YouTube. While reports realized that she was unconsciously watching videos and learned nothing. So she deletes YouTube on her phone and will only watch necessary videos by laptop. Since she thinks using laptop can prevent her from being addicted to watch videos one after another. The transformative changes she made shows she is on the R3 level.

\subsection{Summary}
In this section, I describe the face-to-face interview I conducted with eight participants to analyze the users' experience of using EmoTrack and the level of reflection they might have when provided with their personal information, including their YouTube viewing behaviour and mood changes. It then presents the results of the interviews and the qualitative discussion of these responses, which shows that EmoTrack can facilitate participants to have levels from R0 to R3.

\section{Self Reflection}
\label{sec:self-reflection}
It is important for developers to use the app from the user's point of view, predict the potential problems that may exist and try to solve them, as well as the iterations. As the developer of EmoTrack, I used it for more than one month. According to the workflow of EmoTrack, users only need to record their mood before and after watching YouTube, so I started using it before it had even been completed. Therefore, I have some feelings about using EmoTrack and reflection about viewing the report.

\subsection{Discussions}
\label{sec:self-discuss}

\subsubsection{About the Bar Chart}
Before having EmoTrack, I actually had some knowledge about myself, I knew how much time I spent on YouTube every day, but I was still surprised by the number of videos that I often watch within a day. Talking about the reason why there can be more than 50 videos in a day is because I watch both long and short videos. According to the user study (refer to Section \ref{sec:user-data}), almost all participants have a clear preference between long and short videos, and only a few users like to watch both. However, to make people aware that time flies when watching short videos one after the other. I think it is still necessary to make a distinction between long and short videos in the bar chart, or add options, such as ``Show results for long videos'', ``Show results for short videos'' and ``Show results for all videos'', in the filter for users to customize.

\subsubsection{About the Detailed Categories}
Initially, I found that there are some categories named ``Unspecified'', ``Uncategorized'', ``Unknown'' and/or ``Undefined'', etc. (See Figure \ref{fig:report-34}). To investigate the reason, I referred to the corresponding videos and found that these videos are private videos or have been deleted. Since YouTube has the feature, authors can upload videos privately or only certain people they prefer with shared links can watch. Additionally, sometimes I watch the live broadcast on YouTube, and if there is no replay video uploaded to the same URL as the live broadcast, then there would be no video valid for that URL. All these situations are the reasons that Python scraper cannot obtain the titles and descriptions of videos and ChatGPT cannot categorize videos correctly.

Next, I found the results of categories given by ChatGPT have shortcomings, even if it is a chatbot with very strong linguistic capabilities, which has been trained on a large amount of language-based data. In categorizing YouTube videos, there are still areas that didn't meet my expectations. For example, when categorizing vlogs, it often generates ‘Food’, ‘Lifestyle’, ‘Study vlog’, ‘Packaging’ and so on; For playlist videos, there will be categories like ``Music'', ``Chill'', ``Spring'' and other words describing styles of playlist or music will appear when categorizing; while classifying TV series, it will be classified as ‘Chinese Drama’, ‘Korean Drama’ because of the language, or ‘Medical’, ‘Crime’, etc. because of the subject matter of the TV series. I think this may be because ChatGPT is not specifically trained to classify YouTube videos. Authors often don't write what kind of video it is in the video synopsis, e.g. the description of a vlog video will often say what is done in the video, and a TV show will summarize the plot, so the more information there is, the more detailed categories ChatGPT will generate, which is different from what I would expect for video classification.

Furthermore, according to Figure \ref{fig:report-34}, I found that the "food" videos, for instance, could make me feel better, feel worse and not make my mood change. While I think that there might be three reasons, firstly, it is inevitable that when I choose my mood it is heavily influenced by reality. For example, maybe watching those videos made me feel better, but not good enough for me to choose "Good''. Second, I often watch a variety of short videos, it might be the situation that I watch a lot of videos in the meantime, so the one or two "Food" videos do not affect my mood as much as others. Third, there are completely different kinds of "food" videos on YouTube, more than just delicious food, so it has the potential to make a difference in mood.

\subsection{Future Extensions and Summary}
In this section, I talked about my feelings as a developer during the trial of EmoTrack and its shortcomings, as well as speculations about the reasons for them.

For future extensions, firstly, enables the report to be presented in more formats for users to choose. Secondly, select or train the NLP model that are more suitable for classifying YouTube videos.

\section{Summary}
This chapter outlines the evaluation approaches used for EmoTrack, including user testing, the SUS questionnaire, interviews and autoethnography. Data collected from the one-week user testing revealed the participants’ daily use habits of YouTube and the relationship with their mood when watching videos. According to the SUS results, EmoTrack was widely perceived as having good usability and performance, with an average SUS score of 79.8. The interviews with participants provided favorable results, most interviewees expressed their acceptability of EmoTrack. I also collected many valuable users’ feedback and suggestions to improve EmoTrack, such as adding the reminder features, generating more summative reports, and increasing more mood selections. In the self-reflection section, I shared my experiences using EmoTrack as the developer. As the main concern mentioned in this stage is the diversification of the reports’ style and the limitations of ChatGPT.


\chapter{Conclusion}
\label{chap:conclusion}

\noindent

\section{Overview}
This dissertation investigated whether EmoTrack can help users to track their online activities on YouTube, facilitate their reflection on their video watching behaviour, and help them develop better skills to engage on the internet. In this chapter, I will discuss the contributions and achievements established during the project. I will also summarize the efforts taken to reach this point and emphasize the findings from the evaluations. This chapter will conclude by discussing the limitations and any potential future work that could be done to improve user experience and reflection through using EmoTrack.

\section{Main Contributions and Achievements}
\subsection{Software Application Development}
In this project, I developed a multi-platform application EmoTrack to record users’ YouTube activities and their changes in mood before and after watching videos. It can also provides graphic visualizations for users to track their daily use of YouTube, as well as reflect on the effect of watching videos on their mood.

The main contributions and achievements of this project are:
\begin{enumerate}
    \item I have designed and built the User Interface of EmoTrack with Flutter, including the following functionality:
    
    (the detailed description of the implementation of the front-end in Section \ref{sec:ui})
    \begin{enumerate}
        \item Built the sign-in and register system with Firebase Authentication, ensuring the security and uniqueness of users’ accounts.
        \item Designed and built an easy-to-use interface for users to record their mood before and after watching YouTube videos.
        \item Built a port for users to upload their YouTube watch history to the Cloud Storage, ensuring the independence and security of users' personal information.
        \item Designed and built a graphic visualization interface with the powerful Flutter graph library for users to view their daily activities on YouTube within an arbitrary time range, as well as the full lists of all videos’ categories related to their changes in mood.
    \end{enumerate}
    \item Implemented the following back-end functionality with Python:

    (the detailed description of the implementation of the back-end in Section \ref{sec:back-end})
    \begin{enumerate}
        \item Extracted the required information from files in Cloud Storage and transmitted data to the database.
        \item Got data from the database.
        \item Filtered the desired data, such as mood records and YouTube videos, according to the user’s selection of the time range.
        \item Implemented two approaches to obtain the information (title, description, category, etc.) of YouTube videos, one is web scrapping, and the other is YouTube Data API.
        \item Integrated with ChatGPT to categorize videos with their titles and descriptions, expecting more accurate and intelligent results.
        \item Processed data to prepare it to be visualized.
    \end{enumerate}
    \item Established a structured database with Firebase.
    \item Constructed the server with Quart RESTful framework, establishing API endpoints for transmitting data and commands between the front-end and the back-end.
    \item Deployed the back-end server on Google App Engine and the web application of the front-end on Firebase Hosting
\end{enumerate}

\subsection{User Usability and Acceptability Evaluation}
I conducted a one-week usability study of EmoTrack with thirteen participants to collect their application usage data for statistical analysis. I also administered questionnaires and interviewed the participants to analyze the usability and acceptability of EmoTrack quantitatively and qualitatively. Furthermore, I evaluated EmoTrack with autoethnographic methods from the perspective of the developer, particularly underscoring areas for future development.

The week-long user testing revealed insights into participants' daily YouTube consumption patterns and their correlation with mood fluctuations. About 58\% of participants watched YouTube for 1 to 3 hours a day and most videos were watched during the evening from 6 pm to 12 am.

The System Usability Scale Survey results show that EmoTrack was rated favourably in terms of usability, evidenced by a high average SUS score of 79.8.The qualitative interviews further confirmed the acceptability of EmoTrack.

EmoTrack facilitated participants to reflect on their YouTube video watching behaviour and the impact on their mood, with reports of different levels of reflections, from R0 to R3.

In addition to affirmation, there was much constructive feedback, such as the need for enhanced mood options, additional reminder functionalities, and more comprehensive reporting features.

Furthermore, in a reflective account, I, as the developer, shared personal experiences with the application, noting the need for more varied report styles and addressing the limitations encountered with ChatGPT’s integrations.

\section{Limitations}
\subsection{Cumbersome Features and User Privacy}
While using EmoTrack, the feature to obtain users’ YouTube watch history might be too complicated and would prevent users from using it continuously for a long period of time. It requires users to download their YouTube watch history from the website of Google Takeout, and users must strictly select the format of the file as JSON, which cannot be completed on their own without the step-by-step guide I provided for them, revealed by some participants during the user test.

With this comes the issue of user privacy, as the YouTube watch history downloaded from Google Takeout contains the entire history since the creation of the YouTube account. If the user does not edit the contents of the watch history file, the entire history will be uploaded to the database. Although the content of each user's data is tightly bound to the user's account and will never be leaked to other users, the content of the database is completely transparent to the developer.

One possible solution to address these issues is to iterate EmoTrack to allow users to sign in with Google Account, which aims to obtain the login credentials and authorization from users for their Google Account. With the authorization, EmoTrack will be able to automatically obtain the user's YouTube viewing history. However, this has to be approved by Google.

\subsection{Mood Recording Reminder}
Some participants pointed out that they usually forget to record their mood, such as forgetting to press “Start” before starting watching or forgetting to press “Stop” after have stopped watching. These issues will make the data for each mood recording useless.

These issues can be solved by adding a notification function to EmoTrack, if the user doesn't press “Stop” within a certain amount of time after starting, EmoTrack will send a notification, which has the added benefit of reminding the user that they've been watching for a long time if they really haven't stopped watching.

Furthermore, EmoTrack could be added with the ability to detect if a user has YouTube turning on and off. However, this feature involves issues with permissions within the phone and will have limitations depending on different devices.

\subsection{Small Size of User Testing}
The evaluation of EmoTrack was limited by the relatively small number of participants, which may limit the generalizability of the results. Future studies of EmoTrack could attempt to invite a bigger and more diverse group of participants and extend the duration of the user testing.

\section{Future Work}
The outcomes of this project, including the findings from the evaluation, my learning progress and problem resolutions during the developing and testing phases, provide a strong basis for future extensions.

The future work should start from solving the limitations mentioned in the last section, then suggestions from users’ interviews (see Section \ref{sec:suggestion}) and self-reflection (see Section \ref{sec:self-reflection}) should be considered. 

In addition, in order to achieve the originally proposed project goals, more social media platforms should be integrated, with comprehensive behavioural analysis of all social media usage, users will be more reflective and improve their online engagement skills more effectively.

Moreover, EmoTrack can be iterated to be suitable for people in different countries since the rapid growth of social media affects people all over the world, developing this new feature requirea a more flexible time zone converter implemented in both the front-end and the back-end.

Furthermore, potential future work may focus on the mental health of users, and EmoTrack can potentially be used by well-being and mental health services.


%
%
%

\backmatter

\bibliography{dissertation}

\begin{thebibliography}{10}

\bibitem{r41}
tz — dateutil 3.9.0 documentation\_2017, 2017.
\newblock URL: \url{https://dateutil.readthedocs.io/en/stable/tz.html}.

\bibitem{r42}
fl\_chart\_2019, Jun 2019.
\newblock URL: \url{https://pub.dev/packages/fl_chart}.

\bibitem{r30}
how to download your google data - google account help, 2019.
\newblock URL: \url{https://support.google.com/accounts/answer/3024190?hl=en}.

\bibitem{r11}
Firebase, 2022.
\newblock URL: \url{https://firebase.google.com/}.

\bibitem{r31}
api reference | youtube data api | google for developers, 2023.
\newblock URL: \url{https://developers.google.com/youtube/v3/docs}.

\bibitem{r16}
app backend and api protection solution | firebase\_2024.
\newblock \url{https://firebase.google.com/products/app-check?_gl=1*6j20on*_up*MQ..*_ga*MzAyODAyMjgxLjE3MTQyNTc2NDQ.*_ga_CW55HF8NVT*MTcxNDI1NzY0NC4xLjAuMTcxNDI1NzY0NC4wLjAuMA..}, 2024.

\bibitem{r29}
app engine application platform | google cloud, 2024.
\newblock URL: \url{https://cloud.google.com/appengine/?utm_source=google&utm_medium=cpc&utm_campaign=emea-gb-all-en-dr-bkws-all-all-trial-%7Bmatchtype%7D-gcp-1707574&utm_content=text-ad-none-any-DEV_%7Bdevice%7D-CRE_%7Bcreative%7D-ADGP_%7B_dsadgroup%7D-KWID_%7B_dstrackerid%7D-%7Btargetid%7D-userloc_%7Bloc_physical_ms%7D&utm_term=KW_%7Bkeyword%7D-NET_%7Bnetwork%7D-PLAC_%7Bplacement%7D&%7B_dsmrktparam%7D%7Bignore%7D&%7B_dsmrktparam%7D&gclsrc=aw.ds&gad_source=1&gclid=Cj0KCQjwudexBhDKARIsAI-GWYXd6QXb5t4wnCTYmsJsvLn1DrHZhARl1IwOpbMbbHqSdGqTAHNCDoQaAkWIEALw_wcB&gclsrc=aw.ds&hl=en}.

\bibitem{r14}
cloud firestore | store and sync app data at global scale\_2024.
\newblock \url{https://firebase.google.com/products/firestore?_gl=1*bug0wa*_up*MQ..*_ga*MTgzMjU5NTM4Mi4xNzE0MjI5MjI2*_ga_CW55HF8NVT*MTcxNDIyOTIyNi4xLjAuMTcxNDIyOTIyNi4wLjAuMA..}, 2024.

\bibitem{r15}
cloud storage for firebase | store and serve content with ease\_2024.
\newblock \url{https://firebase.google.com/products/storage?_gl=1*134gmgm*_up*MQ..*_ga*MTgzMjU5NTM4Mi4xNzE0MjI5MjI2*_ga_CW55HF8NVT*MTcxNDIzNjU4OS4yLjAuMTcxNDIzNjU4OS4wLjAuMA..}, 2024.

\bibitem{r13}
firebase authentication | simple, no-cost multi-platform sign-in\_2024.
\newblock \url{https://firebase.google.com/products/auth?_gl=1*1fi586i*_up*MQ..*_ga*MTgzMjU5NTM4Mi4xNzE0MjI5MjI2*_ga_CW55HF8NVT*MTcxNDIyOTIyNi4xLjAuMTcxNDIyOTIyNi4wLjAuMA..}, 2024.

\bibitem{r27}
flutter - build apps for any screen, 2024.
\newblock URL: \url{https://flutter.dev/}.

\bibitem{r32}
what is an api (application programming interface)? | ibm\_2024, Apr 2024.
\newblock URL: \url{https://www.ibm.com/topics/api}.

\bibitem{r46}
Alexandra Bailin, Ruth Milanaik, and Andrew Adesman.
\newblock Health implications of new age technologies for adolescents: a review of the research.
\newblock {\em Current opinion in pediatrics}, 26(5):605--619, 2014.

\bibitem{r34}
Anna~Costanza Baldry, Anna Sorrentino, and David~P Farrington.
\newblock Cyberbullying and cybervictimization versus parental supervision, monitoring and control of adolescents' online activities.
\newblock {\em Children and Youth Services Review}, 96:302--307, 2019.

\bibitem{r17}
Aaron Bangor, Philip~T Kortum, and James~T Miller.
\newblock An empirical evaluation of the system usability scale.
\newblock {\em Intl. Journal of Human--Computer Interaction}, 24(6):574--594, 2008.

\bibitem{r48}
Jan Boehmer, Robert LaRose, Nora Rifon, Saleem Alhabash, and Shelia Cotten.
\newblock Determinants of online safety behaviour: Towards an intervention strategy for college students.
\newblock {\em Behaviour \& Information Technology}, 34(10):1022--1035, 2015.

\bibitem{r28}
Bruce Chen.
\newblock Showcase - flutter apps in production, 2024.
\newblock URL: \url{https://flutter.dev/showcase?_gl=1*1mtf1i*_up*MQ..*_ga*MTk5MzI5NzkxLjE3MTQ4Mzg2MTA.*_ga_04YGWK0175*MTcxNDgzODYwOS4xLjAuMTcxNDgzODYwOS4wLjAuMA..}

\bibitem{r9}
Paul Covington, Jay Adams, and Emre Sargin.
\newblock Deep neural networks for youtube recommendations.
\newblock In {\em Proceedings of the 10th ACM conference on recommender systems}, pages 191--198, 2016.

\bibitem{r2}
Lusiana~Citra Dewi, Alvin Chandra, et~al.
\newblock Social media web scraping using social media developers api and regex.
\newblock {\em Procedia Computer Science}, 157:444--449, 2019.

\bibitem{r44}
Aiman El~Asam and Adrienne Katz.
\newblock Vulnerable young people and their experience of online risks.
\newblock {\em Human--Computer Interaction}, 33(4):281--304, 2018.

\bibitem{r18}
Rowanne Fleck and Geraldine Fitzpatrick.
\newblock Reflecting on reflection: framing a design landscape.
\newblock In {\em Proceedings of the 22nd conference of the computer-human interaction special interest group of australia on computer-human interaction}, pages 216--223, 2010.

\bibitem{r39}
Arup~Kumar Ghosh, Karla Badillo-Urquiola, Shion Guha, Joseph~J LaViola~Jr, and Pamela~J Wisniewski.
\newblock Safety vs. surveillance: what children have to say about mobile apps for parental control.
\newblock In {\em Proceedings of the 2018 CHI Conference on Human Factors in Computing Systems}, pages 1--14, 2018.

\bibitem{r3}
Digital 2024: Global Overview Report — DataReportal – Global~Digital Insights.
\newblock Datareportal – global digital insights, Jan 2024.
\newblock URL: \url{https://datareportal.com/reports/digital-2024-global-overview-report}.

\bibitem{r23}
Wenxiang Jiao, Wenxuan Wang, Jen-tse Huang, Xing Wang, Shuming Shi, and Zhaopeng Tu.
\newblock Is chatgpt a good translator? yes with gpt-4 as the engine.
\newblock {\em arXiv preprint arXiv:2301.08745}, 2023.

\bibitem{r38}
Betul Keles, Niall McCrae, and Annmarie Grealish.
\newblock A systematic review: the influence of social media on depression, anxiety and psychological distress in adolescents.
\newblock {\em International journal of adolescence and youth}, 25(1):79--93, 2020.

\bibitem{r1}
Ian Li, Anind Dey, and Jodi Forlizzi.
\newblock A stage-based model of personal informatics systems.
\newblock In {\em Proceedings of the SIGCHI conference on human factors in computing systems}, pages 557--566, 2010.

\bibitem{r8}
Ying~Chieh Liu and Min~Qi Huang.
\newblock Examining the matthew effect on youtube recommendation system.
\newblock In {\em 2021 International Conference on Technologies and Applications of Artificial Intelligence (TAAI)}, pages 146--148. IEEE, 2021.

\bibitem{r6}
Kai Lukoff, Ulrik Lyngs, Himanshu Zade, J~Vera Liao, James Choi, Kaiyue Fan, Sean~A Munson, and Alexis Hiniker.
\newblock How the design of youtube influences user sense of agency.
\newblock In {\em Proceedings of the 2021 CHI Conference on Human Factors in Computing Systems}, pages 1--17, 2021.

\bibitem{r26}
Zheheng Luo, Qianqian Xie, and Sophia Ananiadou.
\newblock Chatgpt as a factual inconsistency evaluator for abstractive text summarization.
\newblock {\em arXiv preprint arXiv:2303.15621}, 2023.

\bibitem{r4}
Christoffer Lysenst{\o}en, Tormod B{\o}e, Gunnhild~Johnsen Hjetland, and Jens~Christoffer Skogen.
\newblock A review of the relationship between social media use and online prosocial behavior among adolescents.
\newblock {\em Frontiers in Psychology}, 12:579347, 2021.

\bibitem{r50}
Mark Matthews, Gavin Doherty, John Sharry, and Carol Fitzpatrick.
\newblock Mobile phone mood charting for adolescents.
\newblock {\em British Journal of Guidance \& Counselling}, 36(2):113--129, 2008.

\bibitem{r12}
Laurence Moroney and Laurence Moroney.
\newblock Using authentication in firebase.
\newblock {\em The Definitive Guide to Firebase: Build Android Apps on Google's Mobile Platform}, pages 25--50, 2017.

\bibitem{r36}
Marie-Theres Nagel, Svenja Sch{\"a}fer, Olga Zlatkin-Troitschanskaia, Christian Schemer, Marcus Maurer, Dimitri Molerov, Susanne Schmidt, and Sebastian Br{\"u}ckner.
\newblock How do university students’ web search behavior, website characteristics, and the interaction of both influence students’ critical online reasoning?
\newblock In {\em Frontiers in Education}, volume~5, page 565062. Frontiers Media SA, 2020.

\bibitem{r33}
Tuong Nguyen-Van, Thanh Nguyen-Van, Tien-Thinh Nguyen, Dong Bui-Huu, Quang Le-Nhat, Tran~Vu Pham, and Khuong Nguyen-An.
\newblock A homomorphic encryption approach for privacy-preserving deep learning in digital health care service.
\newblock In {\em Asian Conference on Intelligent Information and Database Systems}, pages 520--533. Springer, 2022.

\bibitem{r37}
Ofcom.
\newblock Ofcom online nation 2023 report, 2023.
\newblock URL: \url{https://www.ofcom.org.uk/__data/assets/pdf_file/0029/272288/online-nation-2023-report.pdf#:~:text=URL%3A%20https%3A%2F%2Fwww.ofcom.org.uk%2F__data%2Fassets%2Fpdf_file%2F0029%2F272288%2Fonline}.

\bibitem{r19}
Aisling~Ann O'Kane, Yvonne Rogers, and Ann~E Blandford.
\newblock Gaining empathy for non-routine mobile device use through autoethnography.
\newblock In {\em Proceedings of the SIGCHI Conference on Human factors in Computing Systems}, pages 987--990, 2014.

\bibitem{r40}
Corina Sas and Alan Dix.
\newblock Designing for reflection on experience.
\newblock In {\em CHI'09 Extended Abstracts on Human Factors in Computing Systems}, pages 4741--4744. 2009.

\bibitem{r20}
Leon~D Segal and Jane~Fulton Suri.
\newblock The empathic practitioner: Measurement and interpretation of user experience.
\newblock In {\em Proceedings of the Human Factors and Ergonomics Society Annual Meeting}, volume~41, pages 451--454. SAGE Publications Sage CA: Los Angeles, CA, 1997.

\bibitem{r7}
Joan~E Solsman.
\newblock Youtube’s ai is the puppet master over most of what you watch.
\newblock {\em CNET, January}, 10, 2018.

\bibitem{r35}
Jiayi Sun, Wensheng Gan, Han-Chieh Chao, S~Yu Philip, and Weiping Ding.
\newblock Internet of behaviors: A survey.
\newblock {\em IEEE Internet of Things Journal}, 2023.

\bibitem{r47}
Teresa Swist, Philippa Collin, Jane McCormack, and Amanda Third.
\newblock Social media and the wellbeing of children and young people: A literature review.
\newblock 2015.

\bibitem{r45}
Melina~A Throuvala, Mark~D Griffiths, Mike Rennoldson, and Daria~J Kuss.
\newblock Perceived challenges and online harms from social media use on a severity continuum: a qualitative psychological stakeholder perspective.
\newblock {\em International journal of environmental research and public health}, 18(6):3227, 2021.

\bibitem{r10}
Karin Van~Es.
\newblock Youtube’s operational logic:“the view” as pervasive category.
\newblock {\em Television \& new media}, 21(3):223--239, 2020.

\bibitem{r43}
Caroline Violot, Tu{\u{g}}rulcan Elmas, Igor Bilogrevic, and Mathias Humbert.
\newblock Shorts vs. regular videos on youtube: A comparative analysis of user engagement and content creation trends.
\newblock {\em arXiv preprint arXiv:2403.00454}, 2024.

\bibitem{r5}
Kirsten Weir.
\newblock Social media brings benefits and risks to teens. here’s how psychology can help identify a path forward.
\newblock {\em Monitor on Psychology}, 54(6), 2023.

\bibitem{r49}
Nancy~E Willard.
\newblock {\em Cyber-safe kids, cyber-savvy teens: Helping young people learn to use the Internet safely and responsibly}.
\newblock John Wiley \& Sons, 2007.

\bibitem{r21}
Peter Wright and John McCarthy.
\newblock Empathy and experience in hci.
\newblock In {\em Proceedings of the SIGCHI conference on human factors in computing systems}, pages 637--646, 2008.

\bibitem{r22}
Tianyu Wu, Shizhu He, Jingping Liu, Siqi Sun, Kang Liu, Qing-Long Han, and Yang Tang.
\newblock A brief overview of chatgpt: The history, status quo and potential future development.
\newblock {\em IEEE/CAA Journal of Automatica Sinica}, 10(5):1122--1136, 2023.

\bibitem{r51}
Antonello Zanini.
\newblock how to scrape youtube with python guide\_2024, May 2024.
\newblock URL: \url{https://brightdata.com/blog/how-tos/how-to-scrape-youtube-in-python}.

\bibitem{r25}
ChengXiang Zhai et~al.
\newblock Statistical language models for information retrieval a critical review.
\newblock {\em Foundations and Trends{\textregistered} in Information Retrieval}, 2(3):137--213, 2008.

\end{thebibliography}



\appendix

\chapter{EmoTrack User Guide}
\label{appx:EmoTrack-guide}

This app is used to record changes in a user's mood before and after watching YouTube videos, as well as the relationship between categories of videos and the effect they have on the user's mood.

When users want to watch YouTube videos, they should open EmoTrack and press the “START” button first to select their mood at that moment, then they can open YouTube and watch no matter how long they want to watch. After watching, users should remember to press the “STOP” button and select the mood.

\begin{figure}[!h]
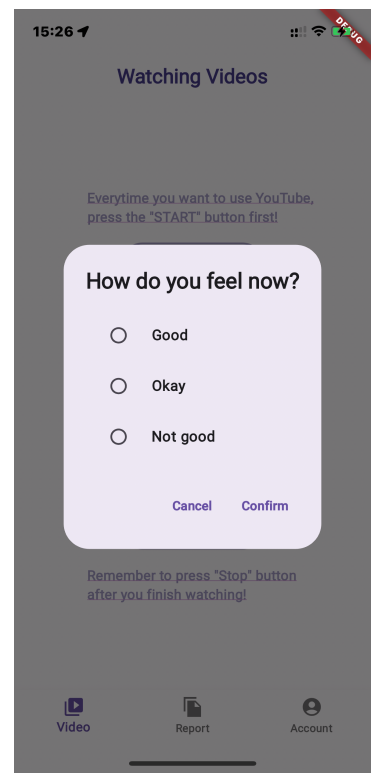

    \centering
    \begin{subfigure}{0.325\textwidth}
        \centering
        \includegraphics[width=0.9\textwidth]{pics/UI_1.PNG}
        \caption{}
        \label{fig:guide-1a}
    \end{subfigure}%
    \hspace*{\fill}
    \begin{subfigure}{0.325\textwidth}
        \centering
        \includegraphics[width=0.9\textwidth]{pics/UI_0.PNG}
        \caption{}
        \label{fig:guide-1b}
    \end{subfigure}
    \caption{User Interface for Mood Recording}
    \label{fig:guide-1}
\end{figure}

When users want to watch their reports (watch the instructions to see how to operate), they need to download their YouTube watching history from Google Takeout, and upload it to the app. Finally, press the “View Report” button, users will see the bar chart for last week by default.

\begin{figure}[!h]
  \centering
  \begin{subfigure}{0.325\textwidth}
    \centering
    \includegraphics[width=0.9\textwidth]{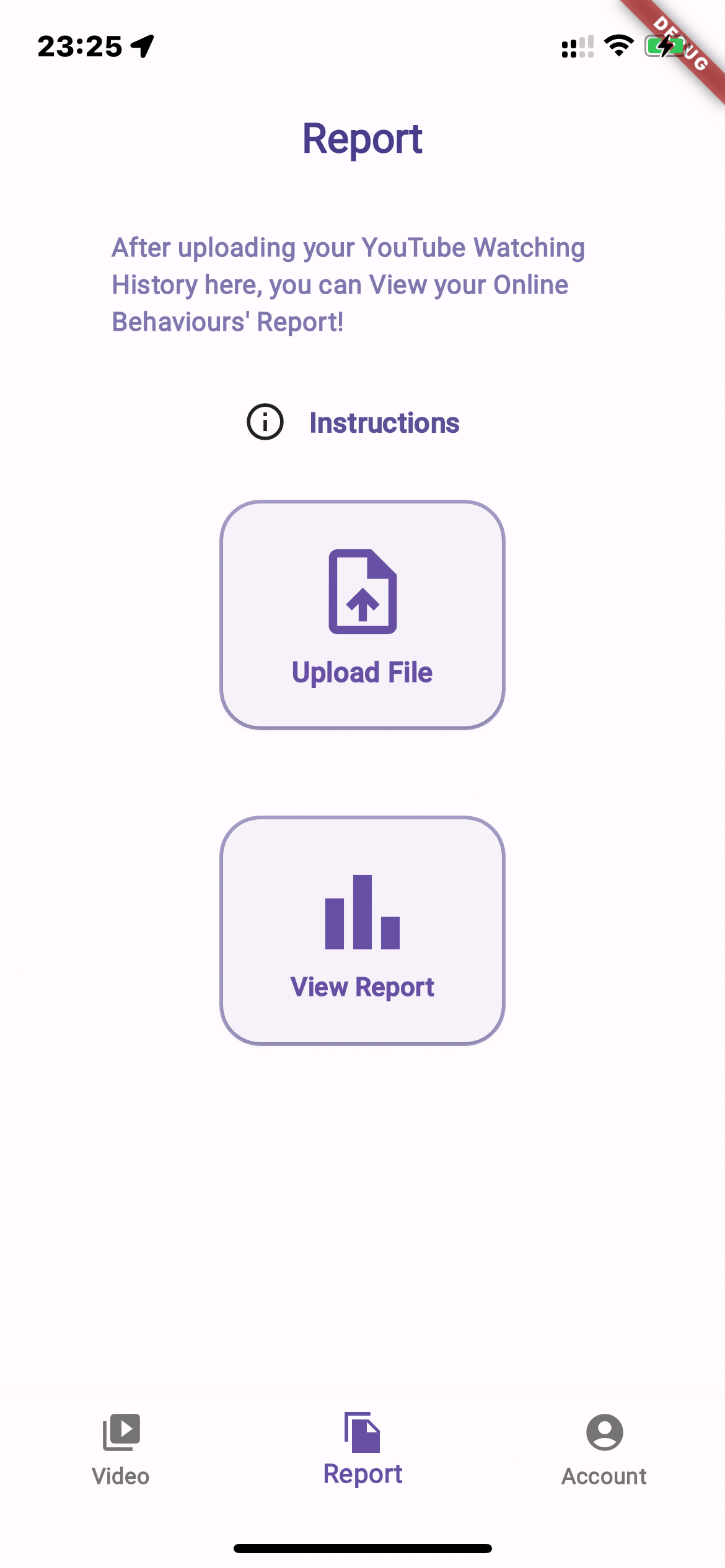}
    \caption{UI of the Report Page}
    \label{fig:guide-2}
   \end{subfigure}%
   \hspace*{\fill}
   \begin{subfigure}{0.325\textwidth}
    \centering
    \includegraphics[width=0.9\textwidth]{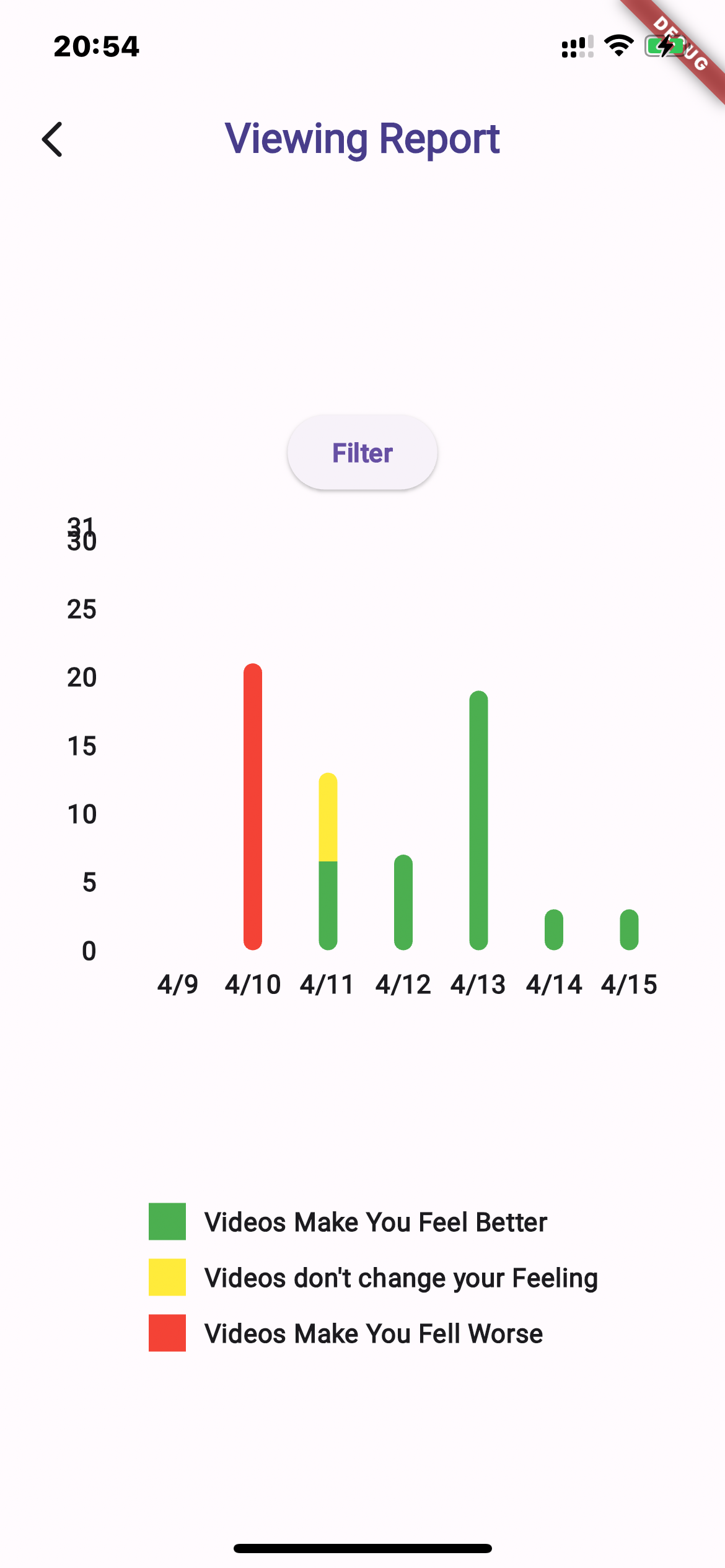}
    \caption{YouTube Watching Report}
    \label{fig:guide-3a}
   \end{subfigure}
   \hspace*{\fill}
   \begin{subfigure}{0.325\textwidth}
    \centering
    \includegraphics[width=0.9\textwidth]{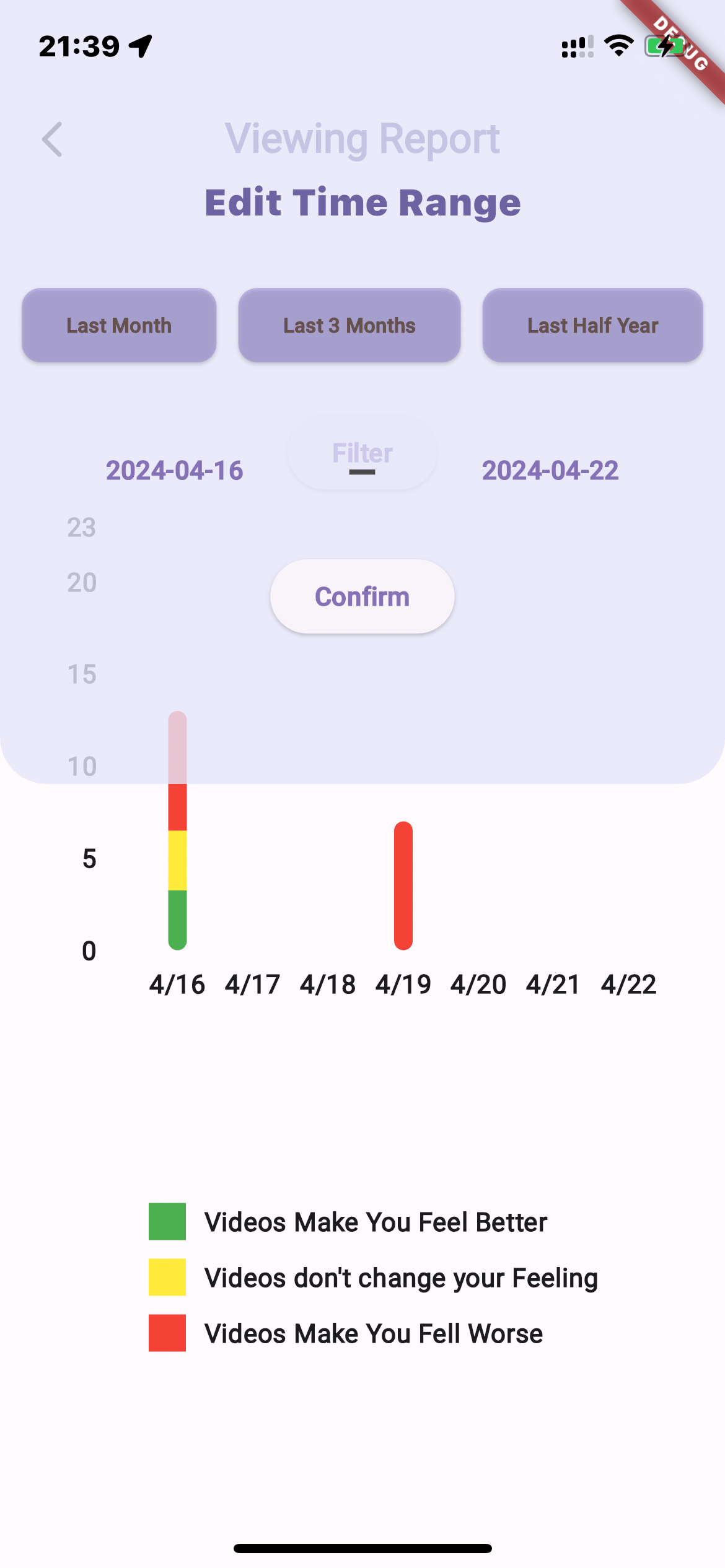}
    \caption{Filter Function to Select Time Range}
    \label{fig:guide-3b}
   \end{subfigure}
   \caption{Example User Interface of the Report Page}
   \label{fig:guide-4}
\end{figure}

Click on the column to show the detailed categories of videos for the specific date.

To view data for different time ranges, users just click the “Filter” button to select the other time ranges or customize the time range themselves.
\begin{figure}[!h]
  \centering
  \begin{subfigure}{0.325\textwidth}
    \centering
    \includegraphics[width=0.9\textwidth]{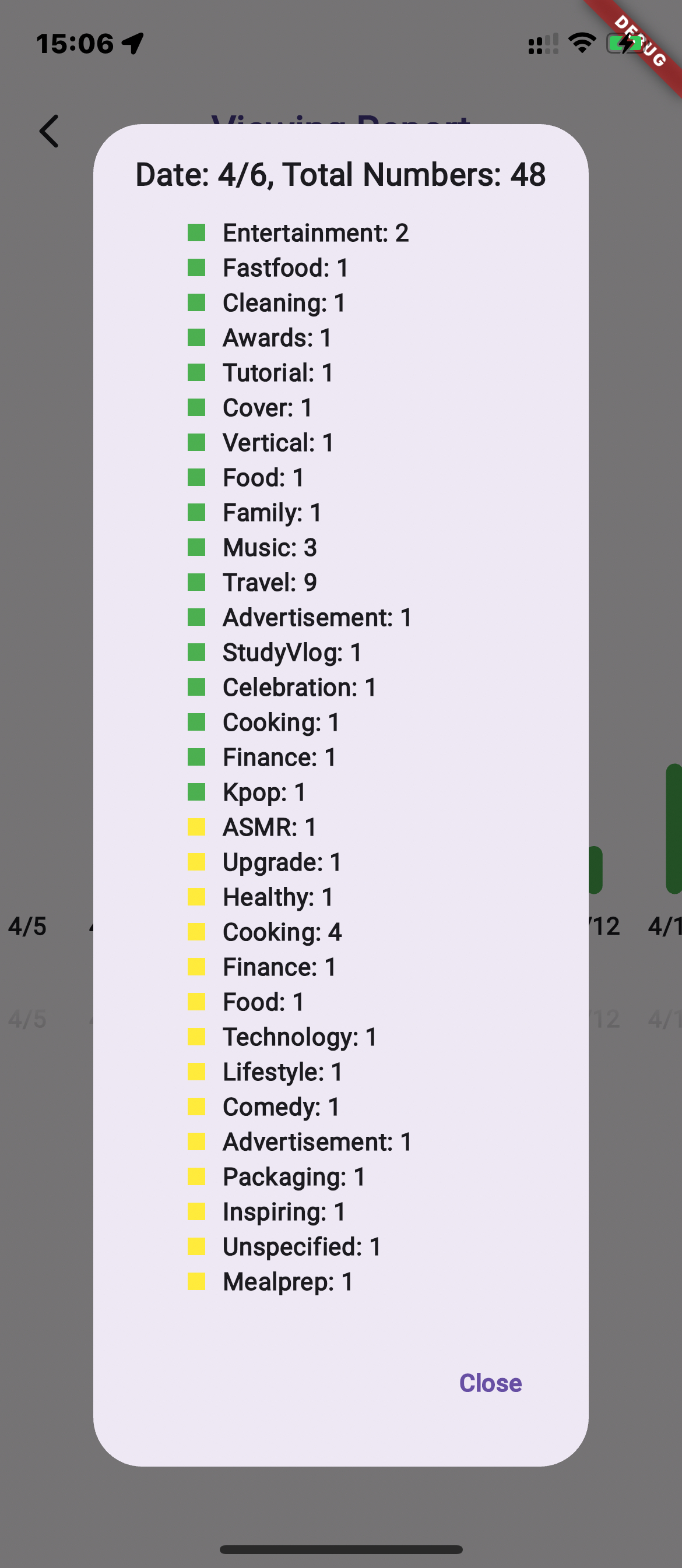}
    \caption{}
    \label{fig:guide-4a}
   \end{subfigure}%
   \hspace*{\fill}
   \begin{subfigure}{0.325\textwidth}
    \centering
    \includegraphics[width=0.9\textwidth]{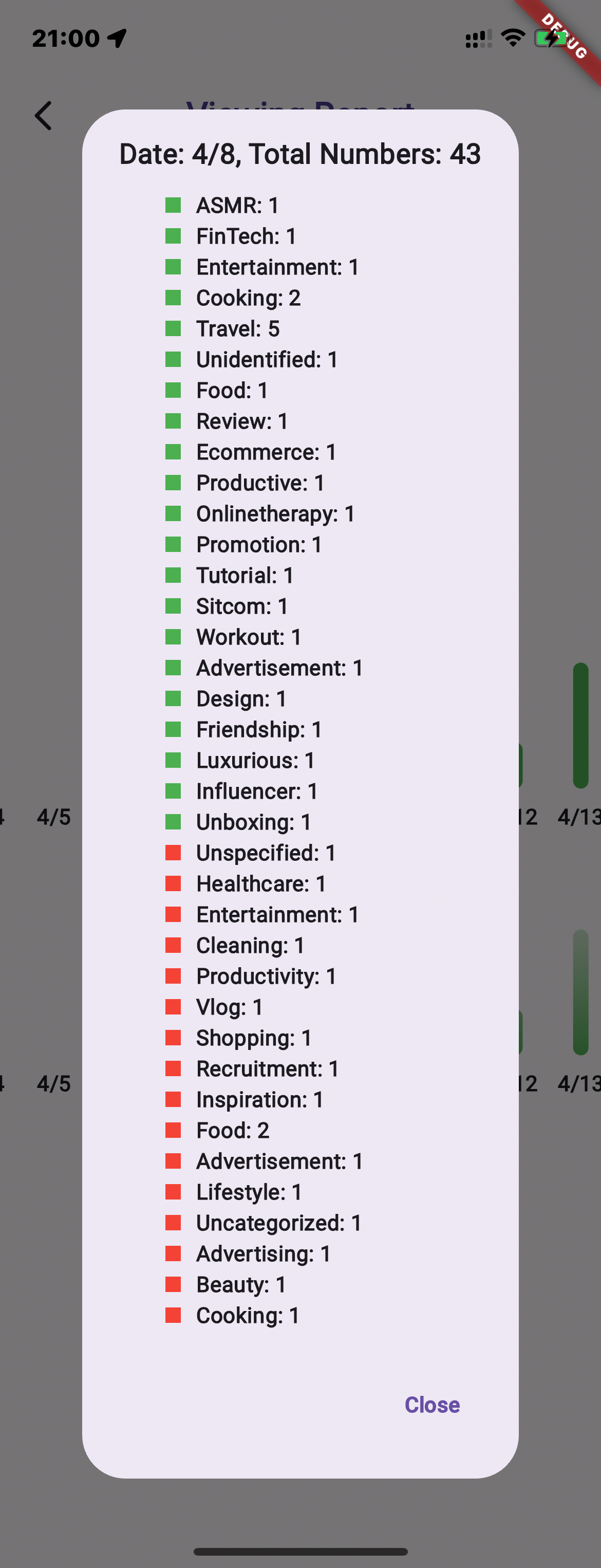}
    \caption{}
    \label{fig:guide-4b}
   \end{subfigure}
   \caption{Example User Interface for viewing video categories watched during a specific time period}
   \label{fig:guide-4}
\end{figure}

\subsubsection{Instructions for Download YouTube History}
\begin{enumerate}
    \item Search Google Takeout. (Figure \ref{fig:guide-5})

    \begin{figure}[!h]
        \centering
        \includegraphics[width=0.5\linewidth]{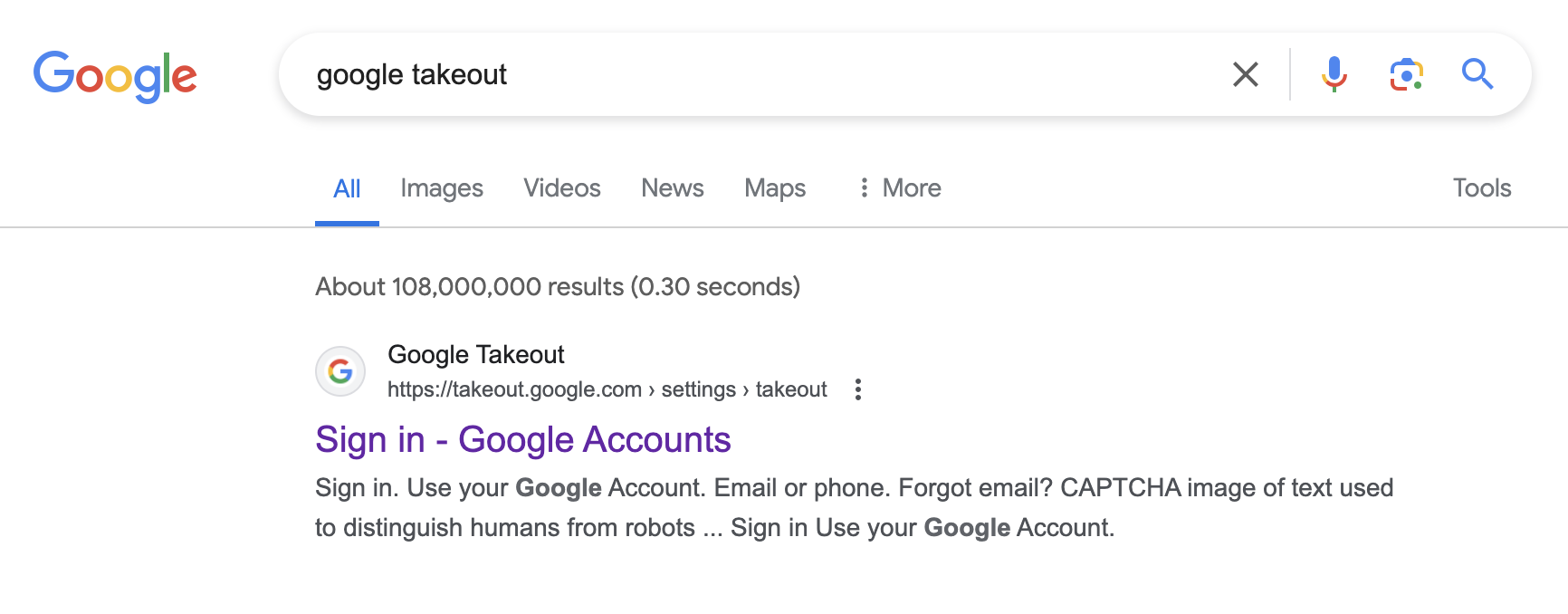}
        \caption{Enter Google Takeout}
        \label{fig:guide-5}
    \end{figure}

    \item Select `YouTube' only. (Figure \ref{fig:guide-6})

    Click the `Format' button on the left, select `JSON' for `history' (Figure \ref{fig:guide-7a}).
    Click the `Type' button on the right, select `history' only (Figure \ref{fig:guide-7b}).
    
    \begin{figure}[!h]
      \centering
      \begin{subfigure}{0.3\textwidth}
        \centering
        \includegraphics[width=0.9\textwidth]{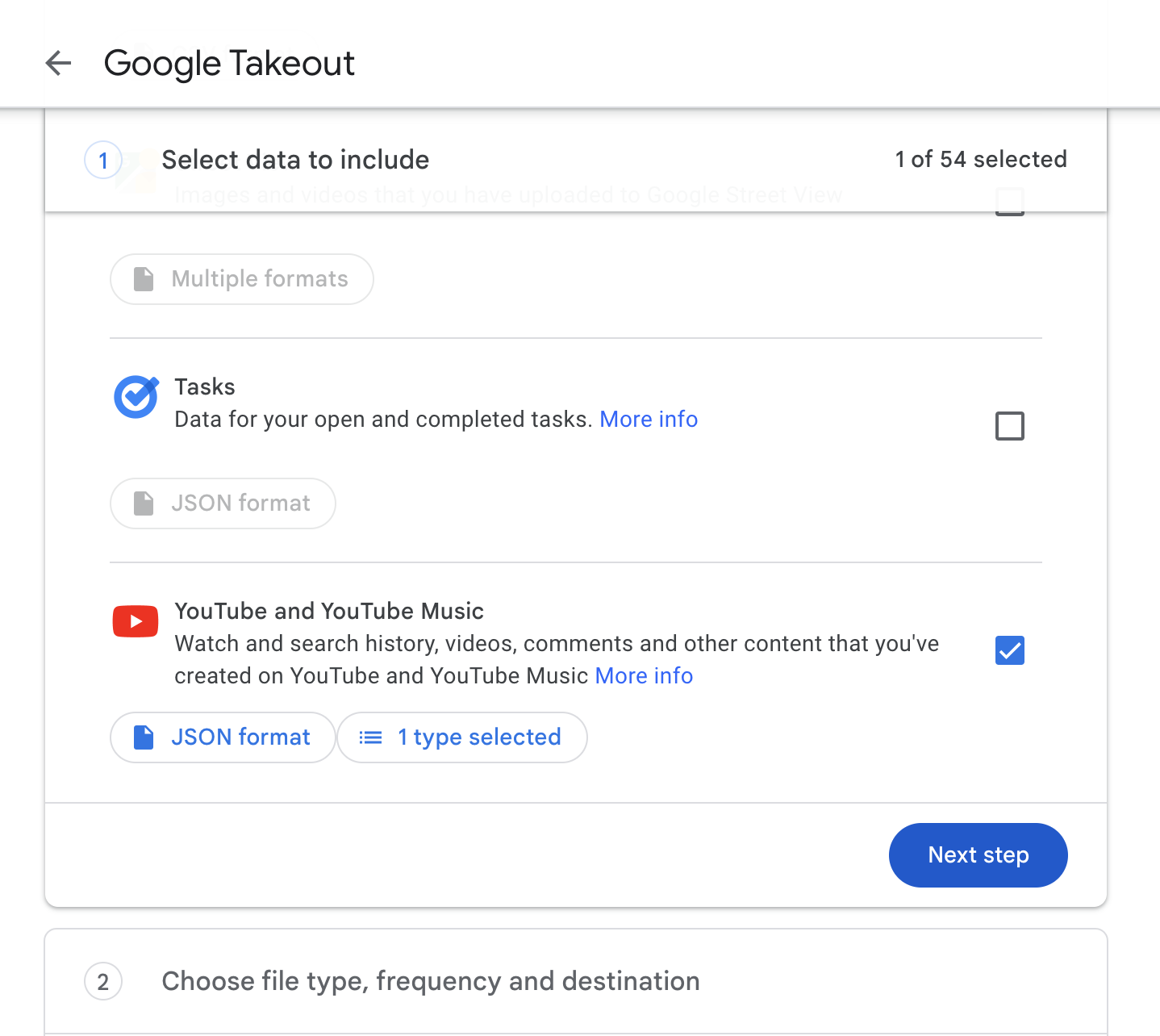}
        \caption{Select `YouTube'}
        \label{fig:guide-6}
       \end{subfigure}
       \hspace*{\fill}
      \begin{subfigure}{0.3\textwidth}
        \centering
        \includegraphics[width=0.9\textwidth]{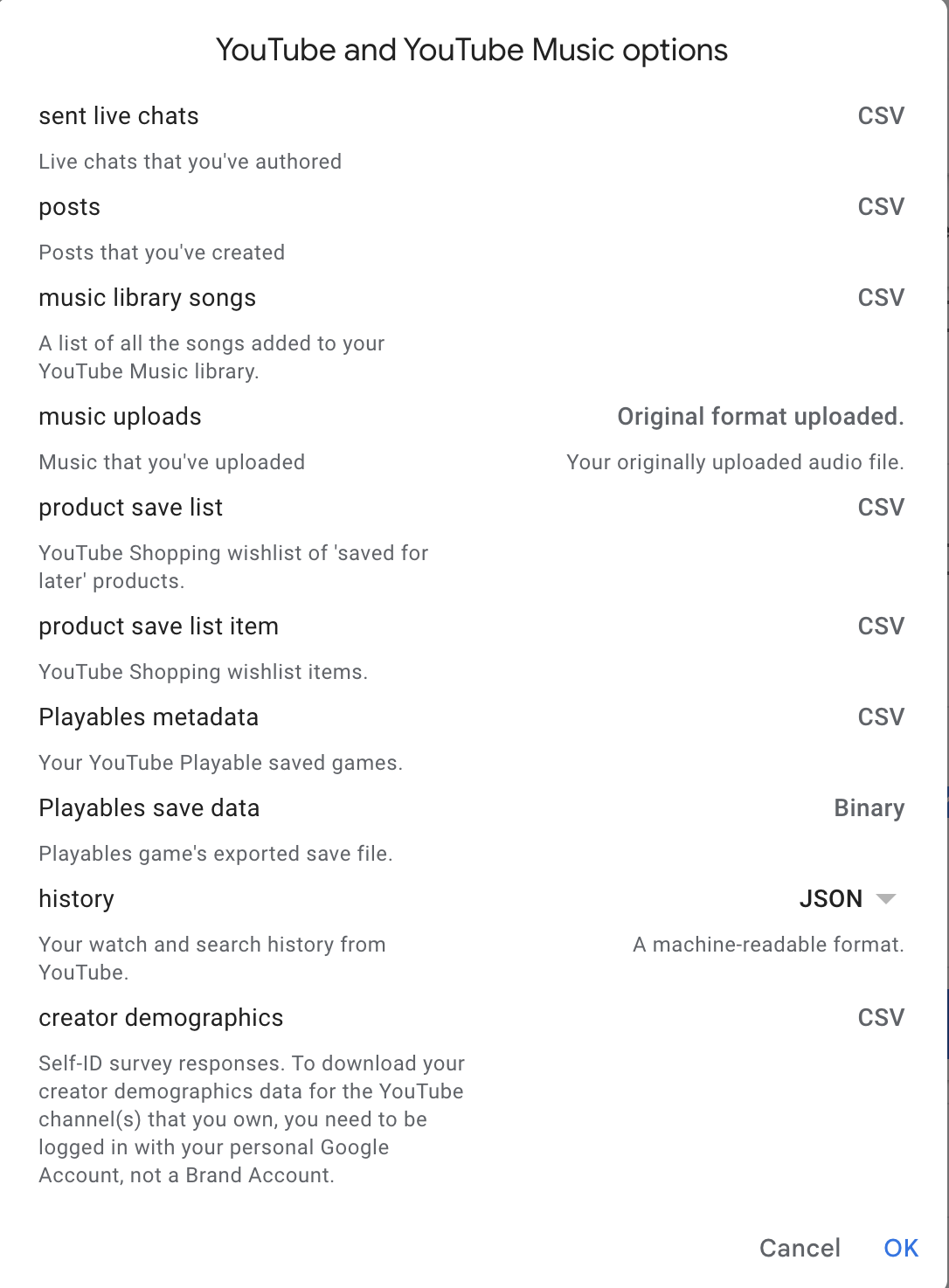}
        \caption{select `JSON' for `history'}
        \label{fig:guide-7a}
       \end{subfigure}
       \hspace*{\fill}
       \begin{subfigure}{0.3\textwidth}
        \centering
        \includegraphics[width=0.9\textwidth]{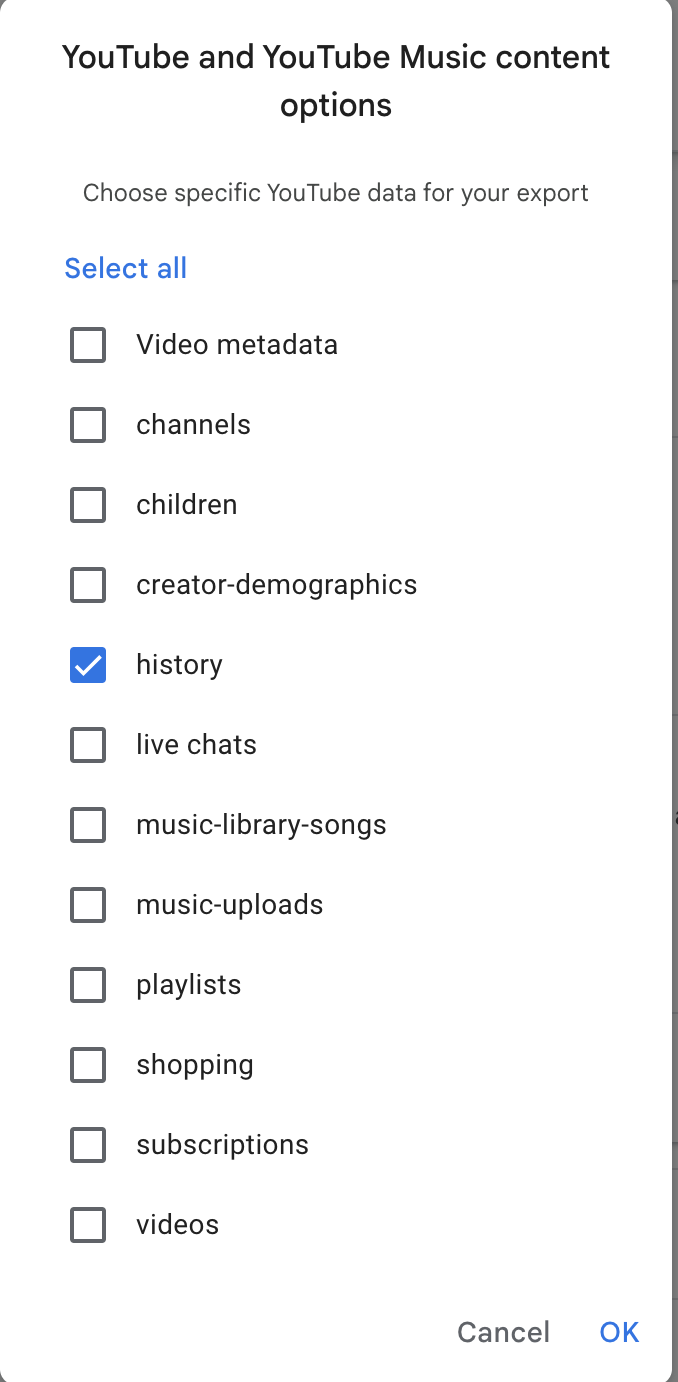}
        \caption{select `history' only}
        \label{fig:guide-7b}
       \end{subfigure}
       \caption{Step for Selection}
       \label{fig:guide-7}
    \end{figure}
    
    \item Click `next step', and click `creat export'.
    \item Click the link received in your email and Download your file.
    \item Unzip the downloaded installation package, find `watch-history.json', and upload it to EmoTrack. (Figure \ref{fig:guide-8})

    \begin{figure}
        \centering
        \includegraphics[width=0.5\linewidth]{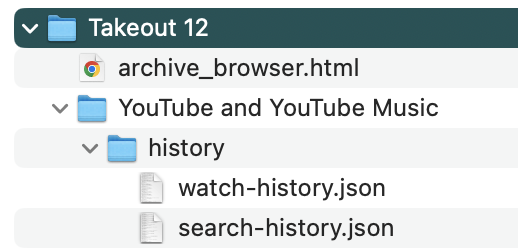}
        \caption{Find `watch-history.json'}
        \label{fig:guide-8}
    \end{figure}
\end{enumerate}

\chapter{Flow Chart of EmoTrack}
\label{appx:workflow}
\begin{figure}
    \centering
    \includegraphics[height=0.9\textheight]{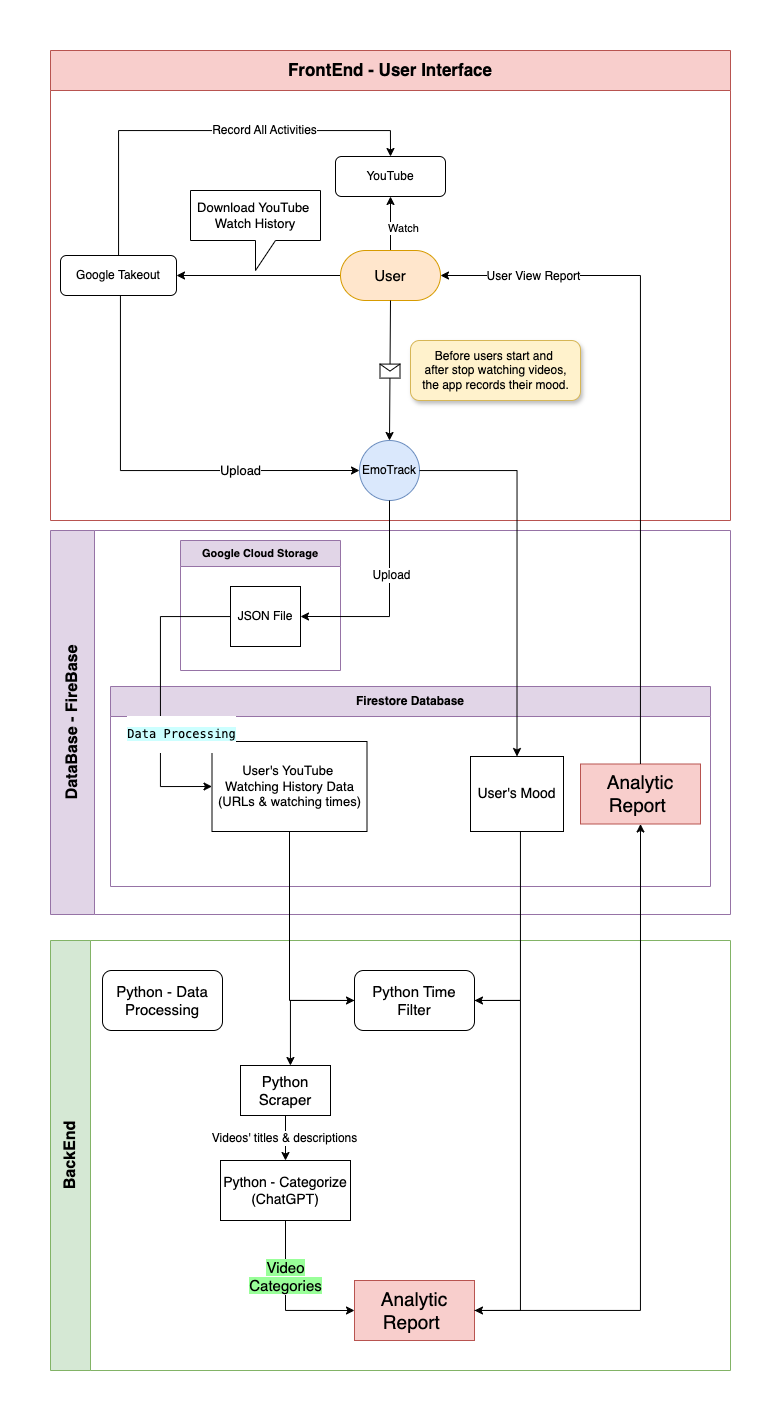}
    \caption{Flow Chart of EmoTrack}
    \label{fig:flowchart-details}
\end{figure}
\end{document}